\pdfoutput=1
\documentclass{aa}  
\usepackage{graphicx}
\usepackage{natbib}
\usepackage{txfonts}

\usepackage[dvipsnames]{xcolor}
\usepackage[colorlinks=true, linkcolor={blue!70!black}, citecolor={blue!70!black}, urlcolor={blue!70!black}]{hyperref}

\begin{document}

   \title{3D tomography of the giant Ly$\alpha$ nebulae of  $z$$\approx$3--5 radio-loud AGN}
   \author{Wuji Wang  \inst{\ref{inst1}}
        \and
        Dominika Wylezalek\inst{\ref{inst1}}
        \and 
        Jo\"{e}l Vernet\inst{\ref{inst2}}
        \and
        Carlos De Breuck\inst{\ref{inst2}}
        \and 
        Bitten Gullberg\inst{\ref{insBG1}}\fnmsep\inst{\ref{insBG2}}
        \and
        Mark Swinbank\inst{\ref{instMS}}
        \and
        Montserrat Villar Mart\'{i}n\inst{\ref{instMVM}}
        \and
        Matthew D. Lehnert\inst{\ref{insML}}
        \and
        Guillaume Drouart\inst{\ref{insAU}}
        \and
        Fabrizio Arrigoni Battaia\inst{\ref{insFAB}}
        \and
        Andrew Humphrey\inst{\ref{instPT}}
        \and
        Ga\"{e}l Noirot\inst{\ref{instGN}}
        \and
        Sthabile Kolwa\inst{\ref{insSK}}
        \and 
        Nick Seymour\inst{\ref{insAU}}
        \and
        Patricio Lagos\inst{\ref{instPT}}
          }

   \institute{   Astronomisches Rechen-Institut, Zentrum f\"{u}r Astronomie der Universit\"{a}t Heidelberg, M\"{o}nchhofstr. 12-14, D-69120 Heidelberg, Germany\label{inst1}\\
                \email{wuji.wang@uni-heidelberg.de}$;\,\,$\email{wuji.wang\_astro@outlook.com}
   \and
   European Southern Observatory, Karl-Schwarzchild-Str. 2, D-85748 Garching, Germany\label{inst2}
    \and
    Cosmic Dawn Center\label{insBG1}
    \and
   DTU Space, Technical University of Denmark, Elektronvej 327, DK-2800 Lyngby, Denmark\label{insBG2}
   \and
   Centre for Extragalactic Astronomy, Department of Physics, Durham University, South Road, Durham DH1 3LE, UK\label{instMS}
   \and
   Centro de Astrobiolog\'{\i}a, CSIC-INTA, Ctra. de Torrej\'{o}n a Ajalvir, km 4, 28850 Torrej\'{o}n de Ardoz, Madrid, Spain\label{instMVM}
   \and
   Univ. Lyon, Univ. Lyon1, ENS de Lyon, CNRS, Centre de Recherche Astrophysique de Lyon UMR5574, 69230 Saint-Genis-Laval, France\label{insML}
   \and
   International Centre for Radio Astronomy Research, Curtin University, 1 Turner Avenue, Bentley, WA 6102, Australia\label{insAU}
   \and
   Max-Planck-Institut fur Astrophysik, Karl-Schwarzschild-Str 1, D-85748 Garching bei M\"{u}nchen, Germany\label{insFAB}
   \and
   Instituto de Astrof\'{\i}sica e Ci\^{e}ncias do Espa\c{c}o, Universidade do Porto, CAUP, Rua das Estrelas, PT4150-762 Porto, Portugal\label{instPT}
   \and
   Institute for Computational Astrophysics and Department of Astronomy \& Physics, Saint Mary’s University, 923 Robie Street, Halifax, NS B3H 3C3, Canada\label{instGN}
   \and
   Physics Department, University of Johannesburg, 5 Kingsway Ave, Rossmore, Johannesburg 2092, South Africa\label{insSK}
             }

   \date{Received xxxx; xxxx}

 
  \abstract{Ly$\alpha$ emission nebulae are ubiquitous around high-redshift galaxies and are tracers of the gaseous environment on scales out to $\gtrsim100$~pkpc (proper kiloparsec). High-redshift radio galaxies (HzRGs, type-2 radio-loud quasars) host large scale nebulae observed in the ionised gas differ from those seen in other types of high-redshift quasars.
  In this work, we exploit MUSE observations of Ly$\alpha$ nebulae around eight HzRGs ($2.92<z<4.51$). All the HzRGs have large scale Ly$\alpha$ emission nebulae with seven of them extended over $100$ pkpc at the observed surface brightness limit ($\sim2-20\times10^{-19}\,\rm{erg\,s^{-1}\,cm^{-2}\,arcsec^{-2}}$). Because the emission line profiles are significantly affected by neutral hydrogen absorbers across the entire nebulae extent, we perform an absorption correction to infer maps of the intrinsic Ly$\alpha$ surface brightness, central velocity and velocity width, all at the last scattering surface of the observed Ly$\alpha$ photons. We find: (i) The intrinsic surface brightness radial profiles of our sample can be described by an inner exponential profile and a power law in the low luminosity extended part; (ii) our HzRGs have higher surface brightness and more asymmetric nebulae than both radio-loud and radio-quiet type-1 quasars; (iii) intrinsic nebula kinematics of four HzRGs show evidence of jet-driven outflows but we find no general trends for the whole sample; (iv) a relation between the maximum spatial extent of the Ly$\alpha$ nebula and the projected distance between the AGN and the centroids of the Ly$\alpha$ nebula; (v) an alignment between radio jet position angles and the Ly$\alpha$ nebula morphology. All of these findings support a scenario in which the orientation of the AGN has an impact on the observed nebular morphologies and resonant scattering may affect the shape of the surface brightness profiles, nebular kinematics and relations between the observed Ly$\alpha$ morphologies. Furthermore, we find evidence showing that the outskirts of the ionised gas nebulae may be `contaminated' by Ly$\alpha$ photons from nearby emission halos and that the radio jet affects the morphology and kinematics of the nebulae. 
  Overall, this work provides results which allow us to compare Ly$\alpha$ nebulae around various classes of quasars at and beyond Cosmic Noon ($z\sim3$).
  
  }



   \maketitle


\section{Introduction}\label{sec:introduction}
Being the most abundant element in the Universe, hydrogen (especially the cold gas, i.e. neutral hydrogen atom and molecular hydrogen, $\rm H_{2}$) is the building block of the baryonic Universe. 
Studying $\rm H_{2}$ directly is difficult due to lack of prominent transition lines. It is often probed using low-$J$ CO transitions as a proxy which unfortunately results in added uncertainties, for example in the conversion factor \citep[e.g.][]{Bolatto_2013}. In contrast, neutral atomic hydrogen can be easily ionised ($E_{\rm H^{0}}=13.6\,\rm{eV}$) and cascade with line emissions being produced. The \ion{H}{i} Ly$\alpha\lambda1216$ (Ly$\alpha$ hereafter) line is the most prominent one among them. For high-redshift galaxies, it is a commonly targeted emission line which can easily be observed in the optical to near-infrared bands  \citep[e.g.][and reference therein]{Hu_McMahon_1996,Cowie_1998,Shimasaku_2006,Dawson_2007,Leclercq_2017,Wisotzki_2018,Umehata_2019,Ono_2021,Ouchi2020}. Ly$\alpha$ emission can be detected on a range of spatial scales, for example at interstellar medium (ISM) to circumgalactic medium \citep[CGM,][]{Tumlinson2017} scales and even beyond the viral radius of the central object out to intergalactic medium (IGM) scales \citep[e.g.][]{Cantalupo2014,cai2019,Ouchi2020}. However, it is non-trivial to identify the origin of Ly$\alpha$ emission (e.g. due to resonant nature of Ly$\alpha$ emission and various potential ionising source acting at once), which is essential to understanding the physics of the emitting gas observed on different scales and around various types of objects \citep[][]{Dijkstra2019,Ouchi2020}. This is further complicated when AGN (active galactic nuclei) are present.

Active galaxies hosting AGN, especially the ones with quasar level activities (bolometric luminosity, $L_{\rm bol}\gtrsim 10^{45}\rm{\,erg\,s^{-1}}$), at high-redshift are known to host Ly$\alpha$ nebulae on scales of a few 100~kpc \citep[e.g.][]{Heckman_1991b,Basu-Zych_2004,Weidinger_2004,Weidinger_2005,Dey_2005,Prescott_2015,Cantalupo2014,FAB2016,borisova2016,cai2019,FAB2019}. The central powerful AGN act as a main ionising mechanism for the surrounding gas which is responsible for the detection of these extended Ly$\alpha$ nebulae \citep[as predicted by theoretical works, e.g.][]{Costa2022}. In addition, the diffuse emission from galaxies nearby to the AGN host can also contribute to the overall profile observed of the central target \citep[e.g.][]{Byrohl_2021}. In some of the giant nebulae, it is natural to find various mechanisms functioning at different scales and positions \citep[e.g.][]{vernet2017}. Therefore, despite leaving internal physics entangled, Ly$\alpha$ acts as a simpler tool for detecting gaseous environment through out cosmic time.

Before wide field integral field spectrographs (IFS) became available, narrow-band imaging and long slit spectroscopy provided effective methods to detect diffuse Ly$\alpha$ nebulae \citep[e.g.][]{Steidel_2000,Francis_2001,Matsuda_2004,Saito_2006,Yang_Yujin2009,Yang_Yujin2010,Cantalupo_2012,Cantalupo2014,Hennawi_2013,Prescott_2015,FAB2016}. However, these observations have been limited by uncertainties in the systemic redshift measurements and limited spatial coverage, respectively. Integral field unit observations (e.g. Multi-Unit Spectroscopic Explorer or MUSE/VLT and Keck Cosmic Web Imager or KCWI/Keck) allow us to measure the extent of the nebulae together with the information of their dynamics. Numerous works of Ly$\alpha$ nebulae around quasars report (10s of kpc to over 100 kpc) extended emission across a large range of redshifts ($z\sim2$ to $z\sim6.3$) and quasar types \citep[e.g. radio-quiet and radio-loud type-1, radio-quiet type-2 and extremely red quasar, ERQ,][]{Christensen_2006,borisova2016,FAB2019,cai2019,Farina2019,denbrok2020,Fossati2021,mackenzie2021,Lau_2022,Vayner_2023,Zhangshiwu_2023}. This diversity in nebula properties suggest a range of driving mechanisms, dependencies on orientation, and demonstrate that well-selected samples are needed. Despite the effort has been done on this topic, a link between the aforementioned types and type-2 radio-loud quasars on CGM scale is missing. 

Among the high-redshift quasar population, high-redshift radio galaxies (HzRGs) are a unique sample despite smaller in number \citep[see ][as a review]{Miley_2008}. They host type-2 quasars and have powerful radio jets. They have been shown to reside in dense protocluster environments \citep[][]{Venemans2007a,Wylezalek2013b,Wylezalek_2014,Noirot_2016,Noirot_2018} which may evolve to modern galaxy clusters. HzRGs were among the first sources where giant Ly$\alpha$ nebulae were discovered \citep[$\sim 10^{44}\,\rm{erg\,s^{-1}}$, $\gtrsim100\,\rm{kpc}$, e.g. ][]{Hippelein_1993, vanojik1996, vanojik1997a, Reuland_2003b,Villarmartin_2006a, Villarmartin_2007b} and observed with the previous generation of IFU instruments \citep[e.g.][]{Adam_1997}.
The Ly$\alpha$ nebulae of HzRGs have been found to have two distinctive parts, namely the high surface brightness kinematically disturbed inner part and the quiescent low surface brightness extended outer nebula \citep[e.g.][]{villarmartin_2002,villarmartin2003,villarmartin_2007a}. The spatial separation of these two parts seem to be consistent with the extent of the radio jets \citep[e.g.][]{villarmartin2003} suggesting that the jet plays a role in disturbing the inner part. Specifically, there is evidence that the Ly$\alpha$ nebulae around HzRGs are related to jet-driven outflows \citep[][]{Humphrey_2006} while some of the quiescent gas may be related to infalling material \citep[][]{Humphrey_2007a}. AGN photoionisation is likely the main mechanism of exciting these nebulae \citep[e.g.][]{villarmartin_2002,villarmartin2003,Morais_2017}, but ionisation by fast shocks might also play a role \citep[e.g.][]{Bicknell_2000,Morais_2017}. Polarisation measurements show that some of the Ly$\alpha$ emission in HzRGs is scattered \citep{Humphrey_2013b}. Despite these works, however, a comparison of the nebulae of HzRGs and other quasar samples has yet to be performed which is the motivation of this work.

The Ly$\alpha$ nebulae of HzRGs are known to be partially absorbed by neutral hydrogen \citep[\ion{H}{i} absorbers, e.g.][]{Rottgering_1995,vanojik1997a,jarvis2003,wilman2004,Humphrey_2008b,kolwa2019}.The absorbing gas is found to be extended on galaxy-wides scales and likely related to outflowing gas from the host galaxy \citep[e.g.][]{Binette_2000,swinkbank2015,Silva_2018b,wang2021}. The correction of these absorption is only possible through spectral observation. Without careful treatment, a considerable amount (a factor of $\gtrsim5$) of flux will be missed, and inaccurate conclusions will be derived. Alternatively, some absorption trough features might potentially be explained by radiative transfer effects \citep[][]{Dijkstra_2014,gronke2015c,gronke2016a,gronke2016b}. Although it is interesting to compare the different treatments of the observed Ly$\alpha$ spectra, it is beyond the scope of this work. 

There was also clear observational evidence that the morphology of the continuum and line emission regions of HzRGs are aligned with jet direction \citep[e.g.][]{Chambers_1987, Pentericci_1999,Miley_2004,Zirm_2005,Duncan_2022} on relatively smaller scale (several kpc to 10s of kpc). Molecular gas detected around HzRGs was reported to be distributed along the jet within and outside the hot spot which may suggest several scenarii \citep[e.g. jet-driven outflow, jet-induced gas cooling and jet propagating into dense molecular gas medium, ][]{Emonts_2014a,Gullberg_2016a,falkendal2021}. On Mpc scale, \citet{West_1991} found that the radio jet often points towards nearby galaxies. \citet{Eales_1992} proposed a model explaining the alignment effect, suggesting that the high-redshift radio emission is often detected when the jet travels close to the major axis of surrounding asymmetrically distributed gas. With the advanced IFS observation and 100s kpc gas tracer of Ly$\alpha$, we are able to probe the intrinsic (i.e. corrected for absorption) gaseous nebula around HzRGs in this work, test its distribution with respect to the radio jets and seek evidence following these pioneering works.

In this paper, we utilise the power of MUSE IFU to fully map the Ly$\alpha$ emission nebulae of a HzRGs sample over a redshift range of $2.92-4.51$ and initiate a comparison with type-1 quasars and study of CGM-scale environments. We introduce our sample of HzRGs, the MUSE observations and data reduction in Sect. \ref{sec:0sam_obs_red}. We present how we measure the maximum extent of the nebulae in Sect. \ref{sec:1nebuseltes} and summarise the spectral fitting procedure in Sect. \ref{sec:1specfitmeth}. We then present the results of surface brightness, kinematics and morphology in Sect. \ref{sec:0lyaradasy} followed by a discussion in Sect. \ref{sec:0discuss}. Finally, we conclude in Sect. \ref{sec:0conclusions}.
In this paper, we assume a flat $\Lambda$CDM cosmology with $H_{0} = 70\, \rm{km\,s^{-1}\,Mpc^{-1}}$ and $\Omega_{m}=0.3$. Following this cosmology, $\rm{1\,arcsec\simeq6.6-7.7\, pkpc}$ for our sample redshifts. Throughout the paper, pkpc stands for proper kiloparsec and ckpc represents comoving kiloparsec, ckpc$=(1+z)$pkpc. In this paper, we use `intrinsic' to refer to the absorption corrected Ly$\alpha$ emission.

\begin{table*}
 \caption{Details of the MUSE observation of the HzRG sample.}\label{tab:sampleobs}
 \centering
\begin{tabular}{c c c l c c l}
\hline
 HzRG  & Redshift   & UT date     & Program ID & Mode & Exp. Time total  & seeing \\
      & $z$\tablefootmark{\rm{\dag}}     & (dd/mm/yyyy)&            &      & hours            & arcsec\tablefootmark{\rm{*}} \\
\hline
MRC 0943-242  & 2.9230 & 21/02/2014            & 60.A-9100(A)  & WFM-NOAO-E & 5.21    & 0.65\tablefootmark{a} \\
     -        &  -     & 15/12/2015-18/01/2016 & 096.B-0752(A) & -          &  -      & -     \\
MRC 0316-257  & 3.1238 & 15/01-17/01/2015      & 094.B-0699(A) & WFM-NOAO-N & 4.24    & 0.61\tablefootmark{b}   \\
TN J0205+2242 & 3.5060 & 03/12-08/12/2015      & 096.B-0752(B) & WFM-NOAO-N & 4.24    & 0.73 \\
TN J0121+1320 & 3.5190 & 06/10/2015            & 096.B-0752(C) & WFM-NOAO-N & 5.30    & 0.83 \\
      -       &  -     & 08/08-28/08/2016      & 097.B-0323(C) & -          & -       & -    \\
4C+03.24       & 3.5828 & 17/06-18/06/2017      & 60.A-9100(G)  & WFM-AO-N   & 1.25    & 0.63  \\
4C+19.71       & 3.5892 & 08/06-02/09/2016      & 097.B-0323(B) & WFM-NOAO-N & 5.83    & 1.03  \\
TN J1338-1942 & 4.0959 & 30/04-06/05/2014      & 60.A-9100(B)  & WFM-NOAO-N & 8.93    & 0.77\tablefootmark{a} \\
    -         &  -     & 30/06/2014            & 60.A-9318(A)  & -          & -       & -     \\      
4C+04.11       & 4.5077 & 03-15/12/2015         & 096.B-0752(F) & WFM-NOAO-N & 4.24    & 0.88 \\
 \hline
\end{tabular}
\tablefoot{
\tablefoottext{\rm{\dag}}{The redshifts are determined from the \ion{He}{ii}$\lambda1640\AA$ or [\ion{C}{i}](1-0) emission line \citep[][]{kolwa2023}. For \object{MRC0316-257} which is not included in \citet{kolwa2023}, we reported its $z_{\rm sys}$ in this paper based on \ion{He}{ii} fit from our MUSE data (Appendix \ref{app:z_sys0316}).}\\
\tablefoottext{\rm{*}}{The seeing reported here is determined from the fitted 2D Moffat FWHM (full width at half maximum) of a star in the white-light image ($5000-9000\,\AA$) produced from the combined cube. We note that the stars used are red in color, i.e. the image quality in the Ly$\alpha$ wavelength should in general be worse than the reported seeing (e.g. larger by 10 to 20\%).}\\
\tablefoottext{a}{The seeing is determined from a star in the overlapping region of the two pointings.}\tablefoottext{b}{There is no available star in the FoV. The seeing is determined from the fit of the most point-like source.}\\
}
\end{table*}

\section{HzRGs sample, observations and data processing}\label{sec:0sam_obs_red}
\subsection{MUSE HzRGs sample}\label{sec:1musesample}
\subsubsection{Sample selection}\label{sec:2sample_sel}
The 8 HzRGs at 2.92$<z<$4.51 (Table \ref{tab:sampleobs}) that we investigate in this paper were selected to (i) be at $z> 2.9$ for Ly$\alpha$ to be covered by MUSE ; (ii) have a known extended bright Ly$\alpha$ ($>10\arcsec$) emission nebula; and (iii) be at DEC $<$ 25$^{\circ}$ to be observable by ground-based telescopes in the southern hemisphere.
This sample also has a wealth of high quality supporting data obtained by our team, including deep \textit{Spitzer}/IRAC and \textit{Spitzer}/MIPS 24 $\mu$m imaging, and \textit{Herschel}/SPIRE detections \citep[][]{seymour2007,debreuck2010}. ALMA Band 3 or 4 data are also available for the sample targeting dust continuum and molecular lines \citep[][]{falkendal2019,kolwa2023}. Being identified as radio galaxies, the radio observations \citep[e.g. VLA,][]{Carilli1997} provide information on the jet morphology and polarisation. Based on these supporting data sets, we have estimates of the total stellar mass of the host galaxies \citep[several $10^{11} \rm M_{\odot}$ for all targets, ][]{debreuck2010} and the star formation rates ranging from uppers limit of $<84\,\rm{M_{\odot}\,yr^{-1}}$ to constraints of 626 $\rm{M_{\odot}\,yr^{-1}}$ \citep[][]{falkendal2019}.


\subsubsection{AGN bolometric luminosity estimation}\label{sec:2Lbol}

To put the HzRGs into context with other quasar species, we plan to link our Ly$\alpha$ nebulae to literature works based on AGN bolometric luminosity. There are different methods for estimating the bolometric luminosity of AGN, $L_{\rm{bol,\,AGN}}$, for example through scaling of the far-IR AGN-heated dust luminosity \citep[e.g.][]{drouart2014}, scaling the IR flux density \citep[e.g. $f_{3.45\,\rm{\mu m}}$ which is used for type-1 quasars,][]{Lau_2022} and through [\ion{O}{iii}] emission \citep[which can be affected by star formation and/or shocks][]{reyes2008, Allen2008}. However, there is a large uncertainty between the values derived through these different methods which makes it non-trivial to directly compare the $L_{\rm{bol,\,AGN}}$ of type-1s and type-2s. For instance, the estimates for type-2 AGN are affected by obscuration by the dusty torus assuming the AGN unification model \citep[e.g.][]{Antonucci1993}. Accounting for this by applying an extinction correction factor would lead to a large uncertainty \citep[e.g.][]{drouart2012} if we were to use the same method for type-1s to estimate the $L_{\rm{bol,\,AGN}}$ for our sample. We report that the $L_{\rm{bol,\,AGN}}$ estimated for our sample using those different methods varies from $10^{45.9}$ to $10^{48.5}\,\mathrm{erg\,s^{-1}}$. Given this large uncertainty, we find it is unreasonable to draw further conclusions from the comparison of $L_{\rm{bol,\,AGN}}$ between type-1s and our HzRGs. However, it is worthwhile to report this estimation procedure and the resulted inconsistency under different assumptions. A systematic study of the $L_{\rm{bol,\,AGN}}$ is beyond the scope of this work and may be done more thoroughly through multi-wavelength approach.

\subsubsection{Jet kinematics}\label{sec:2jetk}
To distinguish between the approaching and receding sides of the jet, we use the kinematics information from [\ion{O}{iii}] as a proxy which is often used for studying quasar outflow \citep[e.g.][]{Veilleux2005,Zakamska2016b,nesvadba2017a,nesvadba2017b,Vayner2021b}. 5 out of 8 of our sample targets have been observed by SINFONI from which the [\ion{O}{iii}] velocity shifts are available \citep[][]{nesvadba2007b,nesvadba2008b,nesvadba2017a}. For \object{MRC0943-242} and \object{TN J1338-1942}, we use the radio hot spot polarisation information as indicator where the more depolarised indicates the far side (receding) of the jet \citep[][]{Carilli1997,Pentericci2000}. These are also consistent with the tentative [\ion{O}{ii}] velocity gradient of \object{TN J1338-1942} found in \citet{nesvadba2017a} \citep[also \ion{He}{ii} kinematics in][]{kolwa2023} and \object{MRC0943-242} \ion{He}{ii}$\lambda1640\AA$ (\ion{He}{ii}) kinematics in \citet{kolwa2019}. For \object{4C+04.11}, \citet{parijskij2014} gives the jet kinematics based on high-resolution radio polarisation. We note here that the reported approaching and receding directions based on the current observations should be treated with caution. The polarisation of the radio lobes could especially be affected by the intervening ionised structures. We also quantified the size of the jets by calculating the angular distance between the jet hot spots on either side to the AGN position (presented in Appendix \ref{app:nebradinfo}). 


\subsection{MUSE observations}\label{sec:1museobs}
In this work, we analyse data from MUSE integral field spectrograph \citep[][]{Bacon_2010, bacon2014} mounted on the ESO Very Large Telescopes (VLT) Yepun (UT4). All observations were carried-out in Wide-Field Mode (WFM) offering a 1$\times$1 arcmin$^{2}$ field of view and spatial sampling of 0.2 arcsec pixel$^{-1}$. MUSE provides two sets of wavelength coverage: a nominal range (N, 480$-$930 nm) and an extended range (E, 465$-$930 nm) without using of the adaptive optics (AO). For observations carried in AO mode, the wavelength coverage of 582$-$597 nm is excluded due to the Na Notch filter. The MUSE spectrograph has the spectral sampling of 0.125 nm pixel$^{-1}$ and resolving power of 1750$-$3750 for 465$-$930 nm which corresponds to $\Delta v \sim 171-90 \,\rm{km\,s^{-1}}$. 

The observations of our sample were carried mostly in service mode under the program IDs 094.B-0699, 096.B-0752 and 097.B-0323 (PI: J. Vernet). For \object{MRC 0943-242}, we also include the data of MUSE commissioning observation under the program ID 60.A-9100(A) \citep[e.g. ][]{Gullberg_2016a}. The extended wavelength coverage was employed for \object{MRC 0943-242}, the lowest redshift sample target, to cover its Ly$\alpha$ emission ($L_{\rm{Ly\alpha,\, obs}} = 4769\,\AA$). We use the MUSE commissioning and science verification data of \object{TN J1338-1942} under the program IDs 60.A9100(B) and 60.A-9318(A) \citep[e.g. ][]{swinkbank2015}. For \object{4C+03.24}, we adopt the data released from the MUSE WFM-AO commissioning observations under the program ID 60.A-9100(G). The information of the observations of our sample, in the order of redshift, is summarised in Table \ref{tab:sampleobs}. For each object, observations consist of 1 (4C+03.24) to 6 (TN J 1338-1942) observing blocks (OBs). Within each OB, the 2 or 3 exposures of 20$-$30 minutes were slightly dithered (with a $<1$\arcsec \,amplitude pattern) and rotated by 90 degrees from each other.

\subsection{Data processing}\label{sec:1datared}
The reduction of the raw MUSE data are carried out following the standard procedure using the MUSE pipeline \citep[][version 2.8.4]{weilbacher2020} executed by EsoRex \citep[ESO Recipe Execution Tool; ][]{esorex2015}. For studying the extended Ly$\alpha$ nebulae to the faintest edge, we reduce the data following the optimised procedure developed in our pilot study of  \object{4C+04.11} \citep[][]{wang2021}. We first reduce each exposure individually with the standard pipeline doing the sky-line subtraction and then using ZAP \citep[Zurich Atmosphere Purge, ][]{soto2016} to remove the sky-line residuals (see below details regarding the ZAP execution). We then combine all exposures to the final data cube using MPDAF \texttt{Cubelist.combine} \citep[MUSE Python Data Analysis Framework][]{bacon2016}. We correct the astrometry of the final combined cubes using star positions from the available \textit{Gaia} EDR3 catalogue \citep[Early Data Release 3, ][]{gaia2021edr3}. Two sources had no GAIA star within the MUSE field-of-view (FoV). For \object{TN J0121+1320} we use the SDSS DR16 \citep[16th Data Release, ][]{ahumada2020} catalogue instead. For \object{MRC 0316-257}, we use \textit{Gaia} EDR3 to first correct the astrometry of the HST/ACS F814W image and then matched the MUSE cube to the HST image.

Using ZAP directly for sky-line residual removal without applying masks may remove faint narrow Ly$\alpha$ line emission at the outskirt of our sample. Since the Ly$\alpha$ nebulae in our sample extend much further beyond the continuum emission regime of the host galaxy and become narrower in line width  \citep[e.g.][]{villarmartin2003, Humphrey_2007a} such that they are mistakenly treated as sky-line residuals and removed. To alleviate this problem \citep[][]{soto2016}, for each source, we (i) generate a first version of the combined data cube without masks in the ZAP step; (ii) construct a Ly$\alpha$ mask that covers most line-emission region\footnote{We note that this mask is only used in this process to eliminate the impact of the Ly$\alpha$ signal on ZAP. The detection map for determining the maximum extent of the Ly$\alpha$ nebula is described in Sect. \ref{sec:2max_neb_ext}.}; (iii) re-run ZAP using this Ly$\alpha$ mask on individual cubes for each exposures; (iv) combine the newly obtained individual cubes to the final version data cube with MPDAF.

We also correct for small residual (mostly) negative background level offsets probably due to a slight over-subtraction of the sky continuum in previous steps. To do so, we (i) extract a median spectrum from an $r\simeq10\arcsec$ circular aperture around the radio galaxy masking all continuum sources falling in the aperture; (ii) mask the Ly$\alpha$ line emission wavelength range and strong sky-lines \citep[$>10^{16}\,\rm{erg\,s^{-1}\,cm^{-2}\,\AA^{-1}\,arcsec^{-2}}$, ][]{hanuschik2003} for this median spectrum; (iii) fit a 6th-order polynomial to this masked spectrum; (iv) subtract this solution from the whole cube. 

Finally, to correct for the known underestimation of the variance in the standard pipeline reduction \citep[see][]{weilbacher2020}, variance scaling is implemented as described in \citet{wang2021}. Specifically, we scale the variance extension propagated by the pipeline based on the scale factor calculated in source-free regions using the variance estimated from the data extension.

\begin{table*}
 \caption{HzRGs MUSE sample properties.}\label{tab:sampleinfo}
 \centering
\begin{tabular}{c l l c r c }

\hline
HzRG  & RA (J2000)\tablefootmark{\rm{\dag}}    & DEC (J2000)\tablefootmark{\rm{\dag}} & SB limit\tablefootmark{\rm{*}} &$L_{\rm{Ly\alpha,\,int}}$\tablefootmark{\rm{\ddag}} &$L_{\rm{Ly\alpha,\,obs}}$\tablefootmark{\rm{$\mathsection$}} \\ 
    & hh:mm:ss      &  dd:mm:ss   & $10^{-19}$ cgs &  $10^{44}\,\rm{erg\,s^{-1}}$ &  $10^{44}\,\rm{erg\,s^{-1}}$\\ 
\hline
MRC 0943-242  &  09:45:32.73 & $-$24:28:49.65 \tablefootmark{a}&17.0 & 4.7$\pm0.1$ & 2.39$\pm$0.03\\
MRC 0316-257  &  03:18:12.07 & $-$25:35:10.22 \tablefootmark{b}&2.62 & 7.3$\pm1.2$ & 1.04$\pm$0.04 \\
TN J0205+2242 &  02:05:10.69 & $+$22:42:50.4  \tablefootmark{a} \tablefootmark{c}&13.3 & 8.3$\pm0.4$ & 3.82$\pm$0.06 \\
TN J0121+1320 & 01:21:42.73  & $+$13:20:58.0  \tablefootmark{a}&5.64 & 2.0$\pm0.1$ & 0.55$\pm$0.01  \\
4C+03.24       & 12:45:38.37  & $+$03:23:21.0  \tablefootmark{d}&11.5 & 28.8$\pm0.8$& 5.43$\pm$0.20  \\
4C+19.71       & 21:44:07.56  & $+$19:29:14.6  \tablefootmark{a}&4.77 & 4.3$\pm0.3$ & 1.48$\pm$0.22  \\
TN J1338-1942 & 13:38:26.10  & $-$19:42:31.1  \tablefootmark{e}&4.34 & 12.3$\pm0.3$& 5.89$\pm$0.10  \\
4C+04.11       & 03:11:47.97  & $+$05:08:03.74 \tablefootmark{f}&9.84 & 20.0$\pm0.7$& 2.89$\pm$0.07  \\
 \hline
\end{tabular}
\tablefoot{\tablefoottext{\rm{\dag}}{Position of the AGN and/or host galaxy.}
\tablefoottext{\rm{*}}{The surface brightness limit is the 2$\sigma$ limit extracted from continuum-source- and Ly$\alpha$-free regions in a narrow band image. The narrow band image is collapsed from $v_{05}$ to $v_{95}$ (see Section \ref{sec:0intmaps}). The cgs unit is  $\rm{erg\,s^{-1}\,cm^{-2}\,arcsec^{-2}}$.}
\tablefoottext{\rm{\ddag}}{Intrinsic Ly$\alpha$ luminosity (i.e. corrected for absorption) of the nebula. It is integrated over the entire area selected in Sect. \ref{sec:1nebuseltes} and multiplied with the luminosity distance using the cosmological parameters (Sect. \ref{sec:introduction}).}\tablefoottext{\rm{$\mathsection$}}{Observed Ly$\alpha$ luminosity of the nebula integrated over the entire area from $v_{05}$ to $v_{95}$ (see text).}\\
References of the positions: \tablefoottext{a}{\textit{Spitzer} \citep[][]{seymour2007, debreuck2010}.} 
\tablefoottext{b}{\textit{Chandra} Obs. ID 5734 (PI:Pentericci). \textit{Chandra} Source Catalog \citep[][]{evans2010}.}
\tablefoottext{c}{Radio \citep[][]{debreuck2002}.} 
\tablefoottext{d}{Radio \citep[][]{vanojik1996}.} 
\tablefoottext{e}{Radio \citep[][]{debreuck1999}.} 
\tablefoottext{f}{Radio \citep[][]{parijskij2014}.}}
\end{table*}

  \begin{figure*}
  \centering
        \includegraphics[width=\textwidth,clip]{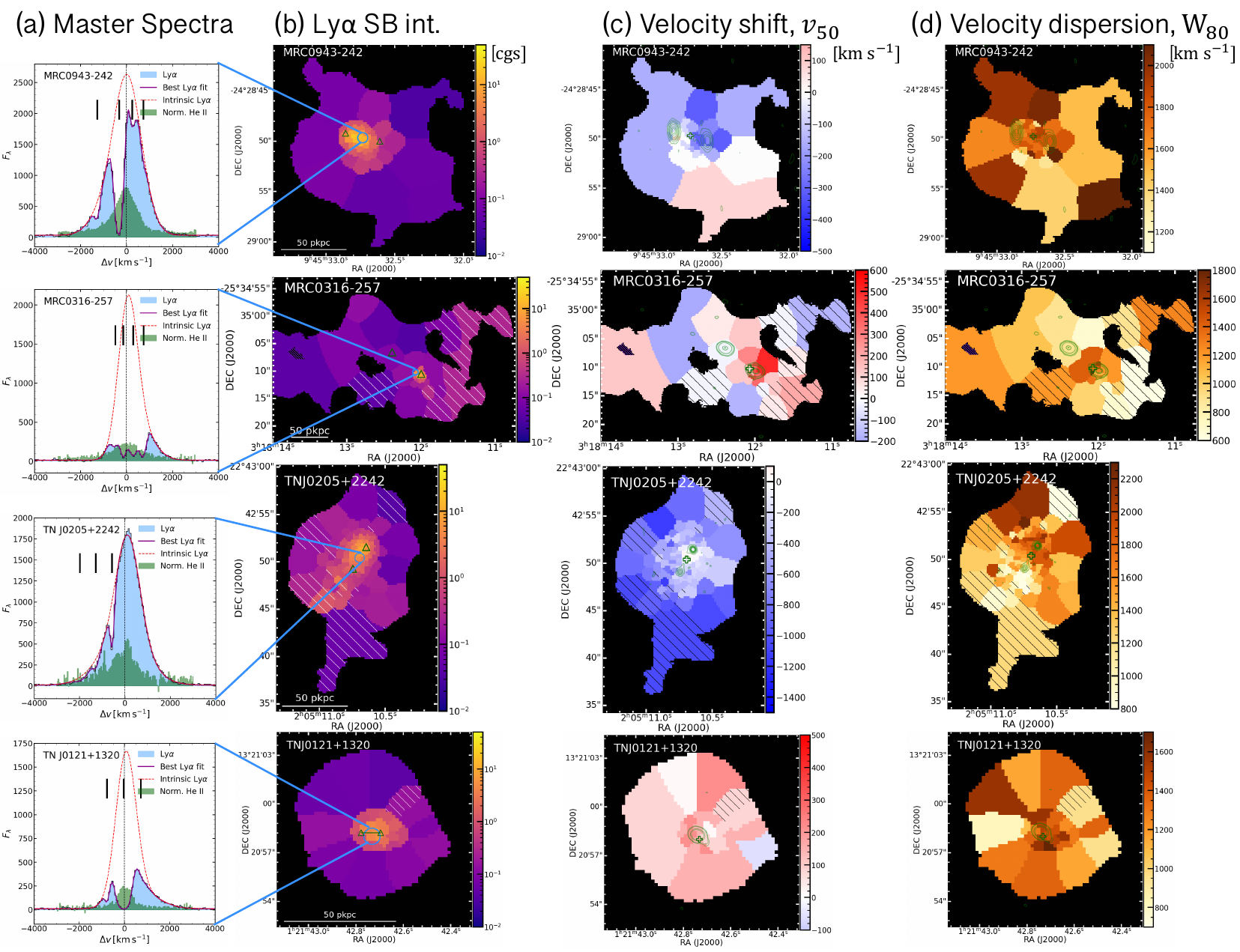}
      \caption{Mapping results of our MUSE HzRGs sample. \textbf{(a)} Master Ly$\alpha$ spectrum (blue shaded histogram) extracted from a $r=0.5\,\mathrm{arcsec}$ aperture at the AGN position with best fit (solid dark magenta line). Red dashed curve shows the intrinsic Ly$\alpha$ from fitting, i.e. corrected for absorption. The vertical black bars above the emission line mark the positions of the \ion{H}{i} absorbers. The yellow shaded region (if any) indicates the 5 wavelength pixel range excluded in the fitting due to the contamination from the $5577\AA$ sky-line. The flux density unit, $F_{\lambda}$, is $10^{-20}\,\mathrm{erg\,s^{-1}\,cm^{-2}\,\AA^{-1}}$. We also show the scaled \ion{He}{ii}$\lambda1640\AA$ spectrum extracted from the same position in green histogram. We scale the peak flux density of \ion{He}{ii} to $0.3-0.7$ (varied for different targets) of the maximum peak flux density of observed Ly$\alpha$ spectrum in $-1000$ to $1000$ $\mathrm{km\,s^{-1}}$. The $\Delta v=0\,\mathrm{km\,s^{-1}}$ is the systemic redshift based on \ion{He}{ii} or [\ion{C}{i}] \citep[Table \ref{tab:sampleobs},][]{kolwa2023}. \textbf{(b)} Intrinsic Ly$\alpha$ surface brightness map. The flux in each tile is the integrated flux of the line emission corrected for absorption, i.e. total flux of the one or two Gaussians, see Sect. \ref{sec:1specfitmeth}. The light blue circle shows the aperture where the Master spectrum is extracted from. Green triangles mark the positions of the radio lobes. We place a green bar linking the triangles on \object{TN J0121+1320} to indicate the unresolved state of its radio emission. The length of the bar represents the linear size of the $3\sigma$ contour along the east-west direction. The white hatched regions are the ones where the flux uncertainty is higher than 50\% of the fitted intrinsic flux. The white bar indicates the 50 pkpc at the redshift of the radio galaxy. The unit of the surface brightness is $10^{-16}\,\mathrm{erg\,s^{-1}\,cm^{-2}\,arcsec^{-2}}$. We apply the same color scale for all targets. \textbf{(c)} $v_{50}$ map of the intrinsic Ly$\alpha$ nebula. The zero velocity used for each target is determined by the systemic redshift (Table \ref{tab:sampleobs}). Green contours show the morphology of the radio jet in arbitrary values. The green cross mark the AGN position (Table \ref{tab:sampleinfo}). \textbf{(d)} $W_{80}$ map of the intrinsic Ly$\alpha$ nebula. The black hatched regions on (c, d) are the same as (b). The purple hatched regions (in \object{4C+03.24} and \object{TN J1338-1942}) are manually excluded due to contamination from either foreground star or known companion \citep[Arrow galaxy in the filed of \object{MRC0316-257}, see][]{vernet2017}. We note that the color scales for panel (c) and (d) are customised. The purple hatched area (if any) indicates the manually excluded region affected by foreground star or known Ly$\alpha$ emitter.}
         \label{fig:map_int_1}
         \addtocounter{figure}{-1}
  \end{figure*}

  \begin{figure*}
  \centering
        \includegraphics[width=\textwidth,clip]{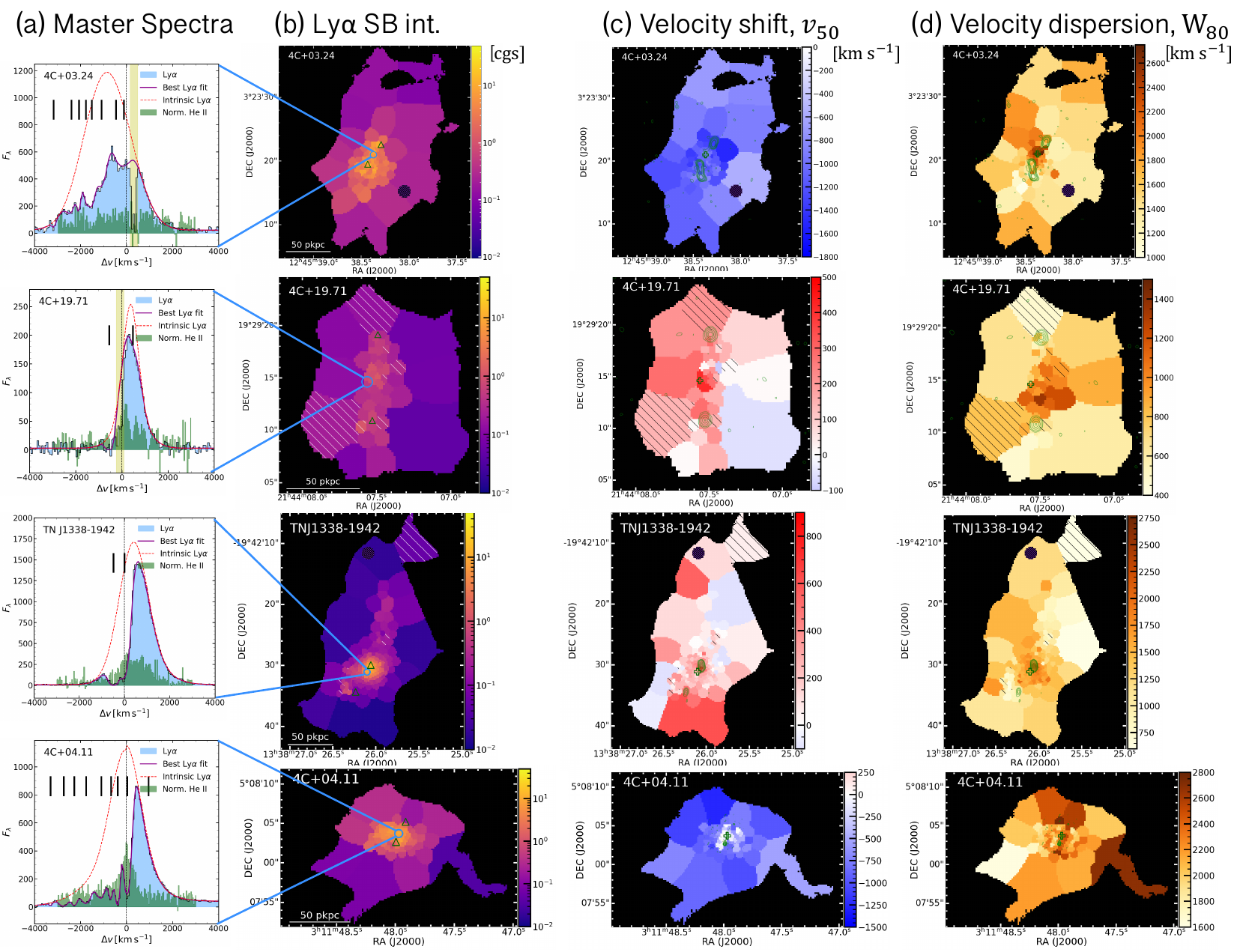}
      \caption{Continued.}
         
  \end{figure*}

  \begin{figure*}
  \centering
        \includegraphics[width=\textwidth,clip]{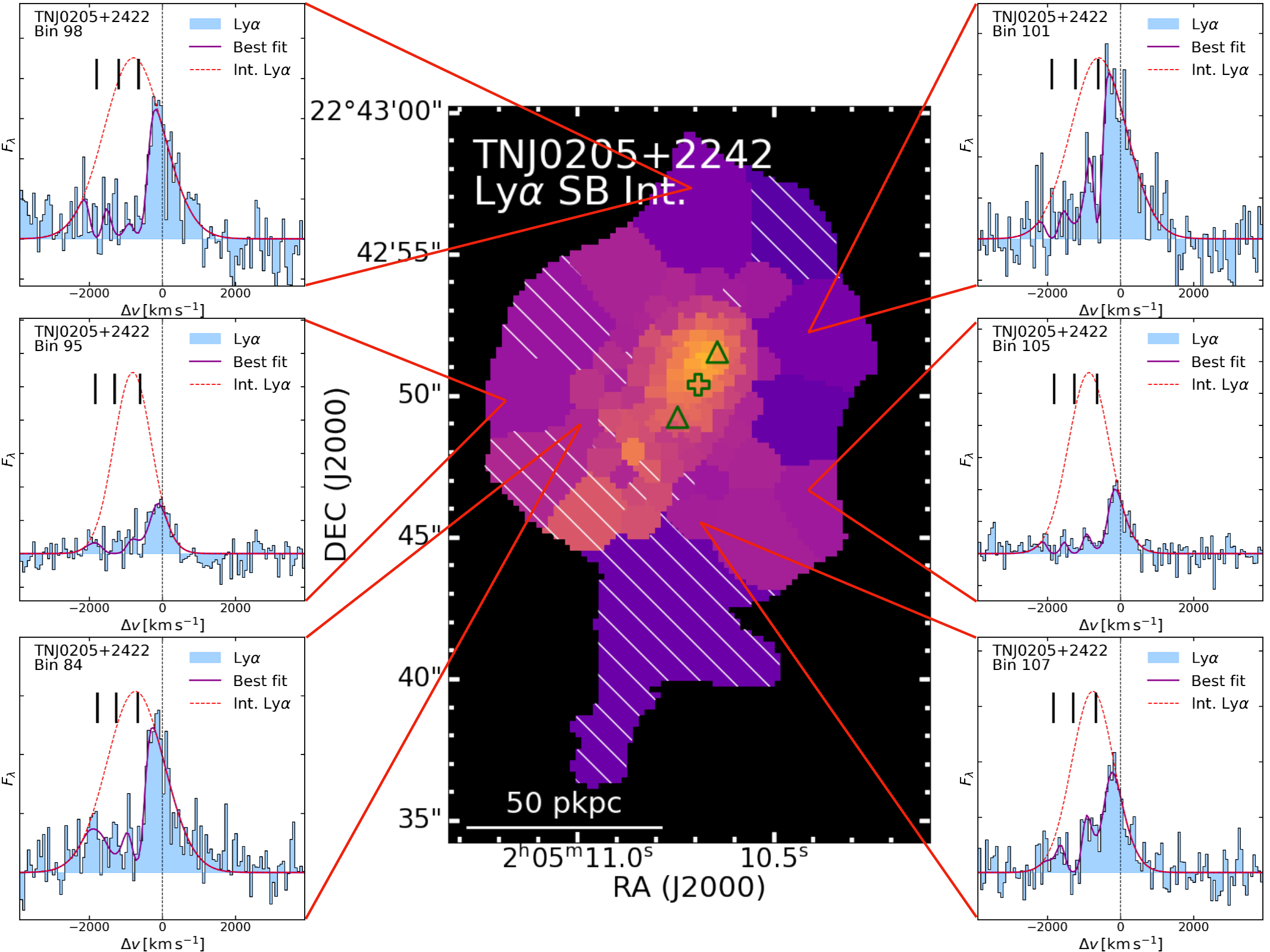}
      \caption{Example for the intrinsic mapping of the Ly$\alpha$ nebula of \object{TNJ0205+2242}. The central panel shows the intrinsic surface brightness map of \object{TNJ0205+2242} which is the same as Fig. \ref{fig:map_int_1}b. The green cross and triangles mark the position of the AGN and jet lobes, respectively. In each of the side panel, we show the spectrum (blue shade histogram in normalised flux unit) extracted from the individual spatial bin whose number is labeled at the top left, and the best fit (dark magenta curve) and recovered intrinsic Ly$\alpha$ (dashed red line). The black vertical bars indicate the positions of the \ion{H}{i} absorbers.}
        \label{fig:tnj0205_spa}
         
  \end{figure*}

  \begin{figure*}
  \centering
        \includegraphics[width=\textwidth,clip]{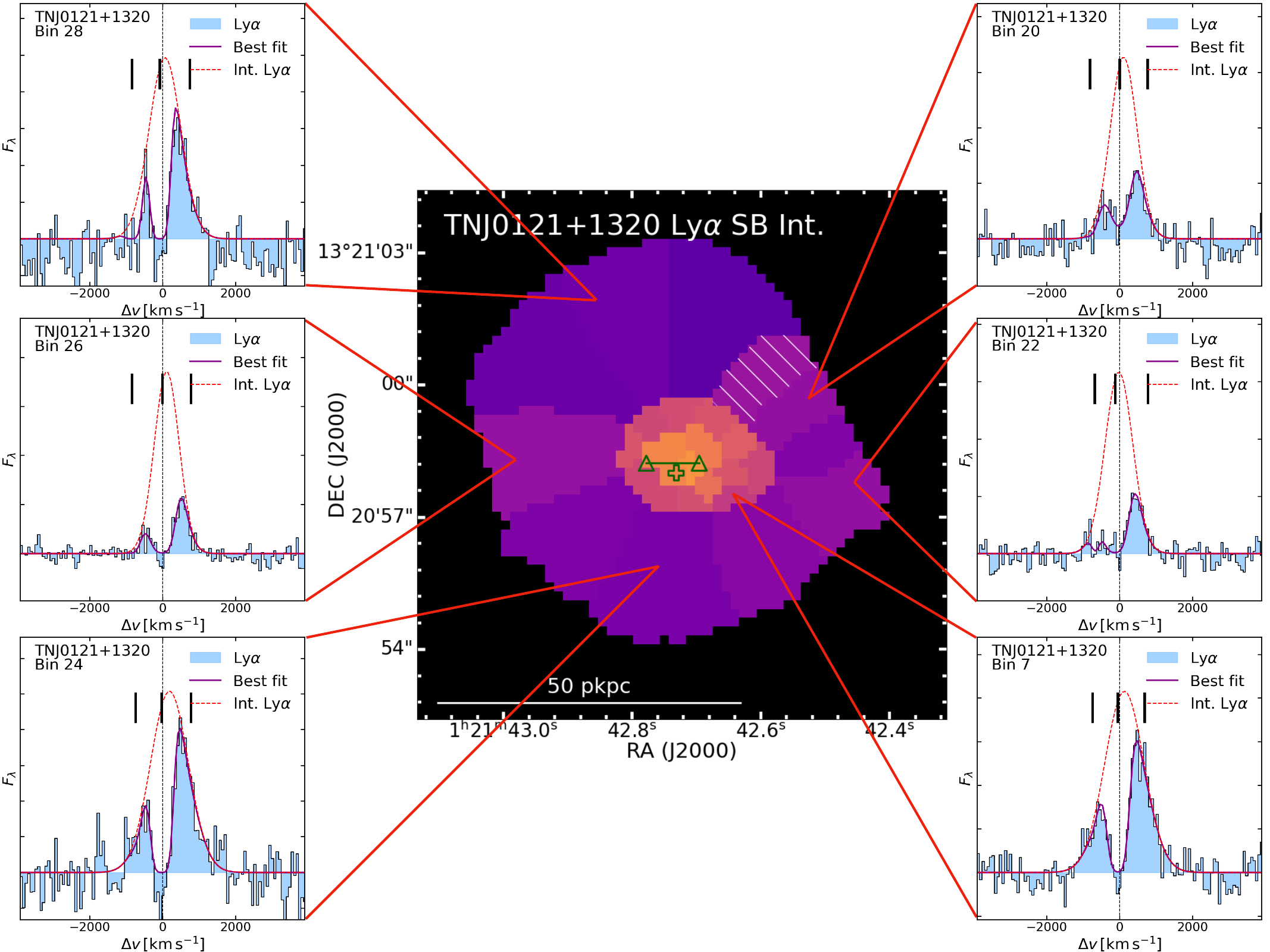}
      \caption{Similar figure  as Fig. \ref{fig:tnj0205_spa} for \object{TNJ0121+1320}.}
        \label{fig:tnj0121_spa}
         
  \end{figure*}
  \begin{figure*}
  \centering
        \includegraphics[width=\textwidth,clip]{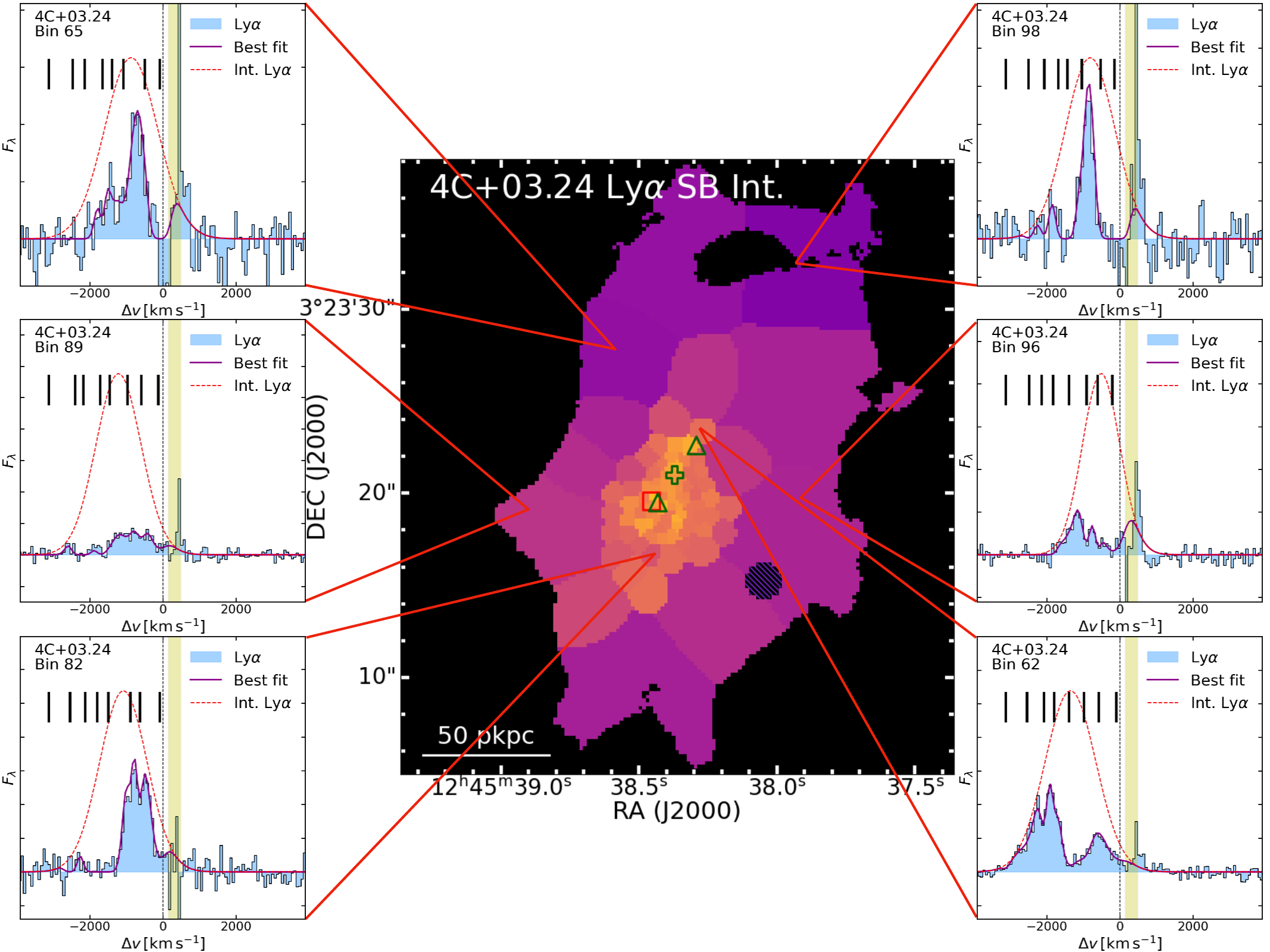}
      \caption{Similar figure  as Fig. \ref{fig:tnj0205_spa} for \object{4C+03.24}. The red box marks the secondary southern $K-$band continuum emission peak \citep[][Sect. \ref{sec:2kin_prof}]{vanBreugel_1998}. The yellow shaded regions show the wavelength range excluded due to the $5577\,\AA$ sky-line.}
        \label{fig:4c03_spa}
         
  \end{figure*}

         
   \begin{figure*}[!htb]
  \centering
        \includegraphics[width=0.95\textwidth,clip]{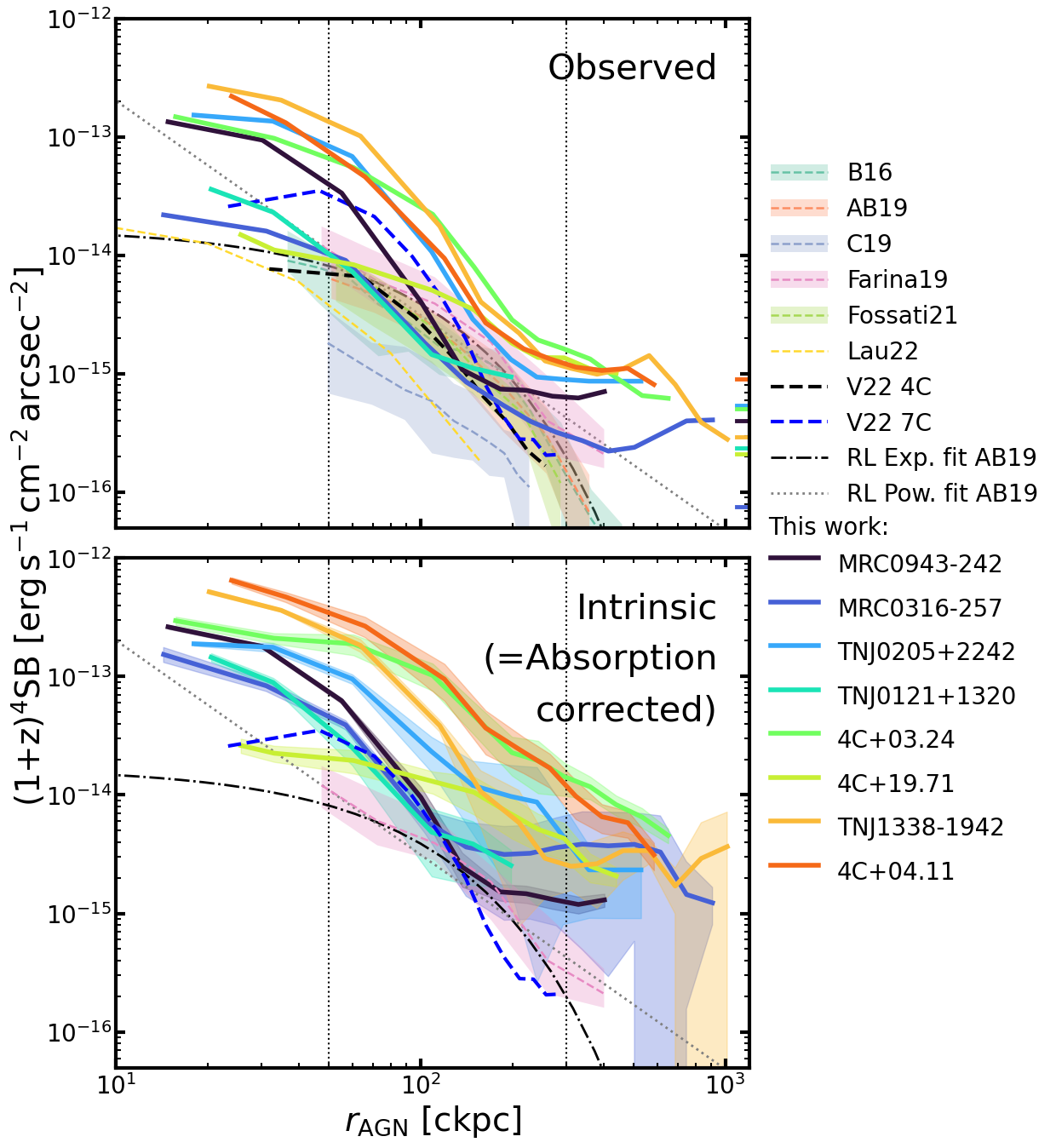}
      \caption{Radial profiles of the Ly$\alpha$ nebulae extracted in circular annuli (Fig.\ref{fig:mapradcir}). For better comparison, we show the radial profile in comoving kpc (ckpc) and take the cosmological dimming into account by a factor of $(1+z)^{4}$, where $z$ is the redshift of the target. The black dot-dashed curve and grey dotted line in both panels are the best fitted exponential and power law profiles of the \citet{FAB2019} radio loud sample, respectively. The two vertical dotted lines mark the 50 and 300 ckpc, respectively. \textit{Upper panel}: Radial profile of observed surface brightness map in thicker solid lines. In this panel, we also include the radial profiles of other quasar samples (dashed lines) for comparison: B16 --- \citet{borisova2016}, AB19 --- \citet{FAB2019}, C19 --- \citet{cai2019}, Farina19 --- \citet{Farina2019}, Fossati21 --- \citet{Fossati2021}, L22 --- \citet{Lau_2022} and V22 --- \citet{Vayner_2023} (two sources, 4C09.17 and 7C 1354+2552). When it is available, we show the range spanned by the 25th and 75th of the comparison sample radial profile as the shaded region around median profile in the same color. The horizontal bars at the right-most indicate the observed surface brightness limits  (scaled by area from Table \ref{tab:sampleinfo}) for each target in the same color.  \textit{Lower panel}: Intrinsic radial profile in thicker solid lines. The shaded regions around each profiles indicates the uncertainty range of the surface brightness from fitting. In this panel, we show again the same profiles of F19 and V22 7C as in the upper panel for comparison. Our HzRGs are extended further with higher surface brightnesses (or flattening in some sources) at larger radii ($\sim 300$ ckpc) compared to other samples.} 
         \label{fig:radcir}
   \end{figure*}


   \begin{figure*}[!htb]
  \centering
        \includegraphics[width=\textwidth,clip]{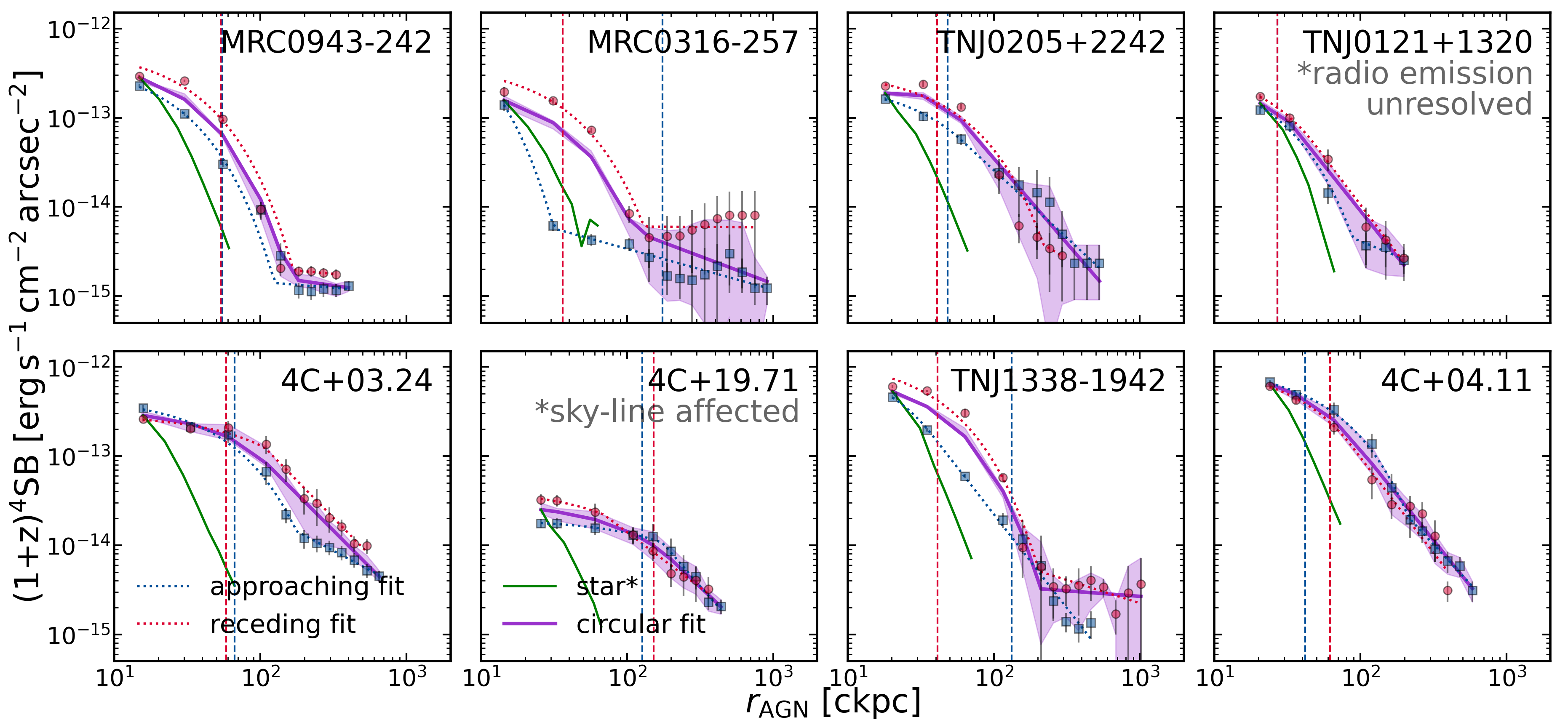}
      \caption{Surface brightness radial profiles for approaching (blue squares) and receding (red circles) directions along the jet axis. The dotted curves in corresponding colors show the exponential$+$power law fits for the two directional profiles. We also include the fits for the circularly averaged profile in solid magenta lines. In each panel, the magenta shaded region mark again the same uncertainty range for the intrinsic surface brightness profile as Fig. \ref{fig:radcir}. The solid green curve is the normalised radial profile of a star extracted up to $2\arcsec$ (the one in the FoV of \object{MRC0316-257} is extracted from a round galaxy due to no available star) showing the PSF (Table \ref{tab:sampleobs}).  The vertical dashed lines indicate the distances of the jet hot spots in corresponding colors. The profile along the receding side of the jet is brighter than along approaching side for most sources within the extent of the jets except \object{4C+03.24} and \object{4C+04.11}. This may indicate different gas density distribution (see Sect. \ref{sec:2dir_sb_prof}). We also identify flatting of the profile at $\gtrsim100$~ckpc for \object{MRC0943-242}, \object{MRC0316-257} and \object{TNJ1338-1942} which may related to nearby companions (see Sect. \ref{sec:1dis_env}).}
         \label{fig:rad_pf_fit}
   \end{figure*}

   \begin{figure*}[!htb]
  \centering
          \includegraphics[width=\textwidth,clip]{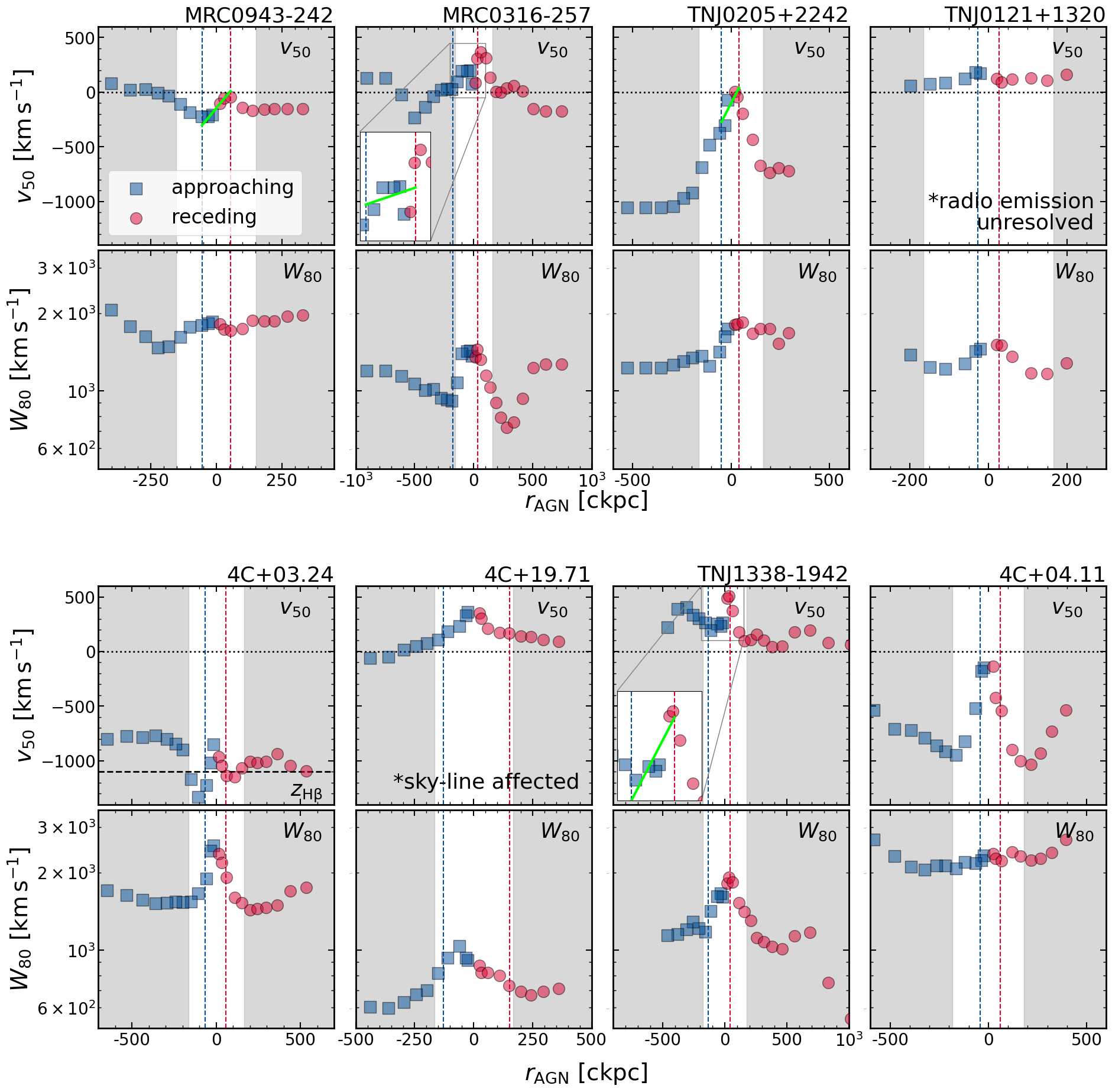}
      \caption{Directional $v_{50}$ and $W_{80}$ profiles for approaching (blue squares) and receding (red circles) sides along the jet axis extracted from the intrinsic maps (i.e. corrected for absorption, Fig. \ref{fig:map_int_1}cd). The vertical dashed lines indicate the distances of the jet hot spots (blue for approaching, red for receding, Appendix \ref{app:nebradinfo}). We note that the radio emission of \object{TN J0121+1320} is unresolved. The grey shaded regions are $>$5 arcsec from the host galaxy. The data points in the shaded regions should be treated with caution given the large tile size may smear kinematic structures. The horizontal black dotted line in the $v_{50}$ panel marks the $0\,\mathrm{km\,s^{-1}}$ derived from systemic redshift. The dashed horizontal black line in the $v_{50}$ panel of \object{4C+03.24} indicates the velocity shift of H$\beta$ redshift ($z_{\rm H\beta}\simeq3.566$, $-1100\,\mathrm{km\,s^{-1}}$ with respect to the systemic redshift) \citep[][]{nesvadba2017b} with respect to its systemic used in this paper (Table. \ref{tab:sampleobs}, see text). We note that the range of the x-axis is customised for each target and that the $W_{80}$ is shown in logarithmic scale. We show a zoom-in view of the central part of the $v_{50}$ profiles of \object{MRC 0316-257} and \object{TN J1338-1942} in the insets. For \object{MRC 0943-242}, \object{MRC 0316-257}, \object{TN J0205+2242} and \object{TN J1338-1942}, we mark the fit to the $v_{50}$ profiles within the jet hot spots (vertical lines) in green lines to guide the eye of the evidence of nebula velocity gradient following jet kinematics. In general, there is no clear evidence of a trend in bulk motion identified for the whole sample. For some targets (\object{4C+03.24}, \object{4C+19.71} and \object{TN J1338-1942}), $W_{80}$ decreases with increasing radial distance which may indicate that the jet is disturbing the gas.}
         \label{fig:v50_W80_r}
   \end{figure*}


\section{Data analysis}\label{sec:0dataana}
\subsection{Ly$\alpha$ nebulae extent and tessellation}\label{sec:1nebuseltes}
 To systematically study the Ly$\alpha$ nebulae of our HzRGs sample, we first need to determine all the voxels (volume pixel) containing usable Ly-alpha signal (Sect. \ref{sec:2max_neb_ext}) and bin the data to a sufficient S/N using a tessellation technique (Sect. \ref{sec:2lyatesmethod}) before fitting the emission feature described in Sect. \ref{sec:1specfitmeth}.

\subsubsection{Maximum extent of the nebulae}\label{sec:2max_neb_ext}
  

To select the Ly$\alpha$ signal with optimised sensitivity and capture the very low surface brightness structures of the nebulae, we used our own version of the adaptive smoothing algorithm described in \cite{martin2014b} \citep[see also][for an application to one of the sources in our sample]{vernet2017}. We first smooth the data cube in the wavelength direction by averaging $n_{\lambda}$ neighbouring pixels. Then for each wavelength plane, the algorithm iteratively smoothes spatially with a growing gaussian kernel selecting pixels passing a given SNR threshold ($T_{\mathrm{S/N}}$) and leaving to the next iterations only spaxels below this S/N threshold, until a maximum smoothing radius is reached ($\sigma_\mathrm{max}$). The spaxels not selected by the end of the iterative process are masked out. To further clean the smoothed data cube from spurious noise features and make sure that a proper line fitting can be made, we mask spatial positions selected by the adaptive smoothing algorithm in less than $n_{\rm c}$ consecutive wavelength bin.

To determine the optimal combination of the 4 parameters ($n_{\lambda}$, $T_{\mathrm{S/N}}$, $\sigma_\mathrm{max}$ and $n_{\rm c}$), we explore a range of possible combinations and select the set that is most sensitive to the extended low-surface brightness emission while at the same time minimising the number of detached `island-like' structures (see Appendix \ref{app:detec_map} for details). We note that the maximum nebulae extents selected by this method are similar to the results from previous studies of individual targets by different procedure \citep[TNJ1338-1942 from][]{swinkbank2015} or pure manual selection \citep[MRC0316-257 in][]{vernet2017}. We then manually clean up this map for the few remaining isolated island-like regions with further checking spectra extracted from these regions. This clean-up is accompanied by signal checking through spectrum extraction and only affects low S/N regions (Appendix \ref{app:detec_map}). Thus, the bulk of the detection map remains unchanged. This resulting detection map defines the pixels that we consider as part of the nebula and that we use in the analysis in this paper (see also Appendix \ref{app:detec_map}).

\subsubsection{Tessellation procedure}\label{sec:2lyatesmethod}

In order to increase the S/N to a level that allows fitting of the Ly$\alpha$ line,  especially close to the detection limit at the periphery of the nebulae, we tessellate the Ly$\alpha$ detection map. To construct the tessellation map, we firstly use a S/N map based on a narrow-band image ($\sim 15\,\AA$ wide) extracted around the Ly$\alpha$ emission peak. We implement a two-step Voronoi binning \citep[][]{cappellari2003} procedure which optimises the performance for both high S/N and low S/N regions by tessellating individually on these two parts. Specifically, the two-step procedure uses different target S/N for inner and outer regions. In this way, we can avoid large size tiles at the low S/N (outer) regions which may unnecessarily smear spatial resolution by imposing too high target S/N. We then combine the tessellated regions from the two-step process into one map. We emphasise that the tessellation is a trade-off between spatial resolution and S/N. The main goal of the work is to study the extend Ly$\alpha$ nebulae to the detection limit. This can only be achieved by sacrificing the spatial information. The details of this tessellation process are described in Appendix \ref{app:tesstll}, and in \ref{app:smimage} we present the resulting the maps.

\subsection{Spectral Fitting}\label{sec:1specfitmeth}
\subsubsection{Ly$\alpha$ absorption modeling}\label{sec:2fitmodel}
In this work we treat the Ly$\alpha$ emission system of HzRGs as an idealised case where several assumptions have been made prior to the analysis: (i) the radio galaxies reside in giant reservoirs of neutral hydrogen ($\sim$100s$\,$kpc); (ii) the neutral hydrogen is rather diffuse with large covering factor; (iii) the geometry of the giant reservoirs is unknown but can be highly asymmetrical due to the influence of the radio jet. Under these assumptions, it is natural that we observe absorption effecting the Ly$\alpha$ profiles. Indeed, such absorption troughs are observed in our Ly$\alpha$ spectra (see Fig. \ref{fig:map_int_1}) that need to be accounted for when drawing conclusions about the intrinsic emission line flux and higher moment measurements. Specifically, high resolution spectrocopy using the UltraViolet and Echelle Spectrograph (UVES) on the VLT exists for seven out of eight of the targets in our sample \citep[][Ritter et al. in prep.]{jarvis2003,wilman2004}. These spectra with $\sim 30$ higher resolution than MUSE display sharp edges which is fully consistent with a well-defined absorption profile rather than radiative transfer effects. We note that the the term `radiative transfer' used in the paper refers to the process where Ly$\alpha$ photons are scattered in frequency (wavelength) but are still captured in the spectrum (i.e. not `lost' in the observer's line of sight). In our assumptions, contrary to this, the photons are `lost' either due to being scattered outside our line of sight or being absorbed by dust and remitted at longer wavelength. We therefore adopt the technique used in our pilot study \citep[][and equations therein]{wang2021} and fit the spectra using a combination of Gaussian emission line profiles and Voigt absorption troughs \citep[e.g.][]{teppergarcia2006,teppergarcia2007,krogager2018}. This procedure has also been implemented successfully in the literature for fitting the Ly$\alpha$ line emission in HzRGs \citep[e.g.][]{swinkbank2015,silva_2018a,kolwa2019}. 

The known degeneracy between the \ion{H}{i} column density and Doppler parameter in our fits \citep[e.g.][]{Silva_2018b} does not affect the reconstructed intrinsic emission which is the focus of this work. We show the `Master Ly$\alpha$ spectrum' extracted from a central 1\arcsec\ aperture in Fig. \ref{fig:map_int_1}a which presents how the intrinsic profile compares to the observed spectrum (see Sect. \ref{sec:0intmaps} for details).

We emphasise that the term `intrinsic Ly$\alpha$ emission' throughout the work refers to the nebula Ly$\alpha$ emission corrected for intervening absorbers. The absorption troughs seen on the spectra (Fig. \ref{fig:map_int_1}a) are due to the Ly$\alpha$ emission being absorbed by these neutral hydrogen gas clouds or shells along the line of sight. Under the aforementioned assumptions, a natural consequence is that the absorbers must be distributed across the whole projected extension of the nebula. The fact that we mostly observe these features continuously across the extent of the nebulae in most HzRGs indeed indicates they are coherent intergalactic-scale structures. This can be found in Fig. \ref{fig:tnj0205_spa}, \ref{fig:tnj0121_spa} and \ref{fig:4c03_spa} where similar absorption features are seen in the selected spectra at larger distance (10s of kpc) away from AGN. Similar maps of the remaining sources are shown in Appendix \ref{app:suppmap} which are the ones have been previously published \citep[][]{swinkbank2015,Gullberg_2016a,vernet2017,falkendal2021,wang2021}. Our approach is a common interpretation in studies of HzRGs. Conversely, such absorbers are not often seen in the Ly$\alpha$ nebulae of other quasars. This reinforces the interpretation that strong (radio-mode) feedback on intergalactic scales is needed to create such `shells' of \ion{H}{i} material. The use of a Gaussian as underlying intrinsic emission profile is supported both by observational and modeling works \citep[e.g.][]{FAB2019,Chang_2022}. This could be a result of prior radiative transfer effects of Ly$\alpha$ \citep[e.g. local scattering or scattering from the broad line region of the AGN][]{Verhamme2006,gronke2016a,gronke2016b,Zhihui_Li_2022}. The radiative transfer modelling requires assumptions about the composition and geometry (and kinematics) of the gas near the AGN which is not the focus of this paper. Hence, we just assume the Gaussian shape of the Ly$\alpha$ (which could be due to the radiative transfer effects) and correct for the absorption along the line of sight to reconstruct the intrinsic emission on CGM scales. Incorporating radiative transfer calculations into the study of HzRGs Ly$\alpha$ nebulae is beyond the scope of this current work. Further developments of theoretical works are required (e.g. adding jet and resolving shells in simulations), and our dataset would be well suited for such studies. We therefore stress that the presented results are only valid for the stated assumptions that absorption rather than radiative transfer is primarily responsible for the line profiles. We will discuss the limitation of this treatment in Sect. \ref{sec:1discabcorr}.

We note that the $\ion{O}{v}]\lambda\lambda1213.8,1218.2$ (\ion{O}{v}]) line underneath the Ly$\alpha$ can affect the obtained flux especially in the nuclear region where the ionisation parameter (and metallicity) is higher \citep[][]{Humphrey_2019}. In our pilot study \citep[][]{wang2021} of \object{4C+04.11}, we found the contribution from \ion{O}{v}] is negligible. Hence, we do not further include \ion{O}{v}] in our line fitting. We leave the inspection to future work when data of metal lines (e.g. \ion{N}{v}$\lambda\lambda1238, 1243$ which is found to be related with \ion{O}{v}]) and high resolution spectra are analysed.

\subsubsection{Fitting procedure}\label{sec:2spatialfit}

To reconstruct the intrinsic Ly$\alpha$ emission across the nebula, we fit each spectrum in each tessellation bin (see Sect.  \ref{sec:2lyatesmethod}) following the procedure described in Sect. \ref{sec:2fitmodel}. We take into account the physical connection between neighbouring tiles by using the fit results of a previous connected bin as the starting parameters for the next bin (see Appendix \ref{app:selmasktess} for details on the ordering). We determine the number of absorbers based on the Master spectrum where the S/N is the highest (Fig. \ref{fig:map_int_1}a). We then use that same number of absorbers across the nebula, where the centroid, column density and Doppler parameter of absorbers are fitted in a given range (Appendix \ref{app:spatialfit}). This assumption is supported by the profile shapes at the largest spatial extents (see Fig. \ref{fig:tnj0205_spa}, \ref{fig:tnj0121_spa}, and \ref{fig:4c03_spa} and also Appendix \ref{app:suppmap}). We note that the number of absorbers selected here may be incomplete but this has minor effects on the results of this paper: (i) the absorbers that impact most the intrinsic flux (i.e. spatially extended $\sim10\arcsec$ and having higher column density and/or larger Doppler parameter) are included; (ii) absorbers that seem to be `superfluous' at the wings have only minor effects on the reconstructed flux where S/N is low (Fig. \ref{fig:tnj0205_spa}, \ref{fig:tnj0121_spa} and \ref{fig:4c03_spa}). Future work using high spectral resolution data will address these issues also taking into account that some of these absorbers have counterparts in metal lines covered by the MUSE data \citep[e.g. \ion{N}{v}$\lambda\lambda1238, 1243$ and \ion{C}{IV}$\lambda\lambda1548,1551$][]{kolwa2019,wang2021}. We perform the fit in each bin using both one and two Gaussian emission line components and we choose the solution that minimises the reduced $\chi^2$. The fit is done using a Least-squares method followed by a Markov chain Monte Carlo sampling \citep[MCMC, \texttt{emcee}][]{foremanmackey2013}. The uncertainties we report are either the direct output of the $1\sigma$ error by the MCMC or the propagated $1\sigma$ error. A detailed description of the fitting procedure is provided in Appendix \ref{app:spatialfit}. We reiterate that we do not report any further parameters on the absorption features which will be analysed in future work in combination with higher spectral resolution data \citep[e.g.][Ritter et al. in prep.]{jarvis2003,wilman2004,kolwa2019}. We present the results of this procedure for all of our sources in Fig. \ref{fig:map_int_1}.

\section{Results}\label{sec:0lyaradasy}
\subsection{Intrinsic mapping}\label{sec:0intmaps}

In this section, we present the intrinsic maps (i.e. corrected for absorption) constructed following the fitting procedure described in Sect. \ref{sec:1specfitmeth}. For each sources we show the Master spectrum together with its best fit in Fig. \ref{fig:map_int_1}a as an example (Sect. \ref{sec:2fitmodel}). We also show the non-resonant \ion{He}{ii} spectrum extracted in the same aperture (green histogram, not continuum subtracted) which is used for systemic redshift ($\Delta v=0\,\mathrm{km\,s^{-1}}$, Table \ref{tab:sampleobs}) determination. We note that there is no \ion{He}{ii} detected at the AGN position for \object{4C+03.24} (Sect. \ref{sec:2kin_prof}). In addition, to illustrate how fitting procedure works spatially (Sect. \ref{sec:2spatialfit}), 
the selected exemplar individual fits are shown in Fig. \ref{fig:tnj0205_spa}, \ref{fig:tnj0121_spa} and \ref{fig:4c03_spa} (also see Appendix \ref{app:suppmap}).


The intrinsic Ly$\alpha$ surface brightness maps are shown in Fig. \ref{fig:map_int_1}b on the same flux scale. Regions with larger fitting uncertainties ($\gtrsim50\%$) that should be treated with caution are indicated by the overlaid hatched tiles. We report the total intrinsic Ly$\alpha$ luminosities ($L_{\mathrm{Ly\alpha,int}}$) of the nebulae in Table \ref{tab:sampleinfo} and their maximum linear extent, $d_{\rm max}$, in Table \ref{tab:nebpropert}. Down to the surface brightness limit (Table \ref{tab:sampleinfo}), seven of our nebulae are extended over $100$ pkpc with the largest being $\sim 347$ pkpc (\object{MRC0316-257}). \object{TNJ0121+1320} is the only target with nebula $<100$ pkpc ($\sim72$ pkpc). The total intrinsic surface brightness ($L_{\mathrm{Ly\alpha,int}}$) of the nebulae ranges from $2$ to $29\times10^{44}\,\mathrm{erg\,s^{-1}}$.

 To characterise the kinematic information of the intrinsic nebulae that are fitted with one or two Gaussians, we use a set of non-parametric emission line measurements \citep[see e.g. ][]{liu2013_b} derived from the cumulative line flux as a function of velocity $\Phi(v)$ defined as: 
\begin{equation}
    \Phi(v) = \int_{-\infty}^{v} f(v')dv'
\end{equation}
where $f(v')$ is the flux density at $v'$. 
The often used $v_{50}$ is the velocity where the cumulative flux reaches 50\% of the total integrated value, $\Phi(v_{50})=0.5\Phi(\infty)$. The $v_{05}$, $v_{10}$, $v_{90}$ and $v_{95}$ are defined similarly. The line width measurement, $W_{80}$, defined in this context is $W_{80}=v_{90}-v_{10}$. In case of single Gaussian fits, $W_{80}$ is directly related to the FWHM and $v_{50}$ is the Gaussian centroid.

The non-parametric velocity shift ($v_{50}$) and line width ($W_{80}$) of the nebulae are shown respectively in panels (c) and (d) of Fig. \ref{fig:map_int_1}. The $v_{50}$ maps do not show clear trend on larger scale (i.e. beyond the jet hot spots) for the whole sample. This is foreseeable given that (i) Ly$\alpha$ is a resonant line which is sensitive to scattering (i.e. it will not necessarily show the bulk velocity of the gas), and we only observe the last scattering surface; (ii) the size of the tile far from the centre is larger which could smear out potential velocity structures; (iii) the line emissions on several 10s of pkpc could trace the inflowing gas \citep[or other gas components not governed by the host galaxy and/or kinematically related to the quasar outflow, e.g.][]{vernet2017}. Within the extent of the radio jets, 3 targets (\object{MRC0943-242}, \object{MRC0316-257} and \object{TN J1338-1942}) show tentative velocity gradients consistent with the jet kinematics (Sect. \ref{sec:2jetk}). For the line width maps, $W_{80}$, 3 targets (\object{4C+03.24}, \object{4C+19.71} and \object{TN J1338-1942}) show a trend with the line being broader near the centre and becoming narrower outwards. There are some tiles on the periphery of the nebulae, for example the southwest tile of \object{4C+04.11}, displaying larger  $W_{80}$ values ($\gtrsim 2500\,\mathrm{km\,s^{-2}}$). 
Except \object{4C+19.71}, all targets show a line width of $\sim 800-2500\,\rm{km\,s^{-1}}$. For \object{4C+19.71}, due to the strong $5577\,\AA$ sky-line located close with the observed Ly$\alpha$ peak wavelength, its line width  should be treated as lower limit ($\gtrsim 600\mathrm{km\,s^{-1}}$) especially for the tiles in the outskirts of the nebula. 

We note that the non-parametric measurements used in this mapping are based on intrinsic (= absorption-corrected) line profiles which are determined through model fitting, same as \citet{wang2021}. In Appendix \ref{app:suppmap}, we present the maps of observed surface brightness and flux ratio as supplementary material.




\subsection{Radial profiles}\label{sec:1rad_prof}

\subsubsection{Circularly averaged surface brightness radial profiles}\label{sec:2rad_sb_pro}

In this section, we present the surface brightness radial profile of the eight Ly$\alpha$ nebulae. In order to compare our HzRGs to other quasar samples, we extract the surface brightness profile centred around the AGN in circular annuli. The annuli over which the profiles extracted are shown in Fig. \ref{fig:mapradcir}. We compute the surface brightness in each annulus as the mean of the surface brightness of each contributing spaxel weighted by the fraction of the spaxel area covered by the annulus. Table \ref{tab:Int_SB} lists the extracted intrinsic profile values.

Figure \ref{fig:radcir} shows the radial profiles after correction for cosmological dimming and in comoving units for observed (upper panel) and intrinsic (lower panel, corrected for absorption) maps. The dashed lines in the upper panel represent the comparison quasar samples or single targets \citep[ERQ and 2 radio-loud quasars from][respectively]{Lau_2022,Vayner_2023}. The selected quasars are all observed by advanced IFU instruments (MUSE or KCWI) and cover a large range of redshift and physical properties. They are luminous radio-quiet quasars at $z\sim3.2$ quasar from \citet{borisova2016} \citep[profiles from ][]{Marino2019}, luminous type-1 quasars at $z\sim3.17$ from \citet{FAB2019}, luminous quasars at $z\sim2.3$ from \citet{cai2019}, high redshift quasars at $z\sim6.28$ from \citet{Farina2019} and luminous quasars at $z\sim3.8$ from \citet{Fossati2021}. The quasar nebulae do not show so many absorption features as in our HzRGs and the studies were preformed without absorption corrections (Sect. \ref{sec:1discabcorr}). Nevertheless, since the comparison quasar samples are not corrected for absorption, we do not show them all again in our intrinsic profile (lower) panel. The two exceptions are \citet{Farina2019} (F19) and \citet{Vayner_2023} (7C 1354+2552, V22 7C). Those two are on the higher surface brightness end of the comparison samples, and we will examine them quantitatively along with both the intrinsic and observed HzRGs profiles. We note that \cite{Vayner_2023} fitted the Ly$\alpha$ absorbers from the spatially integrated 1D spectrum and found $\sim 10^{13.5}\,\mathrm{cm^{-2}}$ for the column densities. For \cite{Farina2019}, there is not much evidence of absorption. Therefore, the comparison is legitimate. The best fit profiles to the observed Ly$\alpha$ nebulae of radio loud quasars in \citet{FAB2019} are included in both panels of Fig. \ref{fig:radcir} which can be used as a reference between the two panels.

Except for the ERQ from \citet{Lau_2022} (which is also highly obscured), type-1 radial profiles are dominated by direct emission from bright AGN point source in the inner regions ($\sim$50 ckpc or $\sim$10 pkpc). Hence, due to point spread function (PSF) subtraction, the inner-most radius covered in the comparison samples is limited to $\sim$50 ckpc in most cases \citep[except][]{Vayner_2023}. At larger radii, the contamination by the PSF should be negligible. Of the three single target profiles, V22 7C \citep[7C 1354+2552 from][]{Vayner_2023} has the highest surface brightness. At a radius of $\sim$50 ckpc, the intrinsic surface brightness of our HzRG sample has a factor of $0.5-7$ compared with V22 7C (7 of our targets are brighter). This source then shows a faster drop off compared with HzRGs.  At the faint end corresponding to $\sim300$ ckpc (except \object{TN J0121+1320}), the HzRGs have a factor of $7-100$ higher surface brightness than V22 7C. The profile of \citet{Farina2019} shows the highest surface brightness among the comparison samples. At $\sim$50 ckpc, the intrinsic HzRG profiles are still a factor of $1.1-15$ (or $4-40$ at $\sim400$ ckpc, except \object{TN J0121+1320}) brighter than the 75th percentile of \citet{Farina2019}. These indicate that our eight observed HzRG have some of the brightest known Ly$\alpha$ nebulae (Sect. \ref{sec:1musesample}). We note that the jet compression is also known to result in high Ly$\alpha$ nebula luminosity \citep[e.g.][]{Heckman_1991b,Heckman_1991a}. Compared to quasars with similarly deep observations (i.e. avoiding the surface brightness detection limit), our HzRG sample generally maintains a high surface brightness out to larger radii (5 out of 8 $>500$ ckpc). We note again that the detected extent of our nebulae will have similar range even if adopting other detection methods than the ones in this paper (Sect. \ref{sec:1nebuseltes}). For example, \citet{Gullberg_2016a} reported the similar extend Ly$\alpha$ nebula in \object{MRC0943-242} with less exposure time. \citet{vernet2017} detected the nebula of \object{MRC0316-257} $>700$ ckpc based on visual detection. \cite{swinkbank2015} found the $>500$ ckpc nebula of \object{TN J1338-1942} with (or even without) a simpler binning algorithm. Hence, we are sure that the detection of the $\gtrsim500$ ckpc nebulae in our sample based on our method is robust. However, we do caution that this sample is not representative since they are selected to have bright and extended Ly$\alpha$ emission.
The profiles of \object{MRC0943-242}, \object{MRC 0316-257} and \object{TN J1338-1942} show a flattening at $r_{\rm AGN}>200$ ckpc. For the comparison samples, their profiles drop off monotonically and drop below detection limit at radii smaller than our HzRGs. The lowest surface brightness of HzRG intrinsic profiles is $\sim1\times10^{-15}\,\rm{erg\,s^{-1}\,cm^{-2}\,arcsec^{-2}}$ (\object{MRC0316-257}, corrected for cosmological dimming) which is higher than the faint end of the quasar samples by a factor of $5-40$ (not at similar comoving distances). These indicate that we are observing some of the most extend Ly$\alpha$ nebulae, in two cases (\object{MRC0316-257} and \object{TN J1338-1942}) even extending beyond the field of view of MUSE. By simply comparing our intrinsic profiles to the exponential and power law fits of \citet{FAB2019}, we find that the inner part of HzRGs profiles are exponential-like (especially \object{MRC0943-242} and \object{TN J1338-1942}) while extended parts show power law decline. We note, however, that the exponential part is affected by seeing smearing. 

If we do not correct for the Ly$\alpha$ absorption and instead measure at the observed radial profiles (Fig. \ref{fig:radcir} upper panel), 5 of the HzRGs are still brighter than the comparison samples, but by a lower factor of $\sim 2-4$ ($\sim 2-6$) at radii of $\sim$400 ($\sim$50) ckpc compared to the 75th percentile of \citet{Farina2019}. Comparing the results from intrinsic and observed profiles, this suggest that the quasar samples may miss a non-negligible amount of flux ($\gtrsim5$) due to uncorrected for absorption. 

The radial surface brightness profiles of the comparison samples are extracted from a fixed velocity or wavelength range, for example $\pm2000\,\rm{km\,s^{-1}}$ in \citet{cai2019}, $30\,\AA$ in \citet{FAB2019} and $\pm500\,\rm{km\,s^{-1}}$ in \citet{Farina2019}. Considering the redshift difference between these samples, the integration range adopted are consistent. For our study, particularly for the observed radial profile, our extraction is based on the $v_{05}$ and $v_{95}$ which are determined based on intrinsic fitting (Sect. \ref{sec:0intmaps}). In this way, we can minimise the uncertainties coming from the observed line width difference, for example between the emission lines in the vicinity of the host and outskirts of the nebula. Our velocity range ($v_{05}$ and $v_{95}$) used is basically the value of $W_{90}$ which has the range of $\sim 800-2700\,\rm{km\,s^{-1}}$ for all tiles of all targets. Nevertheless, we conduct a check by extracting observed circular radial profiles through integration of $30\,\AA$ around the systemic redshift of our targets for comparison. The results vary by $\sim \pm 10\%$ in each annulus to the profiles in Fig. \ref{fig:radcir} (from $v_{05}$ and $v_{95}$), especially for emissions at $>50\,\rm{ckpc}$ where the line width is narrower comparing to the centre. The $30\,\AA$ extracted profile could be 40\% less than the $v_{05}-v_{95}$ extraction in the centre regions (wider line width) for high-redshift targets where the fixed wavelength range in observed frame corresponds to a narrower rest frame range. Hence, to alleviate this problem brought by the difference in line width and redshift range spanned by our sample, we keep the $v_{05}$-to-$v_{95}$ extraction. As for the intrinsic radial profile, it is redundant to integrate from a narrower range when we can have the direct fit results for the integrated Ly$\alpha$ line. We show the flux ratio between the intrinsic and observed maps in the same velocity range in Appendix \ref{app:suppmap} which can be used as a proxy for scaling between the two profiles. Therefore, the different wavelength (velocity) ranges used when extracting radial profiles for our study and comparison quasar samples will not bring additional discrepancy besides the relatively large surface brightness value in our sample.

\subsubsection{Directional surface brightness profiles}\label{sec:2dir_sb_prof}

Since the shapes of our Ly$\alpha$ nebulae are asymmetric (Sect. \ref{sec:1nebulaasy}), the radial profiles extracted in Sect. \ref{sec:2rad_sb_pro} smear out direction-dependent features. For instance, several HzRG Ly$\alpha$ nebulae display features aligned with their radio jets, such as having higher line width and elongated morphology along the jet axis \citep[e.g.][]{vanojik1997a,villarmartin2003,Miley_2004,Zirm_2005,Humphrey_2007a,Morais_2017}. Hence, in this section, we study the radial profile of the intrinsic Ly$\alpha$ emission along the direction of the radio jet which could exert a kinematical and/or electromagnetic influence on the surrounding gas. Due to the limited S/N, we split our nebulae into two half parts (approaching and receding, Appendix \ref{app:nebradinfo}) along the jet direction and extract the surface brightness profile in each direction using the same annuli as Sect. \ref{sec:2rad_sb_pro}. Figure \ref{fig:rad_pf_fit} shows these directional profiles. We also show the position of the jet hot spot for the receding (red) and approaching (blue) side with vertical dashed lines.\footnote{We note that the radio emission of \object{TN J0121+1320} is unresolved. The `jet size' represented by the vertical lines are linear size of the $3\sigma$ contour along the east-west elongation of the radio map.} Qualitatively speaking, the surface brightness on the receding side is higher than on the approaching side within the radio jet extent for most of our sources (except \object{4C+03.24} and \object{4C+04.11}). In three sources, the receding jet hot spot is closer to the AGN: \object{MRC0316-257}, \object{TN J0205+2242} and \object{TN J1338-1942}. This result was first reported by \citet{McCarthy_1991} where the authors found that the line emission is brighter in HzGRs on the side with shorter radio jet. They interpreted this as a large scale asymmetry in the density of gas on either side of the nucleus: the denser gas absorbs more ionising radiation resulting in brighter emission lines, while the radio jet is more contained as it travels more slowly through the denser medium.

\subsubsection{Fitting the surface brightness profiles}\label{sec:2sb_prof_fit}
To quantify the shape of the profiles, we fit the circularly averaged intrinsic profile and two directional intrinsic profiles with a piecewise function split into an exponential for the inner part and power law for the outer part. This can be mathematically represented by
\begin{equation}\label{eq:exp_pl_piece}
    \text{SB}(r) = \left \{
    \begin{array}{lr}
    C_{\rm e}\mathrm{exp}(-r/r_{\rm h}) &\, r\leq r_{\rm b}\\
    C_{\rm p}r^{\alpha} & \, r>r_{\rm b}
    \end{array}
    \right . \text{,}
\end{equation} where $r_{\rm h}$ is the scale length of the exponential profile, $r_{\rm b}$ is the distance at which the inner and outer profiles separate and $C_{\rm e}$ and $C_{\rm p}$ are normalisation parameters for exponential and power law profiles, respectively ($C_{\rm p}=C_{\rm e} \mathrm{exp(-r_{\rm b}}/r_{\rm h})/r^{\alpha}_{\rm b}$).  The determination of the piecewise function is motivated by previous studies of quasar Ly$\alpha$ nebula \citep[e.g.][]{FAB2019,cai2019,denbrok2020} which fit the profile by either power law or exponential. We also test to fit our profiles use only one of the two functions. The single profile, however, cannot fit some targets well. For example, the reduced $\chi^{2}$ are high ($>20$) for \object{MRC0943-242}, \object{MRC0316-257} and \object{TN J1338-1942} with the single-function fit. We therefore fit all of the profiles with the piecewise function for consistency. Figure \ref{fig:rad_pf_fit} shows the fits and Table \ref{tab:SBfit} presents the fitted parameters.  

For most of our targets, the two directional surface brightness profiles are similar to the circularly averaged profile. One exception is the approaching side of \object{MRC0316-257} which has $\sim1$ dex lower than the receding side. This could be an extreme case of uneven Ly$\alpha$ emitting which may trace the different gas distribution. In Fig. \ref{fig:rad_pf_fit}, we also show the distance of the jet hot spots on both directions (Appendix \ref{app:nebradinfo}). There is no correlation between the distance of the jet hot spot and $r_{\rm b}$ (nor $r_{\rm h}$). As Sect. \ref{sec:2rad_sb_pro} described, our HzRGs are high in surface brightness (large $C_{\rm e}$); the reasons for this include (i) our sample is composed of HzRGs with bright Ly$\alpha$ emission, (ii) our profiles are absorption corrected, (iii) the quasar surface brightness is extracted from a fixed wavelength range (Sect. \ref{sec:2rad_sb_pro}).  The exponential shape is also seen in other quasar samples, for example \citet{FAB2019}, \citet{Farina2019}, \citet{denbrok2020} and \citet{Lau_2022}. The $r_{\rm h}$ values derived for our sample are mostly $<20$ pkpc (Table \ref{tab:SBfit}) which is consistent with the quasars. This suggests a similarity between the central (high surface brightness) part of HzRGs to other quasars (type-1 radio-loud and radio-quiet, type-2 radio-quiet), despite the high surface brightness in our sample. We note that the PSF-subtraction of quasar samples and resolution effects will impact the inner part to the profile. We will further discuss the power law declining (flattening) part of our nebula in Sect. \ref{sec:0discuss} combining the information from nebular morphology (Sect. \ref{sec:1nebulaasy}).

\subsubsection{Radial profiles of kinematic tracers}\label{sec:2kin_prof}
It is of interest to study how the nebula kinematics changes radially which may offer evidence of outflow and/or inflow \citep[e.g.][]{Humphrey_2007a,swinkbank2015,vernet2017}. We stress that it is beyond the scope of this work to separate different Ly$\alpha$ kinematics emission components (e.g. systemic and outflow) which will be inspected through high resolution spectroscopic data. Hence, we adopt the $v_{50}$ and $W_{80}$ parameters to measure the overall kinematics of the line emitting gas (Sect. \ref{sec:0intmaps}). We caution that the kinematics derived in this way may be biased, for example if there are several gas components with different kinematics but on similar flux levels.

In Fig. \ref{fig:v50_W80_r}, we show the directional radial profiles of $v_{50}$ (Fig. \ref{fig:map_int_1}c) and $W_{80}$ (Fig. \ref{fig:map_int_1}d), respectively. The profiles are extracted in a similar way as the directional surface brightness in Sect. \ref{sec:2dir_sb_prof} by splitting the map into two halves (approaching and receding). The $v_{50}$ ($W_{80}$) value shown at each radius is averaged in the corresponding annulus. Within the extent of the radio jet hot spots (vertical dashed lines), \object{MRC0943-242}, \object{MRC0316-257}, \object{TN J0205+2242} and \object{TN J1338+1942} show evidence of jet-driven outflows \citep[e.g.][]{nesvadba2008b,nesvadba2017a} if we ignore the absolute $v_{50}$ value but focus on the relative gradient. That is to say the velocity shift at the approaching side is higher than the receding side. For these four targets, we overplot a solid green line to show the fit of the velocity radial profile within the radio jet extent in Fig.\ref{fig:v50_W80_r}. The same velocity gradient is also identified in \ion{He}{ii} for \object{MRC0943-242} \citep[][]{kolwa2019}, \object{MRC0316-257} (Appendix \ref{app:z_sys0316}) and \object{TN J1338+1942} \citep[][]{swinkbank2015}. There is no other evidence from the $v_{50}$ of ordered gas bulk motion for the overall sample. This further suggests that Ly$\alpha$ is an unreliable tracer of kinematics at least on 10s to $100$s~pkpc scale in HzRGs. We note that the tessellation implemented, especially for tiles with larger size ($\sim 5\,\mathrm{arcsec^{2}}$) which are usually located in the low S/N region away from the host galaxy, may smear out potential kinematic features. One possible consequence of combining different kinematic components is broadening of the line width. This may be the case for the receding side of \object{MRC0316-257} and both sides of \object{TNJ0121+1320} and \object{4C+04.11}. In general, the $W_{80}$ does not show an increasing trend toward larger $r_{\rm AGN}$. However, if the line width decreases intrinsically away from the AGN, this will counteract the broadening which makes it difficult to check the impact of smearing. Therefore, we mark the regions with $r_{\rm AGN}>5\arcsec$ on the kinematic radial profile using grey shade to flag the possible high uncertainty in Fig. \ref{fig:v50_W80_r}. 

If we assume that the bulk of the gas resides in the potential well of the radio galaxy, we expect to see the Ly$\alpha$ emission gas centred around systemic velocity, at least in the vicinity of the AGN. Offsets of the $v_{50}$ levels at $r_{\rm AGN}\sim0$~ckpc from $0\,\mathrm{km\,s^{-1}}$ (based on systemic redshift, Table \ref{tab:sampleobs}) are  identified for most of our targets which may be due to the aforementioned bias from different kinematic components and scattering of Ly$\alpha$ photons. The most noticeable case is \object{4C+03.24} which has an offset of $\sim900\,\mathrm{km\,s^{-1}}$. We note that its systemic redshift (Table \ref{tab:sampleobs}) is based on [\ion{C}{i}](1-0) emission \citep[][]{kolwa2023} due to lack of \ion{He}{ii} from the AGN position in the MUSE data (Fig. \ref{fig:map_int_1}a). This offset can be eased if we use the redshift of H$\beta$, $z_{\rm H\beta}\simeq3.566$, reported by \citet{nesvadba2017b} as zero velocity. It is marked in black dashed horizontal line in the $v_{50}$ panel of \object{4C+03.24} in Fig. \ref{fig:v50_W80_r}. This corresponds to $-1100\,\mathrm{km\,s^{-1}}$ with respective to the systemic velocity shift used in this paper. We caution that, however, the H$\beta$ was not exclusively extracted at the AGN position (radio core). There is also a known jet-gas interaction in the south of the AGN \citep[see bend of the radio jet contours in Fig. \ref{fig:map_int_1}b and also][]{vanojik1996}. From the $K-$band image \citep[][]{vanBreugel_1998}, we can find a second continuum emission peak in the south. The position of this emission peak is marked by the red square in Fig. \ref{fig:4c03_spa}. Given these pieces of evidence, we propose that there is a companion galaxy at $z_{\rm H\beta} \simeq3.566$ in the south of our radio galaxy ($z=3.5828$). If there is (or was) an interaction between these two galaxies, the companion may have deprived gas from the AGN resulting in a gas poor AGN host \citep[no detection of \ion{He}{ii} and less molecular gas detected at the AGN position,][]{kolwa2023}. The companion then becomes sufficiently massive and gas-rich to deflect the jet. Therefore, the Ly$\alpha$ nebula of \object{4C+03.24} may trace the CGM of the companion galaxy. Scheduled \textit{JWST} data \citep[][]{Wang_2021JWST} will offer a clearer view of this particular situation.

For the $W_{80}$ radial profiles in Fig. \ref{fig:v50_W80_r}, we can first identify that  most of the HzRGs have high $W_{80}$ even at larger radii ($\sim 1500\,\mathrm{km\,s^{-1}}$). The exception is \object{4C+19.71} whose measurement is affected by sky-line residuals (Sect. \ref{sec:0intmaps}). The $W_{80}$ reported here is similar to FWHM (especially at large radii, $>100$ ckpc or $\gtrsim 22$ pkpc) where most of our fit are done with a single Gaussian (Sect. \ref{sec:1specfitmeth}). In \object{4C+03.24}, \object{4C+19.71} and \object{TN J1338-1942}, we can see a clear radial decrease of $W_{80}$ along both directions. This may be related to results found in \citet{villarmartin2003} who observed a Ly$\alpha$ FWHM drop off at distance beyond the extent of the radio jets in a sample of HzRGs (including \object{MRC0943-242}) using deep Keck long slit spectroscopy. In our study, however, we firstly do not find such a decrease in all targets. The grey shaded regions ($r_{\rm AGN}>5\arcsec$) should be treated with caution. We note that the FWHM in \citet{villarmartin2003} was derived without correction for absorption. Since we see a high spatial coverage of absorbers (Fig. \ref{fig:tnj0205_spa}, \ref{fig:tnj0121_spa} and \ref{fig:4c03_spa}), the correction indeed helps with recovering the close-to-intrinsic gas kinematics at large radii. In Fig. \ref{fig:pvdiagram}, we show the position-velocity diagram extracted based on our observed and intrinsic surface brightness maps along the jet as a direct comparison with the long slit spectroscopic study. Although it resembles the detection of \citet{villarmartin2003} at first glance, we note that this is due to the tessellation and the contrast between high surface brightness and low surface brightness part. 

By considering both the radial profiles from $v_{50}$ and $W_{80}$, we can generally find that the profiles within the jet extents have behave differently compared to the profiles outside the jet extent. This again suggests the jet is disturbing or entraining the Ly$\alpha$ emitting gas. There are unclear signs of kinematics other than outflows or inflows seen mostly at larger radial distance $\sim300$~ckpc. For example, \object{MRC0943-242} stays relatively flat (for both $v_{50}$ and $W_{80}$), while \object{MRC0316-257} has a decrease in $W_{80}$ followed by an increase beyond the jet extent on the receding side. We reiterate that in this analysis we do not distinguish between (potential) different kinematics components by using $v_{50}$ and $W_{80}$ to quantify the overall velocity shift and line width. This may bring bias of the measured values. Additionally, the measured kinematics farther from the AGN are averaged from larger annulus (e.g. projected area of $\sim4\times10^{4}$~pkpc$^2$ at $\sim60$~pkpc or $\sim300$~ckpc) which will bring another bias. We point out again that we use grey shade to mark the data $>5\arcsec$ from the center which has larger tile size. Nevertheless, we note that the detected $W_{80}$ of $\sim 10^{3}\,\mathrm{km\,s^{-1}}$ (and abrupt velocity shift) at large radii ($\sim300$~ckpc) in some of the profiles could be caused by the fact that the detected Ly$\alpha$ emission is dominated by emission halos of nearby companions \citep[e.g.][]{Byrohl_2021}. 

   \begin{figure*}[!htb]
  \centering
        \includegraphics[width=\textwidth,clip]{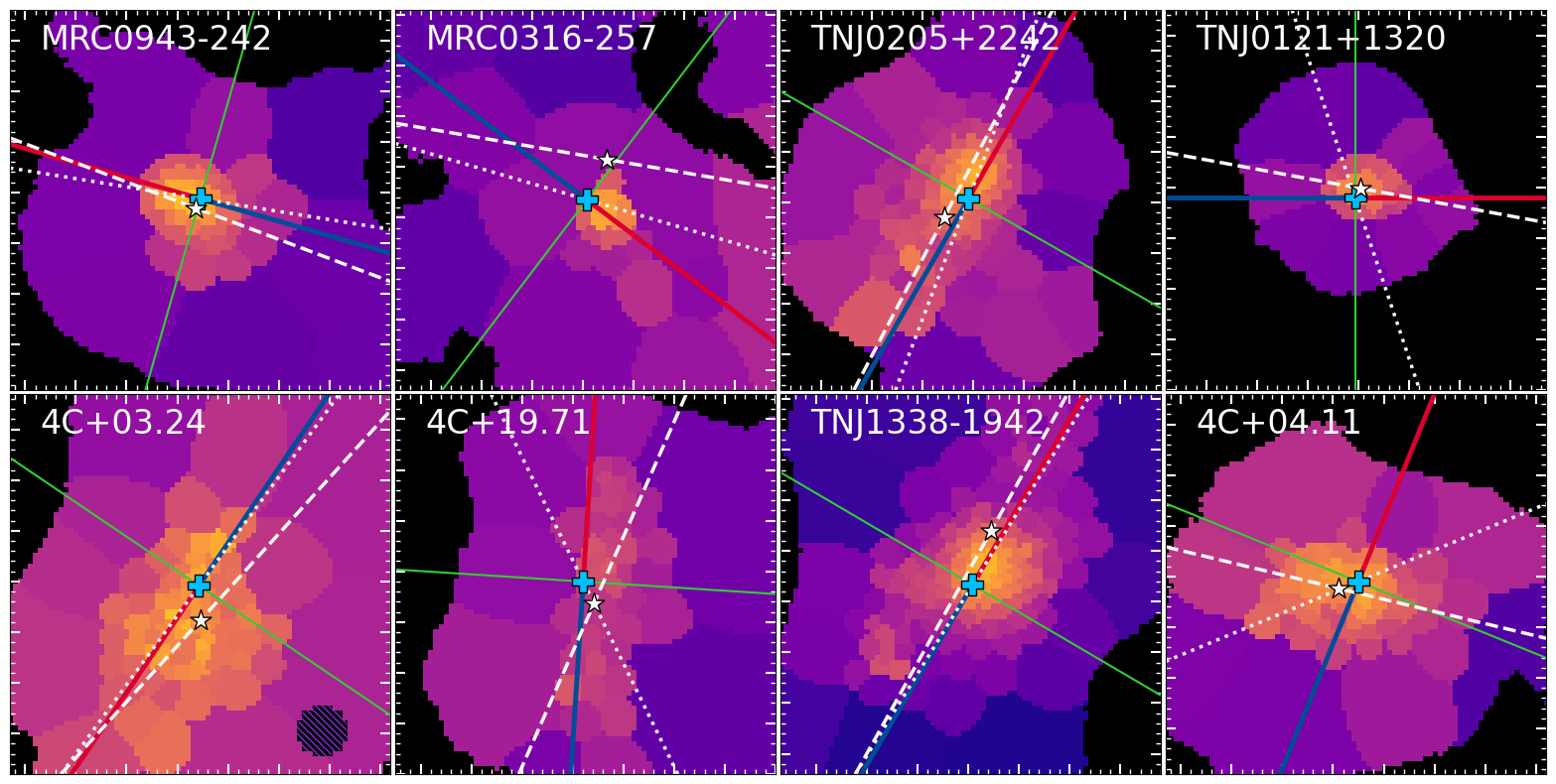}
      \caption{Zoom-in intrinsic surface brightness maps (Fig. \ref{fig:map_int_1}b) of the HzRGs sample to $15\times 15\,\rm{arcsec^{2}}$ (or $\sim 110\times 110\,\rm{pkpc^{2}}$) around the central AGN (blue cross). In each panel, the blue$+$red and green solid lines indicate the direction of, and perpendicular to, the radio jet. The blue (red) color represents the direction of the approaching (receding) jet. The white dashed line shows the flux-weighted position angle of the nebula ($\theta_{\rm weight}$). The white dotted line shows the unweighted position angle of the nebula ($\theta_{\rm unweight}$). The white star indicates the intrinsic flux-weighted centroid of the Ly$\alpha$ nebula. The flux weighted measurement is sensitive to the morphology of the high surface brightness part of the nebula. The unweighted measurement quantifies the morphology of the whole nebula. We find that nebula is elongated along the jet axis for most of HzRGs.}
         \label{fig:mapradind}
   \end{figure*}


   \begin{figure}[h!]
   \centering
   \includegraphics[width=\hsize]{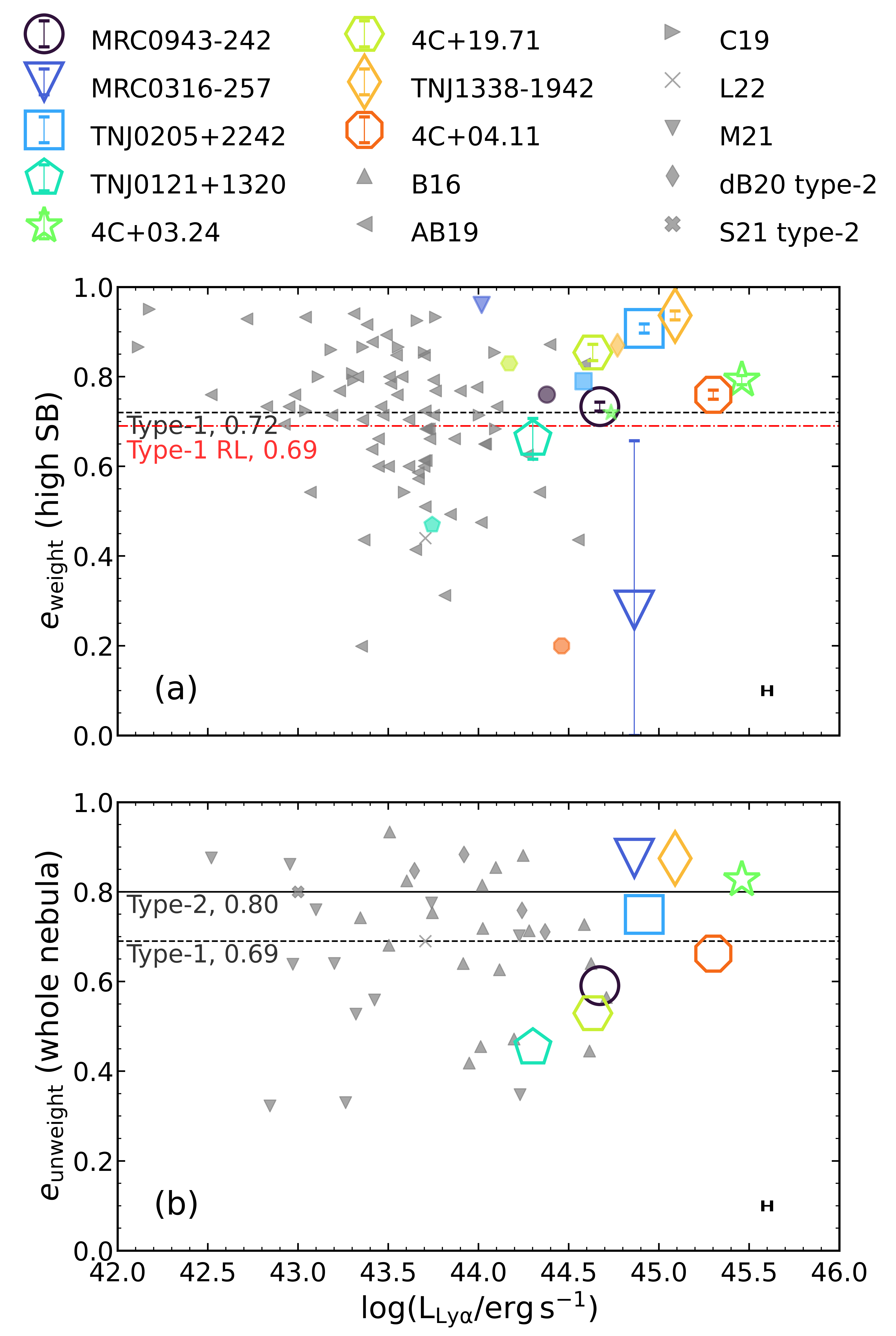}
      \caption{\textbf{(a)} Flux-weighted Ly$\alpha$ nebula elliptical asymmetry measurement versus nebula luminosity, $L_{\rm Ly\alpha}$.  We show the intrinsic flux-weighted ellipticity ($e_{\rm{weight}}$) in larger open symbols for our targets versus their intrinsic Ly$\alpha$ luminosity. We also show the observed flux-weighted ellipticity $e_{\rm{weight,obs}}$ for our targets versus their observed Ly$\alpha$ luminosity in smaller filler symbols. The small grey symbols are data of comparison targets (AB19 --- \citet{FAB2019}, C19 --- \citet{cai2019} and L22 --- \citet{Lau_2022}).  We mark the median flux-weighted (not corrected for absorption) ellipticity, 0.72, of type-1s with the horizontal dashed line. We also show the median $e_{\rm{weight}}$, 0.69, of radio-loud type-1 quasars from \citet{FAB2019} in red horizontal dash-dotted line. \textbf{(b)} Flux-unweighted Ly$\alpha$ nebula elliptical asymmetry measurement versus $L_{\rm Ly\alpha}$. The larger symbol are measurements for our HzRGs while the grey symbols are comparison targets (type-1s: B16 --- \citet{borisova2016}, L22 --- \citet{Lau_2022} and M21 --- \citet{mackenzie2021}; type-2s: dB20 --- \citet{denbrok2020} and S21 ---\citet{sanderson2021}). We mark the median $e_{\rm{unweight}}$ for type-1s (0.69) and type-2s (0.80) in solid and dashed horizontal lines, separately. The $e_{\rm{weight}}$ is sensitive to the morphology of the high surface brightness part of the nebula while the $e_{\rm{unweight}}$ quantifies the morphology of the whole nebula. We note that for $e\to0$, the nebula is closer to round shape and vice versa. At the bottom right, we show the median uncertainty of the intrinsic $L_{\rm Ly\alpha}$ for our sample in logarithmic scale, 0.04. The ellipticity for our sample are higher compared to the other quasars for both high surface brightness part and whole nebula. There is no clear evidence that the nebula ellipticity correlates with luminosity. }\label{fig:asye_vs_llya}
   \end{figure}


  \begin{figure*}
  \centering
  \includegraphics[width=\textwidth]{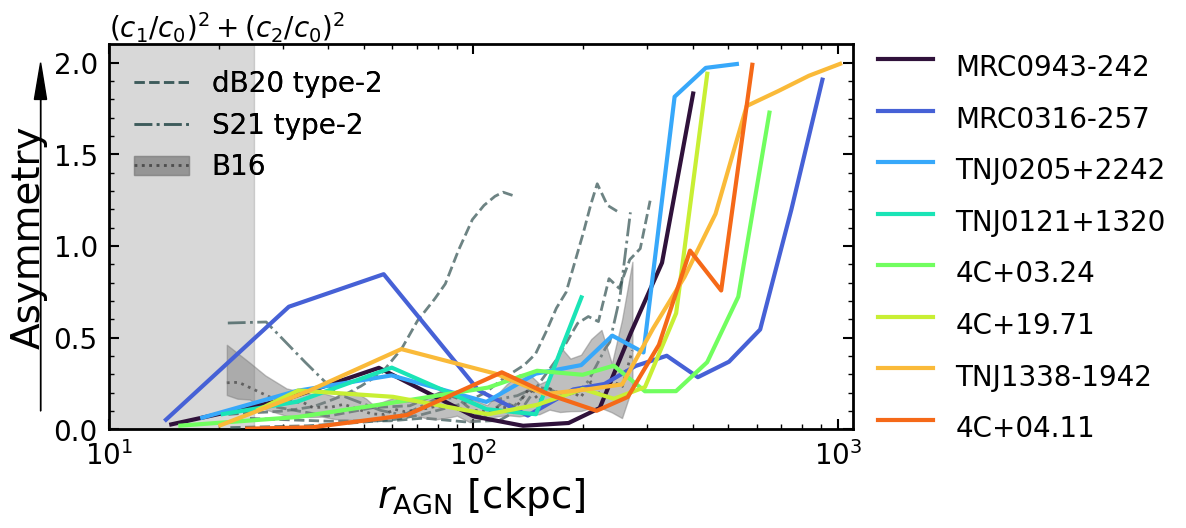}
      \caption{Radial profiles of the surface brightness Fourier decomposition (asymmetry measurement). The $c_{0}$, $c_{1}$ and $c_{2}$ are the 0th, 1st and 2nd modes Fourier decomposition coefficients of the surface brightness radial profile, respectively \citep[see][for definition]{denbrok2020}. The $(c_{1}/c_{0})^{2}+(c_{2}/c_{0})^{2}$ is a measurement of nebula asymmetry along the radial distance from the AGN. Our HzRGs are shown in solid color lines. $r_{\rm AGN}$ is the radial distance measured from the central AGN. For comparison, we include the same measurements for the 4 nebulae of type-2 quasars from \citet{denbrok2020} (grey dashed lines) and the type-2 from \citet{sanderson2021} (grey dot-dashed line). We also include the type-1 measurements from \citet{borisova2016} \citep[dotted line represents the median and shaded region marks the 25th and 75th percentile, quantified by][]{denbrok2020}. The vertical shaded region is the 0.75 arcsec ($\sim 25$ ckpc) range affected by median seeing of our sample \citep[the radial distance where the type-1 PSF is affected is $\sim50$ ckpc, see][]{denbrok2020}. The morphologies for most of the HzRGs nebulae are round (symmetric) $\lesssim100$~ckpc and become asymmetric at larger radial distances $\sim100$~ckpc (see text). }\label{fig:c_radial}
  \end{figure*}

   \begin{figure*}
   \centering
   \includegraphics[width=\textwidth]{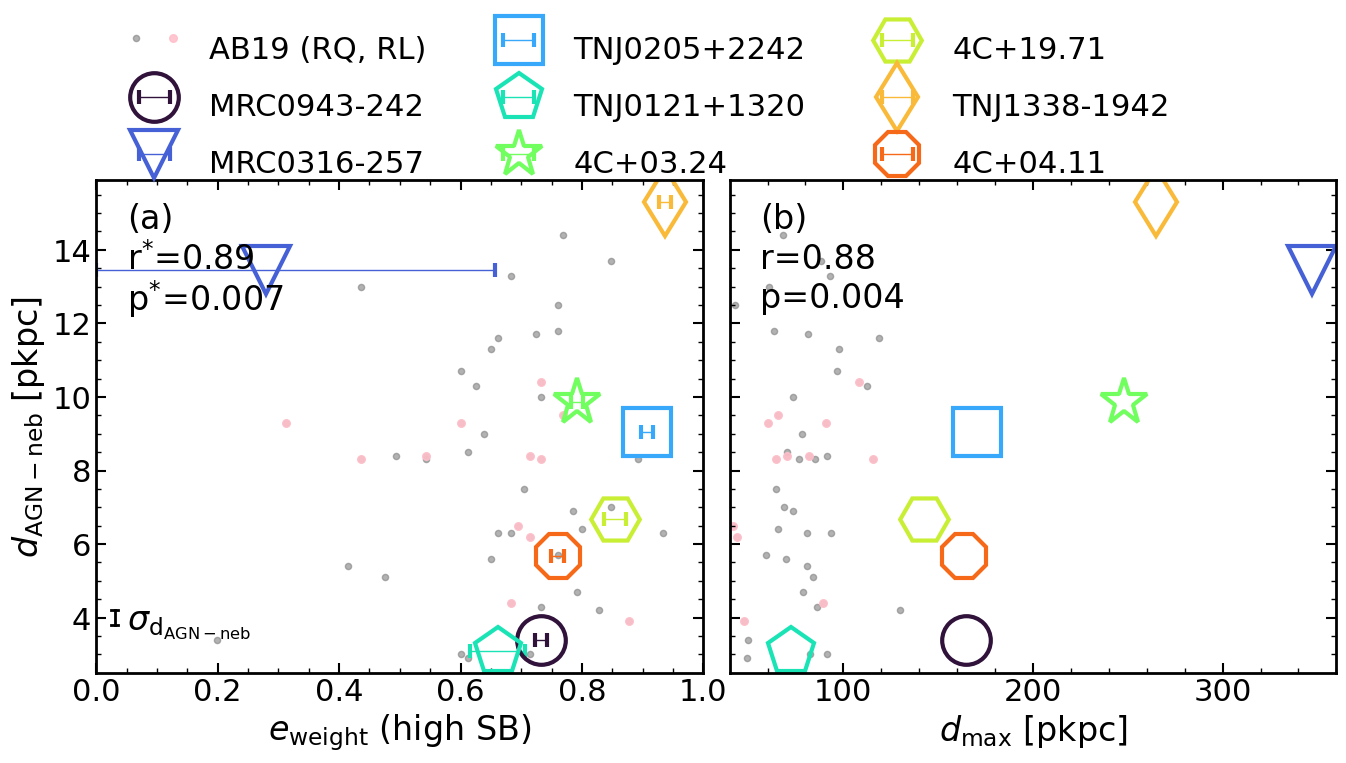}
      \caption{\textbf{(a)} Intrinsic flux-weighted ellipticity (sensitive to the morphology of the high surface brightness part of the nebula), $e_{\rm weight}$, versus distance offset between nebular centroid and AGN position, $d_{\rm AGN-neb}$. The typical uncertainty of $d_{\rm AGN-neb}$ ($\sigma_{\rm d_{\rm AGN-neb}}=0.4$~pkpc, Table \ref{tab:nebpropert}) is shown at bottom left. \textbf{(b)} Maximum Ly$\alpha$ nebula linear extent, $d_{\rm max}$, versus $d_{\rm AGN-neb}$. In both panels, we give the Spearman correlation measurements for our sample at the top left (the star superscript indicates the correlation is calculated excluding \object{MRC 0316-257} in panel (a)). We also include the data from \citet{FAB2019} in both panels in grey (radio quiet) and pink (radio loud) dots. We note that the \citet{FAB2019} sample shown in this figure is incomplete to concentrate on the relation for our targets. There are positive correlations detected for $e_{\rm weight}$-$d_{\rm AGN-neb}$ and $d_{\rm max}$-$d_{\rm AGN-neb}$.}\label{fig:asy_2comp}
   \end{figure*}

   \begin{figure}
   \centering
   \includegraphics[width=\hsize]{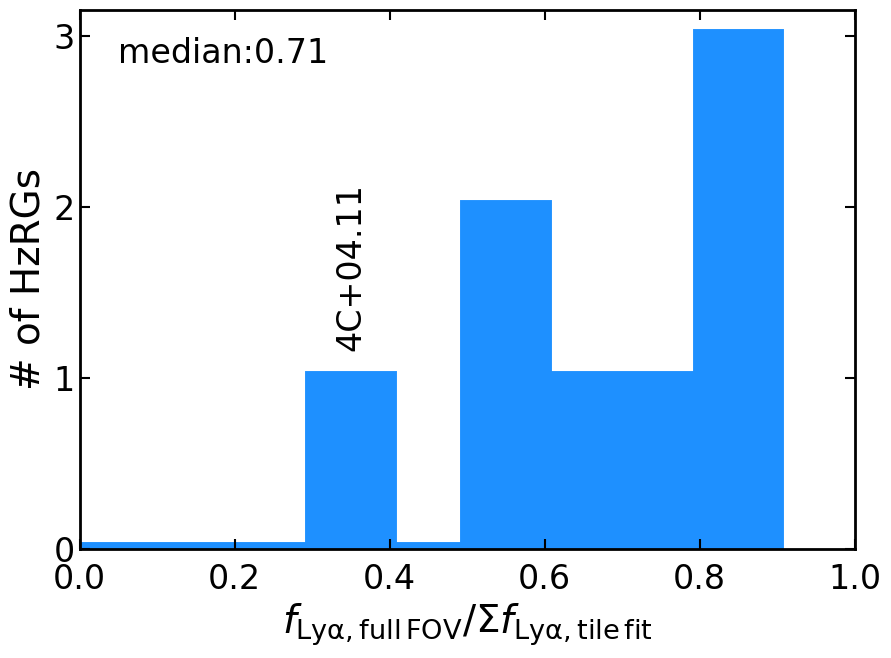}
      \caption{Distribution of $f_{\mathrm{Ly \alpha, full\,FOV}}/\Sigma f_{\mathrm{Ly \alpha, tile\,fit}}$ for our HzRGs. The $f_{\mathrm{Ly \alpha, full\,FOV}}$ is the intrinsic Ly$\alpha$ flux resulted from fitting the spectrum summed over the entire FOV. The $\Sigma f_{\mathrm{Ly \alpha, tile\,fit}}$ is the summation of the intrinsic Ly$\alpha$ flux in each tessellation bin (Fig. \ref{fig:map_int_1}b). The smaller the value, the more likely the Ly$\alpha$ photon is being double-counted when correcting for absorption. \object{4C+04.11} is the one with the smallest ratio which may also indicate over-correction (Appendix \ref{app:spatialfit}).  }\label{fig:hist_FOV_tile}
   \end{figure}

   \begin{figure}
   \centering
   \includegraphics[width=\hsize]{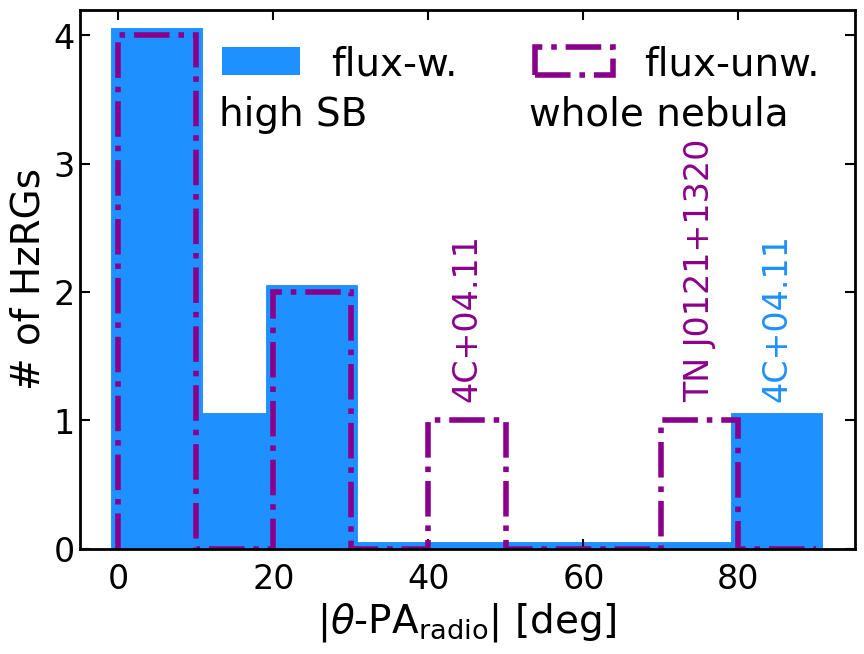}
      \caption{Distribution of the angle difference between nebular position angle and radio jet position angle, $|\theta-\rm{PA_{radio}}|$. The blue histogram shows the number distribution of flux-weighted angle difference, $|\theta_{\rm{weight}}-\rm{PA_{radio}}|$ (sensitive to high surface brightness nebula morphology). The magenta histogram represents the unweighted measurements, $|\theta_{\rm{unweight}}-\rm{PA_{radio}}|$ (quantifying the morphology of the whole nebula). The values are presented in Table \ref{tab:nebpropert}. We mark the obvious outliers in the two distributions in corresponding colors. This shows that most of our HzRGs have their Ly$\alpha$ nebula elongated along the jet axis.}\label{fig:PA_dis_hist}
   \end{figure}


\begin{table*}
 \caption{HzRGs nebulae properties. }\label{tab:nebpropert}
 \centering
\begin{tabular}{c c c c c c r r}

\hline
HzRG & $e_{\rm{weight}}$\tablefootmark{\rm{\dag}} & $e_{\rm{weight,\,obs}}$\tablefootmark{\rm{\dag}}  & $e_{\rm{unweight}}$\tablefootmark{\rm{\dag}} & $|\theta_{\rm{w.}}-\rm{PA_{radio}}|$\tablefootmark{\rm{*}}    & $|\theta_{\rm{unw.}}-\rm{PA_{radio}}|$\tablefootmark{\rm{*}} & $d_{\rm AGN-neb}$\tablefootmark{\rm{\ddag}} & $d_{\rm max}$\tablefootmark{\rm{$\mathsection$}} \\
   &  & & & deg & deg & \multicolumn{1}{c}{pkpc} & \multicolumn{1}{c}{pkpc}\\
\hline
MRC 0943-242  & 0.73$\pm$0.01 & 0.76 & 0.62 & 4.7$\pm$1.8  & 7.2    & 3.4$\pm$0.1 & 164\\
MRC 0316-257  & 0.28$\pm$0.38 & 0.96 & 0.87 & 28$\pm$54    & 20.8   & 13.5$\pm$0.2& 347\\
TN J0205+2242 & 0.91$\pm$0.01 & 0.79 & 0.75 & 2.1$\pm$2.3  & 8.6    & 9.0$\pm$0.1 & 170\\
TN J0121+1320 & 0.66$\pm$0.05 & 0.74 & 0.48 & 10.5$\pm$8.7 & 72.1  & 3.1$\pm$0.1  & 72\\
4C+03.24       & 0.79$\pm$0.01 & 0.72 & 0.82 & 8.1$\pm$1.2  & 2.0  & 9.9$\pm$0.1 & 248\\
4C+19.71       & 0.85$\pm$0.02 & 0.83 & 0.51 & 20.1$\pm$9.1 & 29.3 & 6.7$\pm$0.1  & 143\\
TN J1338-1942 & 0.94$\pm$0.01 & 0.87 & 0.90 & 1.3$\pm$1.4  & 1.2   & 15.3$\pm$0.1& 265\\
4C+04.11       & 0.76$\pm$0.01 & 0.20 & 0.59 & 81.4$\pm$1.3 & 45.3 & 5.7$\pm$0.1 & 163\\
 \hline
\end{tabular}
\tablefoot{\tablefoottext{\rm{\dag}}{Intrinsic flux-weighted ($e_{\rm{weight}}$), observed flux-weighted ($e_{\rm{weight,\,obs}}$) and unweighted elliptical asymmetric measurements. The flux-weighted ellipticity is sensitive to the morphology of the high surface brightness part of the nebula. The unweighted ellipticity quantifies the morphology of the whole nebula. We note that for $e\to0$, the nebula is closer to round shape and vice versa. }
\tablefoottext{\rm{*}}{Absolute difference between flux-weighted (unweighted) nebula position angle, $\theta_{\rm{weight}}$ ($\theta_{\rm{unweight}}$), and radio axis position angle, $\rm{PA_{radio}}$.} 
\tablefoottext{\rm{\ddag}}{Offset between the intrinsic flux-weighted centroid of Ly$\alpha$ nebula and AGN position.}
\tablefoottext{\rm{$\mathsection$}}{Maximum linear extent spanned by the Ly$\alpha$ nebula.} 
}

\end{table*}


\subsection{Morphology of the nebulae}\label{sec:1nebulaasy}

The nebula morphology is related to the ionising sources, gas dynamics and galaxy environment \citep[][]{Byrohl_2021,Costa2022,Nelson_2016}. Especially when the Ly$\alpha$ nebulae (in our sample) can probe the CGM gas beyond 100~pkpc. By visual inspection, we observe that the shape of our Ly$\alpha$ nebulae are asymmetric (e.g. Fig. \ref{fig:map_int_1}). In this section, we study quantitatively the nebula morphology. We first focus on the whole nebula by introducing the morphology quantification measurements (ellipticity, nebula orientation and offset between nebula centroid and AGN position) in Sect. \ref{sec:2neb_mor_quan} and compare with other samples in Sect. \ref{sec:2neb_asymm}. Then, in Sect. \ref{sec:2asym_rad}, we study how the nebula asymmetry changes with radial distance for individual targets. We also report the detected morphology correlations between different measurements (i.e. offset between AGN and nebula centroid position, nebula ellipticity and nebula linear size) in Sect. \ref{sec:2mor_rela}. These shed light on how the central quasar and nearby companions can affect the observed nebula morphology. Finally, in Sect. \ref{sec:2padif_dist}, we show the non-random oration of jet axis and its relation with the elongated direction of nebula which hints at the CGM gas distribution.

\subsubsection{Morphology quantification measurements}\label{sec:2neb_mor_quan}
To quantify the asymmetry, we introduce a set of morphology measurements. \citet{FAB2019} used flux-weighted asymmetry measurements for the Ly$\alpha$ nebulae which is sensitive to the high surface brightness part. In other works \citep[e.g.][]{denbrok2020}, an unweighted asymmetry measurement was adopted for better studying the extended structure of the nebulae (sensitive to the morphology of the whole nebula). To better characterise the morphology of our HzRGs and perform comparison with other samples, we analyse the asymmetry with both the flux-weighted and flux-unweighted methods.

First, we follow \citet{FAB2019} and calculate the flux-weighted asymmetry, $\alpha_{\rm{weight}}$ \citep[see][for the definition]{FAB2019}. This quantifies the asymmetry of the nebula in two perpendicular directions. Together with the asymmetry measurement, we also obtain the flux-weighted position angle $\theta_{\rm{weight}}$, which we use as an indicator for the elongation direction of the nebula after converting it to the same reference system as the radio jet axis (i.e. angle measured east from north). The flux-weighted nebula centroid (centre of the nebula) is also computed. We note that the intrinsic flux and its uncertainty are used as weight to measure these three parameters and to calculate the corresponding uncertainties, respectively. We also derive the asymmetry measurement weighted by observed flux for comparison. Second, to compare with flux-unweighted asymmetry reported for other quasar samples \citep[e.g.][]{borisova2016,denbrok2020}, we calculate the $\alpha_{\rm{unweight}}$ following \citet{denbrok2020}. In this context, we also derive $\theta_{\rm{unweight}}$ (flux-unweighted position angle) to examine the jet-nebula relation with respect to the entire nebula. Fig. \ref{fig:mapradind} visualises the weighted (nebula centroids and $\theta_{\rm{weight}}$) and unweighted ($\theta_{\rm{unweight}}$) parameters on the eight nebulae. 

In \citet{Lau_2022}, the authors compared morphology of different quasar samples. Following their comparison, we convert the aforementioned asymmetry measurement (for both flux-weighted and unweighted), $\alpha$, to an intuitive elliptical asymmetry measurement (or ellipticity) $e = \sqrt{1-\alpha^{2}}$. For $e_{\rm{weight}}\to0$, the nebula is closer to round shape and vice versa. Table \ref{tab:nebpropert} reports the morphological parameters of our sample. Since the absolute flux-weighted centroid position and $\theta_{\rm weight}$ (and $\theta_{\rm unweight}$) are irrelevant, we report the projected distance between the nebula centroid and the AGN position ($d_{\rm AGN-neb}$) and the difference in angles between $\theta_{\rm weight}$ (and $\theta_{\rm unweight}$) and the jet position angle ($|\theta_{\rm weight}-\rm{PA_{radio}}|$ and $|\theta_{\rm unweight}-\rm{PA_{radio}}|$), respectively. The jet position angle ($\rm{PA_{radio}}$) is shown in Fig. \ref{fig:mapradind} and listed in Appendix \ref{app:nebradinfo}. 

\subsubsection{Comparison of nebula asymmetry with other quasar samples}\label{sec:2neb_asymm}
Fig. \ref{fig:asye_vs_llya} presents the ellipticity measurements as a function of their nebula Ly$\alpha$ luminosity for our targets and other quasars \citep[][]{borisova2016,FAB2019,cai2019,mackenzie2021,denbrok2020,sanderson2021, Lau_2022}. We note that the $L_{\rm Ly\alpha}$ for comparison samples are not corrected for absorption. Part of the comparison samples are also used in Sect. \ref{sec:2rad_sb_pro} for surface brightness radial profile analysis. We point out the newly included ones here: faint $z\sim3.0$ type-1 from \citet{mackenzie2021} and type-2 AGN at $z\sim3.4$ from \citet{denbrok2020} and $z\sim3.2$ from \citet{sanderson2021}. The reason why they are not included in radial profile analysis is that they do not add new information. 

The HzRGs from our sample are measured to be asymmetric for their high surface brightness emission region (median $e_{\rm{weight}}\approx 0.78$). Compared to the \citet{FAB2019} and \citet{cai2019} samples, our HzRGs are consistent in asymmetry measurements and on the higher end of their distribution (type-1 median $e_{\rm{weight}}\approx 0.72$, dashed horizontal in Fig. \ref{fig:asye_vs_llya}a). The $e_{\rm{weight}}$ of radio-loud type-1s from \citep[][]{FAB2019} have a median of 0.69 which is even lower than the value of all type-1 targets \citep[][]{FAB2019,cai2019}. This indicates that the radio emission in type-1 does not disturb the gaseous nebula as in our HzRGs at least along the plane of the sky. This further suggests that orientation is a critical factor (Sect. \ref{sec:1scatter_dis}). By comparing the intrinsic flux-weighted and observed flux-weighted elliptical asymmetry measurements, we find the $e_{\rm{weight}}$ can vary significantly (e.g. \object{MRC 0316-257} and \object{4C+04.11}). For \object{MRC0316-257}, we can already identify its asymmetric morphology through visual checking (Fig. \ref{fig:map_int_1}). There is also a large error bar associated with the intrinsic flux-weighted ellipticity. Hence, the morphology of \object{MRC0316-257} is more towards the asymmetric end. The large difference between its intrinsic and observed $e_{\rm{weight}}$ could be due to the absorption correction elevates the flux difference between the high and low S/N regions thus gives more weight to the central nebula. As for \object{4C+04.11}, this could be due to the potential over-correction of the absorption in the low S/N regions given we use nine absorbers across the nebula (Sect. \ref{sec:1specfitmeth}, \ref{sec:1discabcorr}). However, we point again that the absorption is necessary to reconstruct the intrinsic flux given that the absorption features across the nebula were observed \citep[e.g.][]{swinkbank2015,wang2021}.

Our HzRGs have a median $e_{\rm{unweight}}\approx 0.70$ which is lower than the measurement of type-2s and relatively consistent with the type-1s \citep[median $e_{\rm{unweight}}$ $\approx0.80$ and $\approx0.69$, respectively, ][]{borisova2016,denbrok2020,sanderson2021,mackenzie2021}. The ellipicity of the comparison type-2s is calculated from five sources which may not be representative. We note that there is a large scatter in the $e_{\rm{unweight}}$ measured for our sample with four clustered around the type-2s and other four below the median $e_{\rm{unweight}}$ of type-1s. This result could be biased by the implemented analysis methods (i.e. detection map and tessellation, Sect. \ref{sec:1nebuseltes}): (i) the construction of the detection map may smooth out the nebula asymmetry at larger radii (lower S/N); (ii) the tessellation results in lager bin sizes along the direction perpendicular to the radio jet (lower surface brightness and S/N, Sect. \ref{sec:1dis_gaslarge}) which shapes the nebula morphology to be more round. This brings more effects of the quantification of the entire nebula. Hence, the HzRGs nebulae will likely have larger $e_{\rm{unweight}}$ (i.e. more asymmetric) than the quantified value. In spite of that, we can conclude that at least half of our sample, together with the type-2s, distribute at the most asymmetric end in terms of the whole nebula. The rest has the possibility to be skewed to higher $e_{\rm{unweight}}$. Our HzRGs have diverse properties and are not necessarily representative for the entire HzRGs population. Inspection for individual targets is required. 

As for the $L_{\rm Ly\alpha}$, our HzRGs host the most luminous Ly$\alpha$ nebula compared with other quasar types. This is due to: (i) sample selection\footnote{We can see that even the absorption uncorrected observed $L_{\rm Ly\alpha}$, smaller symbols in Fig. \ref{fig:asye_vs_llya}, is on the higher end (Sect. \ref{sec:2sample_sel}).}; (ii) absorption correction; (iii) quasar PSF subtraction of comparison samples (Sect \ref{sec:2sb_prof_fit}). 

We further compare our HzRGs to the ERQ from \citet{Lau_2022}. In Fig. \ref{fig:asye_vs_llya}, both the $e_{\rm{weight}}$ (0.44) and $e_{\rm{unweight}}$ (0.69) of L22 are lower than the measurements from HzRGs. Its $L_{\rm Ly\alpha}$ is also lower than our HzRGs by $\sim2$ orders of magnitude. This confirms that the nebula of this EQR is type-1 like but highly obscured \citep[e.g.][]{Lau_2022}. Since it is the only source of this type, we do not further discuss it.

\subsubsection{Asymmetry radial profile}\label{sec:2asym_rad}
 To further quantify the morphology of individual nebula as a function of distance from the AGN, we follow \citet{denbrok2020} and decompose our HzRGs intrinsic surface brightness using Fourier coefficients, $a_{k}(r)$ and $b_{k}(r)$:
 \begin{equation}
     \mathrm{SB_{Ly\alpha}}(r,\theta)= \sum_{k=0}^{\infty}\left[a_{k}(r)\cdot \cos(k\theta)+b_{k}(r)\cdot \sin(k\theta)\right]\text{,}
 \end{equation} where $\mathrm{SB_{Ly\alpha}}(r,\theta)$ is the surface brightness at given polar coordinate $(r,\theta)$. The coefficients $a_{k}(r)$ and $b_{k}(r)$ are defined as:
  \begin{equation}
      \begin{split}
    a_{k}(r) & = \frac{1}{2\pi} \int^{2\pi}_ {0} \mathrm{SB_{Ly\alpha}}(r,\theta) \cdot \cos(k\theta) d \theta \\
    b_{k}(r) & = \frac{1}{2\pi} \int^{2\pi}_ {0} \mathrm{SB_{Ly\alpha}}(r,\theta) \cdot \sin(k\theta) d \theta\text{.}
    \end{split}
 \end{equation}
The detailed description can be found in  \citet{denbrok2020}. Then we combine $a_{k}(r)$ and $b_{k}(r)$:
\begin{equation}
    c_{k}(r) = \sqrt{a_{k}(r)^{2} + b_{k}(r)^{2}}.
\end{equation}
We measure the radial dependence of the asymmetry (i.e. how much it deviate from a circular shape) of the nebulae by using the ratio between $k$th and 0th modes, $c_{k}(r)/c_{0}(r)$. In practice, only the first three modes are used since the higher mode coefficients are much smaller \citep[][]{denbrok2020}. Fig. \ref{fig:c_radial} shows the radial profiles of this asymmetry measurement (represented by $(c_{1}/c_{0})^{2}+(c_{2}/c_{0})^{2}$ on the y-axis) for our HzRGs. The larger the value, the more asymmetric the morphology shift from circular shape at a given radial distance. We also include type-2 measurements from \citet{denbrok2020} (4 quasars) and \citet{sanderson2021} as comparison. The type-1 sample from \citet{borisova2016} \citep[which is quantified in][]{denbrok2020} is also shown. Our HzRGs generally show an increase of the asymmetry as a function of radial distance (some have a smaller secondary peak at $\lesssim 50$ ckpc) and a steep rise to $>1.5$ (7 out of 8) at $\sim 250$ ckpc. The most noticeable exception is \object{MRC0316-257} which has a secondary peak ($\sim0.8$) at $\sim50$ ckpc ($\sim14$ pkpc). This flux-weighted measurement confirms the large difference we observed in the high surface brightness part of the directional profiles of \object{MRC0316-257} in Fig. \ref{fig:rad_pf_fit}. As we already stated in Sect. \ref{sec:1rad_prof}, the detected extent of our HzRGs ($\gtrsim400$ ckpc) are larger than the comparison quasars ($\sim 300$ ckpc). If we limit the comparison to the largest extent ($\sim 250$ ckpc) reached by the type-1s from \citep{borisova2016}, we find that our HzRGs are similar in radial asymmetry measurements. Specifically, both have a relatively flat profile to $\sim 200$ ckpc which followed by a shallow rise to the value of $\sim0.5$. However, in Sect. \ref{sec:2neb_asymm} (Fig. \ref{fig:asye_vs_llya}) we find that the HzRGs ellipticity is larger than type-1s. Together with the radial asymmetry profile in this section, we can conclude that the asymmetry of the nebulae associated with HzRGs and type-1s are different due to structures at larger radial distance ($>250$ ckpc) from AGN. This is also suggested by \citet{denbrok2020}. Although we caution the large $W_{80}$ at large distance in Sect \ref{sec:2kin_prof} and mark the region with $r>5$ arcsec in grey in Fig. \ref{fig:v50_W80_r}, we can still find that most targets have at least $W_{80}\sim10^{3}\,\mathrm{km\,s^{-1}}.$ This may indicate that these structures are likely disturbed and not `quiescent' gas in the Cosmic Web (Sect. \ref{sec:0discuss}). From the comparison with type-2s \citep[][which also have a steep rise]{denbrok2020,sanderson2021}, we find that the radial asymmetry of our HzRGs rises at larger radial distance ($\gtrsim 250$ ckpc) and reaches higher asymmetry value ($>1.5$ compared with $\sim1.4$ of type-2s). At least one of the radial asymmetry measurement (Cdfs 15, with the furthest extent $r_{\rm AGN}\sim 300$ ckpc) from \citet{denbrok2020} shows an indication of continuous rising up to the detection limit (1 hour integration with MUSE); we cannot rule out the possibility that this target may show a similar trend as the HzRGs with deeper observations.
 
 \subsubsection{Morphological correlations}\label{sec:2mor_rela}
 
 In Fig. \ref{fig:asy_2comp}, we compare the $e_{\rm weight}$ (ellipticity that is sensitive to high surface brightness nebula) and $d_{\rm max}$ (maximum nebula extent) against $d_{\rm AGN-neb}$ (offset between AGN and nebula flux-weighted centroid). For a consistent comparison for the unweighted measurements, we should use the unweighted offset (centroid) and the unweighted ellipticity. We note that the calculation of unweighted ellipticities, $e_{\rm unweight}$, assumes that the centre of nebula corresponds to the AGN position \citep[][]{denbrok2020}. It would involve large uncertainties if we calculated the flux-unweighted nebula centroids (i.e. the geometric center of the nebula) which are entirely depended on the spaxel distribution from our detection maps (Sect. \ref{sec:2max_neb_ext}).  Hence, we report only the correlation between flux-weighted ellipticity and offset, and use it as a proxy of the nebula even though they are more sensitive to the high surface brightness part. 
 
 The (r-value, p-value) of Spearman's correlation coefficients are (0.89, 0.007) and (0.88, 0.004) for the $e_{\rm weight} \, vs. \, d_{\rm AGN-neb}$ and $d_{\rm max} \, vs. \, d_{\rm AGN-neb}$ relations. We note that due to the large uncertainties of $e_{\rm weight}$ for \object{MRC0316-257}, we exclude it from the correlation measurement. From the r-values and p-values, we can conclude that $e_{\rm weight}$ and $d_{\rm max}$ are both correlated with $d_{\rm AGN-neb}$. This suggests that the more asymmetric and more extended nebulae have larger offsets between AGN position and nebular centroid. We also include the \citet{FAB2019} sample for comparison. The (r-values, p-values) of the $e_{\rm weight}-d_{\rm AGN-neb}$ relation for the \citet{FAB2019} whole sample and radio-loud targets are (0.3, 0.02) and (0.3, 0.4), respectively, implying no strong correlations. The  (r-values, p-values) of the $d_{\rm max}-d_{\rm AGN-neb}$ relation for the \citet{FAB2019} sample and the radio-loud targets are (0.4, 0.003) and (0.4, 0.1), respectively, suggesting a moderate correlation in their whole sample but not in their radio-loud target. This indicates that the radio-loud type-1s are different from our radio-loud type-2s HzRGs. We will revisit these correlations in Sect. \ref{sec:0discuss}. 

 \subsubsection{Jet-nebula position angle distribution}\label{sec:2padif_dist}
 We present the distribution of the $|\theta-\rm{PA_{radio}}|$ (Sect. \ref{sec:2neb_mor_quan}, Table \ref{tab:nebpropert}) in Fig. \ref{fig:PA_dis_hist}. Both the flux-weighted and unweighted measurements are shown which are sensitive to high-surface brightness parts and the entire nebulae, respectively. Fig. \ref{fig:PA_dis_hist} shows that most HzRGs have $|\theta-\rm{PA_{radio}}| <30^{\circ}$. A similar result was reported by \citet{Heckman_1991a}. This observation is consistent with the scenario proposed by \citet{Heckman_1991a,Heckman_1991b} that the compression of the gas by the radio jet results in brighter emission along the jet. We will discuss the indications behind the alignments further in \ref{sec:1dis_gaslarge}. The exception is \object{4C+04.11} which has both large $|\theta_{\rm weight}-\rm{PA_{radio}}|$ ($81.4^{\circ}$) and $|\theta_{\rm{unweight}}-\rm{PA_{radio}}|$ ($45.3^{\circ}$), indicating different conditions that in the other targets (such as inflows, Sect. \ref{sec:1discinflow}). In \object{TN J0121+1320} we find a flux-unweighted angle difference of $72.1^{\circ}$. Given its round Ly$\alpha$ nebula morphology and compact radio emission, there is a large uncertainty in angle difference measurement and we do not discuss this source separately. If we assume the observed angle difference is equally distributed from $0-90^{\circ}$, the probability for us to find 7 (or 6) targets in a sample of 8 with $|\theta-\rm{PA_{radio}}|<30^{\circ}$ is only 0.2\% (1.7\%) using bimodal distribution.

\section{Discussion}\label{sec:0discuss}
\subsection{Absorption correction and radiative transfer}\label{sec:1discabcorr}
We stress that the analysis of our sampled Ly$\alpha$ nebulae in this paper is under the idealised assumptions stated in Sect. \ref{sec:2fitmodel}. Specifically, we interpret the trough features in the Ly$\alpha$ spectra as absorption features by the neutral hydrogen gas surrounding the radio galaxy. In this section, we discuss the limitations and possible caveats of this treatment for reconstructing the intrinsic Ly$\alpha$ flux along with the possible effects brought by radiative transfer.

Under our assumptions, the absorbing gas is located in between the observer and the last scattering surface of Ly$\alpha$ photons along the line of sight. The intrinsic Ly$\alpha$ flux reconstructed under our treatment is thus assumed to be the one after radiative transfer processes have shaped the Ly$\alpha$ nebula, and is approximated by a Gaussian profile. We also assume the absorbed Ly$\alpha$ photons in the absorbers to be `lost' rather than continuing their resonant scattering path within the nebula; this may happen when photons are absorbed by dust and re-emitted as infrared thermal emission, or preferentially scattered away from the line of sight due to a particular geometry (see Sect. \ref{sec:2fitmodel}). The latter may occur because photons originating from the backside of the galaxy have a higher chance to be absorbed by dust when transiting the host galaxy \citep{liu2013_b}.  These absorbers can be interpreted as intervening low column density gaseous shells \citep[e.g.][]{vanojik1997a,swinkbank2015,kolwa2019}. The fact that most of the absorbers are located in the blue wing of the Ly$\alpha$ profile as well as the fact the trough of the main absorber is often seen across several tens of kpc scales supports this idea \citep[see Fig. \ref{fig:map_int_1}a, and \object{TNJ0205+2242} in Fig. \ref{fig:tnj0205_spa}, see also \object{TNJ1338-1942} in Fig. \ref{fig:tnj1338_spa} and e.g.][]{wang2021}.

The situation may be different when we encounter embedded absorbers which are supposed to be closer to the source of Ly$\alpha$ photons (i.e. central AGN in our case). This leaves the trough (or `main absorber') at around the systemic redshift of the AGN as predicted by radiative transfer studies \citep[e.g.][]{Verhamme2006}. Even though the scales probed in those simulations are different (CGM scale versus sub-kpc), this might be the case in some of our targets, for example \object{MRC0943-242} and \object{TNJ0121+1320} (Fig. \ref{fig:tnj0121_spa} and \ref{fig:mrc0943_spa}). Therefore, one consequence of our absorption treatment (with `lost' photons) is that we may double-count Ly$\alpha$ photons that are resonantly scattered to the wing and/or other directions by redundantly adding more when correcting for absorption. We implemented a simple test for checking the double-counting effect, which takes advantage of the IFU nature where slit or spaxel loss effects are compensated by photons resonantly scattered in the neighbouring spaxels. Specifically, we summed all individual spectra into one and reconstruct the intrinsic flux for this single spectrum ($f_{\mathrm{Ly \alpha, full\,FOV}}$); this value removes the IFU information, but should be a good measure of the total value emanating from the nebula in the observer's direction. This is then compared to the sum of individually constructed intrinsic flux in each tile (Fig. \ref{fig:map_int_1}b, $\Sigma f_{\mathrm{Ly \alpha, tile\,fit}}$). Fig. \ref{fig:hist_FOV_tile} shows the result, where a value of 1.0 is expected if our treatment is fully accurate. Instead, we observe a median value of 0.71 for our sample, indicating that there may be double-counting of photons. However, several effects may also contribute to this. First, our assumption of a Gaussian shape for the intrinsic profile may not be accurate, as prior radiative transfer effects may have created troughs and boosted the line wings \citep[][]{Verhamme2006}. A future paper presenting high spectral resolution UVES  observations of our sample will help to better model the profiles (Ritter et al. in prep.). Second, the assumption that all absorbers extend across the entire nebula may also be oversimplified. While many absorbers are indeed seen on 10s (or 100s) of kpc scales for our targets (Fig. \ref{fig:tnj0205_spa}, \ref{fig:tnj0121_spa} and \ref{fig:4c03_spa} and also Appendix \ref{app:suppmap}), there may be exceptions.



The Ly$\alpha$ nebula properties of the type-1 (and radio-quiet type-2) sources in the comparison samples are derived without correcting for absorption. They are still good comparison samples given that not many absorption features are detected in them \citep[e.g.][]{FAB2019}. This fact alone is already an indication that there may be an environmental differences between the nebulae of HzRGs and other quasars. Alternatively, this difference may related to intrinsic differences between different AGN types.

Overall, it is likely that both the absorption and radiative transfer effect are working together shaping the the Ly$\alpha$ profile. Our analysis assumes that the absorption correction is the dominant factor.

\subsection{Ly$\alpha$ nebula and AGN unification model}\label{sec:1dis_nebtype12}
The unification model of AGN \citep[e.g.][]{Antonucci1993} proposed that the fundamental difference between type-1 and type-2 is due to the different orientation of the obscuring dusty torus (ionisation cone). Specifically, the ionisation cone of type-1s is more aligned with the observed line of sight than in type-2s. 

Within this unification model, we assume that the power of AGN is on a similar level for type-1s and type-2s and that their gaseous nebulae therefore have similar distributions and masses. If we further assume that the ionising photons are primarily produced by the AGN, then the nebulae should have similar morphologies and luminosities. In this picture, the Ly$\alpha$ nebulae are elongated along the direction of the ionisation cone and have a `rugby-ball' shape. For type-1s, the ionisation cone would be pointing towards the observer resulting in a rounder nebula morphology. For type-2s, the ionisation cone would aligned along the plane of the sky resulting in the observed elliptical morphology. Evidence for such a scenario was indeed reported in \citet{zhicheng_he2018} using ionized gas nebulae but for local AGN and on small scales (sub-kpc to kpc).

Using both the $e_{\rm weight}$ (sensitive to the high surface brightness nebula) and $e_{\rm unweight}$ (whole nebula) quantifying the ellipticity of the nebula, we find that the HzRGs (and other radio quiet type-2s) are more asymmetric than the type-1s in Sect. \ref{sec:2neb_mor_quan}. This observation agrees with the AGN orientation unification model. The orientation of type-1s probably still vary in a range which causes the large range of the ellipticities. The AGN luminosity and dark matter halo mass (gas distribution) can also play an important role in shaping the observed morphology (Fig. \ref{fig:asye_vs_llya}).

With the jet axis indicating the direction of the ionisation cone \citep[][]{drouart2012} and the alignment between the jet axis and the Ly$\alpha$ nebula (at least in the high surface brightness part where the photons of AGN are presumably dominating the ionisation, Sect. \ref{sec:2padif_dist} and Fig.\ref{fig:PA_dis_hist}), we argue that the AGN orientation between type-1s and type-2s (HzRGs) can explain the observed morphological difference. The evidence for this claim comes mostly from the observed ellipticity distribution (Fig. \ref{fig:asye_vs_llya}): the $e_{\rm weight}$ (median 0.69, Sect. \ref{sec:2neb_asymm}) for radio-loud type-1s \citep[][]{FAB2019} are lower than our HzRGs (median 0.78), which is consistent with the jet (ionisation cone)  pointing more towards observers in type-1s. By checking the available radio maps of these radio-loud type-1s \citep[e.g. Fig. B1 in][]{FAB2019}, we indeed find that they have compact radio morphology (i.e. not the two-side jet like HzRGs) suggesting that the jets are aligned more perpendicular to the plane of the sky. We note that \object{TNJ0121+1320} also has a compact radio morphology (i.e. not having two-side jet) and the lowest $e_{\rm weight}$ (0.66) and $e_{\rm unweight}$ (0.48) in our sample (more consistent with type-1 results, see Fig. \ref{fig:map_int_1} and \ref{fig:asye_vs_llya}) which again fits the unification model.

A similar explanation was also suggested by \citet{denbrok2020} based on Ly$\alpha$ nebula studies of type-2 quasars (also included as comparison sample in this paper). \citet{denbrok2020} suggested that the orientation difference based on unification model can explain the nebula morphology at radial distance $\gtrsim 40$~pkpc. Using the same radial asymmetry measurement in Fig.\ref{fig:c_radial} (Sect. \ref{sec:2asym_rad}), we also find large nebula asymmetries at $\gtrsim 40$~pkpc ($\sim 200$~ckpc) in our HzRGs following \citet{denbrok2020} type-2s. \citet{denbrok2020} found more symmetrical morphologies for their type-2s at small radial distances $<30$~pkpc that are more similar to type-1s. \citet{denbrok2020} suggested that additional ionising sources other than the AGN (e.g. star forming processes) may contribute to this observation and smear out the differences between type-1 and type-2 at such small radii. As for our HzRGs, the jet-gas interaction at $\lesssim 20-30$~pkpc ($\sim100$~ckpc) could be a reason for the observed high $e_{\rm weight}$ shown in Fig. \ref{fig:asye_vs_llya}a (compared to type-1s) and the reason of the small rise (showing higher asymmetry compared to type-2s) in the radial asymmetry measurement at $\sim60$~ckpc in Fig. \ref{fig:c_radial}.

\subsection{The role of Ly$\alpha$ resonant scattering}\label{sec:1scatter_dis}
As presented in Sect. \ref{sec:0intmaps} and \ref{sec:1nebulaasy}, Ly$\alpha$ nebulae around our HzRGs are extended in size ($\gtrsim150$~pkpc) and are asymmetric in shape. Interestingly, there is a correlation between the maximum nebula extent $d_{\rm max}$ ($e_{\rm weight}$: ellipticity measurement that is sensitive to high surface brightness nebula) and the offset between AGN position and nebulae's flux-weighted centroid $d_{\rm AGN-neb}$ (Sect. \ref{sec:2mor_rela}). This correlation may also reflect our findings regarding the surface brightness and kinematic radial profiles in Sect. \ref{sec:2rad_sb_pro} and \ref{sec:2dir_sb_prof}, such as the exponential shape of the surface brightness in the inner nebula and high $W_{80}$ value. The resonant nature of Ly$\alpha$ photons may be related to this observation.

\citet{Villar-martin_1996} first proposed the idea of resonant scattering being related to the observed extended nebula emission around HzRGs. In simulations, \citet{Costa2022} found that the scattering of Ly$\alpha$ photons, regardless of the powering source, could result in an offset between the luminosity centroid of the nebula and the quasar position ($d_{\rm AGN-neb}$). This offset can vary depending on the line of sight and can reach $\sim15$ pkpc. Such offsets are consistent with what we find (Table \ref{tab:nebpropert}). The authors also found that scattering can shape the nebula into a more asymmetric morphology ($e_{\rm weight}\to1$). This depends on the gas distribution of neutral hydrogen and observed line of sight as described in \citet{Costa2022}. Given the common case that gas density is the highest on galaxy scales, and that we target type-2 AGN, we expect the Ly$\alpha$ photons to scatter out to larger projected distances rather than travelling directly towards the observer. Such a scenario may explain our observed correlation in Fig. \ref{fig:asy_2comp} between $e_{\rm weight}$ and $d_{\rm AGN-neb}$. Specifically, when the scattering is more efficient, we may observe the nebula to be more asymmetric and more extended, at least in the high surface brightness regime. 

The inner part (luminous) of the surface brightness radial profiles have an exponential shape (Sect. \ref{sec:2rad_sb_pro}). This gradual change in surface brightness and profile slope at smaller radii is suggested to be due to scattering \citep[][]{Costa2022}. The high $W_{80}$ values out to radius of $\sim50$~pkpc (or $\sim230$~ckpc, Sect. \ref{sec:2kin_prof}) could also be related with efficient scattering \citep[Fig. B1 in][]{Costa2022}. The velocity shift of the nebulae can also be complicated due to scattering at the similar radial distance \citep[Fig. B1 in][]{Costa2022} which is the case of our $v_{50}$ (Fig. \ref{fig:map_int_1}c, Sect. \ref{sec:0intmaps}). As shown by the stellar radial profiles (Fig. \ref{fig:rad_pf_fit}), the impact of seeing cannot be fully responsible for the exponential shape at smaller radii. For the observed kinematics, the jet (which is not included in the  simulations) may also play a role here. We overlay the jet hot spot distances on the radial profiles (Fig. \ref{fig:rad_pf_fit} and \ref{fig:v50_W80_r}). Qualitatively, the $v_{50}$ and $W_{80}$ show different behaviours within and outside the extent of the jet hot spots at least for some targets (e.g. \object{MRC0316-257}). We note that the distances marked by the vertical lines are measured from the brightest radio emission positions, i.e. the full extent of jet is even larger (Appendix \ref{app:nebradinfo}). The decrease of $W_{80}$ at  large radii ($\gtrsim100$~pkpc) observed in some targets (e.g. \object{TN J1338-1942}) is, beyond the scope of the \citet{Costa2022} simulations but could be related to the (unvirialised) cosmic web. 

 In the framework where scattering significantly impacting the observed Ly$\alpha$ properties, \citet{Costa2022} furthermore predicts that Ly$\alpha$ nebulae of edge-on AGN should have lower surface brightness, larger $d_{\rm AGN-neb}$, more asymmetric shape and flatter surface brightness profiles in the centre. Most of the predictions agree with our observations except for lower surface brightnesses, which could be due to our selection criteria and/or jet impact. We note that the analysis of \citet[][]{Costa2022} has been done without correcting for absorption which is reason of the discrepancy.
 
 The orientation (Sect. \ref{sec:1dis_nebtype12}) may be the reason for the moderate $d_{\rm max}-d_{\rm AGN-neb}$ relation seen in the sample of \citep{FAB2019} (i.e. the orientation spans a large range within the sample). The larger the viewing angle \citep[i.e. the more edge-on we are observing the AGN, assuming unification model,][]{Antonucci1993}, the more extended the nebula is expected to be, because both sides of the nebula will be observed. The nebula is expected to become more asymmetric and have larger $d_{\rm AGN-neb}$ as \citet{Costa2022} predicted. The study of type-2 quasars \citep[e.g.][]{denbrok2020,sanderson2021} also found that the difference in nebula morphology when comparing to type-1s is related to AGN orientation. Interestingly, for our HzRGs, we do find a relatively strong correlation between $d_{\rm max}$ and $d_{\rm AGN-neb}$ despite the expectation that most HzRG are observed under similar, large viewing angles, (i.e. close to edge-on). All of our targets have clear two-sided radio lobe morphology (except \object{TN J0121+1320}) and none of them shows clear signs of broad-line emission or significant contamination by a bright point-like source in the centre \citep[see also][and Sect. \ref{sec:2jetk}]{drouart2012}. 
 
Resonant scattering has the potential to shape the observed inner parts of the radial profiles (surface brightness and kinematics) and the morphology correlations.
 

\subsection{Large scale environment: nearby Ly$\alpha$ emission halos}\label{sec:1dis_env}
\citet{Byrohl_2021} studied Ly$\alpha$ emission halos and their relation with environmental factors using TNG50 simulation. They found a flattening in the Ly$\alpha$ halo surface brightness radial profile at large radial distances ($\gtrsim30$ pkpc). The authors attributed this to the contribution of scattered photons from nearby halos (companions). In Sect. \ref{sec:1rad_prof} (and \ref{sec:2sb_prof_fit}), we show that the profiles of three of our targets (\object{MRC0943-242}, \object{MRC0316-257} and \object{TN J1338-1942}) also have an obvious flattening at large radii (Fig. \ref{fig:rad_pf_fit}). Coincidentally, these targets are known to live in dense environments \citep[on Mpc scale, e.g.][]{Venemans2007a}. All of this indicates that our nebulae are `contaminated' by Ly$\alpha$ halos associated with nearby companions \citep[e.g.][]{Gullberg_2016a}. In addition, the observed surface brightness radial profile of \object{MRC0316-257} (Fig. \ref{fig:radcir}) at large radii, shows a decline followed of a rise, and is a factor of five brighter than the $2\sigma$ surface brightness limit (16 for the intrinsic) which indicate an extra source of Ly$\alpha$ photons. The filamentary Ly$\alpha$ emitting comic structures are observed on Mpc scales \citep[e.g.][]{Umehata_2019,Bacon_2021}. \citet{Bacon_2021} discussed the possibility that the diffuse Ly-alpha emission is powered by low mass galaxies not directly detectable in the deep (140 hours) MUSE observation.

These results provide additional evidence that the detected nebula connects with the emission halo of companions. If this is the case, we should revisit the $d_{\rm max}-d_{\rm AGN-neb}$ in Sect. \ref{sec:2mor_rela}. When a quasar resides in a dense environment, its apparent size will likely be impacted by neighboring Ly$\alpha$ nebulae. This `contamination' from nearby companions will likely be more related to the large-scale structure and independent of the orientation of the central (quasar) halo. In other words, the hidden parameters behind the $d_{\rm max}-d_{\rm AGN-neb}$ ($e_{\rm weight}-d_{\rm AGN-neb}$, Sect. \ref{sec:2mor_rela}, Fig. \ref{fig:asy_2comp}) relations may be related to the distribution of the companion emission halos. Contamination from neighboring halos will result in the centroid of the nebula being offset from the AGN position and cause a more asymmetric nebula morphology. The nearby companions can be the disturbing sources resulting in the $\sim1000~\mathrm{km,s^{-1}}$ line width seen at $\sim100$~pkpc ($W_{80}$, Fig. \ref{fig:v50_W80_r}).

The radial asymmetry profiles of our targets have a larger value than type-1s and type-2s (Fig. \ref{fig:c_radial}) at larger radii, which is unlikely to be entirely due to the AGN orientation (Sect. \ref{sec:2asym_rad}). This can now be explained by the impact and contamination of nearby companions at 100s of pkcp. We point out that the type-2 from \citet{sanderson2021} has a projected size of $173$ pkpc, a nebula centre offsets from the AGN of $\sim17$ pkpc and high ellipticity (0.8). This source is also found to have nearby companions ($\sim 60$ pkpc). The difference of the radial asymmetry profiles (Fig. \ref{fig:c_radial}) and surface brightness profiles between this source and our HzRGs may be related to the jet (or large scale gas distribution, Sect. \ref{sec:1dis_gaslarge}).

The stellar masses of the galaxies studied in \citet{Byrohl_2021} are in the range of $8.5< \log_{10}(M_{\star}/\mathrm{M_{\odot}})<10$ which is approximately $1-2$ orders of magnitude lower than the stellar masses of our HzRGs, implying lower dark matter halo masses, as well. Such lower masses may explain the difference in the level of Ly$\alpha$ surface brightness where the flattening of the profiles is observed: in the simulations by \citet{Byrohl_2021} the flattening happens at a surface brightness level of $\sim 10^{-20}\,\mathrm{erg\,s^{-1}\,cm^{-2}\,arcsec^{-2}}$ (e.g. their Fig. 7) while we observe it to happen at a level of $\sim 10^{-18}\,\mathrm{erg\,s^{-1}\,cm^{-2}\,arcsec^{-2}}$). The dark matter halo mass difference can also explain the different radial distance at which the companion emission dominates \citep[$\sim 30$ pkpc in][versus $\sim 40$ pkpc, Table \ref{tab:SBfit}]{Byrohl_2021}. We note that we use the $r_{\rm b}$, the radius at which surface brightness radial profile changes from exponential to power law (Sect. \ref{sec:2sb_prof_fit}), as the distance where nearby halos start to significantly impact the surface brightness.

The $\gtrsim100$~pkpc extents of the Ly$\alpha$ and the dense environments of HzRGs \citep{Wylezalek2013b} suggest that nearby halos may contribute to our observations \citep[][]{Byrohl_2021}. However, this also suggests that our CGM study is `contaminated' by sources other than the radio galaxy. For this paper, we exclude obvious emission of clearly detected companions \citep[e.g. `Arrow galaxy'][]{vernet2017}. A systematic census of the companion galaxies in the MUSE HzRG fields will be reported in a future paper.

\subsection{Gas distribution on large scales }\label{sec:1dis_gaslarge}

So far, we have focused our discussion on morphological measurements sensitive to the high-surface brightness part of the Ly$\alpha$ nebulae and plausible explanation of environmental effect on $\sim100$~pkpc scales. It is likely that the jet shapes the nebula in the vicinity of the quasar to align with the jet axis (skewed distribution of $|\theta_{\rm weight}-\rm{PA_{radio}}|$ towards values $<30^{\circ}$, Fig.\ref{fig:PA_dis_hist}) through interaction with the gaseous medium. Beyond the extent of the jets, we use $|\theta_{\rm unweight}-\rm{PA_{radio}}|$ (which is equally sensitive to the whole nebula) to inspect the relation between the jet axis and halo asymmetry.

While it is unlikely that the jet plays a major role shaping the morphology of the large-scale halo, \citet{Eales_1992} suggested that the observed direction of the radio jet is the result of an inhomogeneous gas density. When a jet is launched along the direction of higher gas density (e.g. $n_{\rm H}\sim10^{-2}\,\mathrm{cm^{-3}}$), the jet `working surface' (hot spots) will leading to brighter radio emission. Such a scenario would introduce a bias in observing jets preferably along directions of higher densities in flux-limited samples and we might expect a correlation between the morphology of a large scale halo and radio jet axis. Our observations presented in Figures \ref{fig:asy_2comp} and \ref{fig:PA_dis_hist} is in agreement with such a scenario.


The detection of molecular gas in HzRGs along the jet axis but beyond hot spots provides further evidence \citep[$<20^{\circ}$, which suggests the jet runs into filament of cold gas][]{Emonts_2014a}. If this direction indeed traces a filament of the cosmic web \citep[e.g.][]{West_1991,Umehata_2019,Bacon_2021} with a higher density of companion galaxies, then this is also in agreement with the discussion presented in Sect. \ref{sec:1dis_env}. \citet[][]{Humphrey_2007a} found evidence of infalling CGM gas on to the HzRGs which bridges the radio galaxy and to the large-scale (IGM) gas. If the CGM gas is being accreted onto the central HzRGs through these higher density distributions, it would reflect the scenario proposed by \citet[][]{Humphrey_2007a}.

\subsection{Inflow from the cosmic web?}\label{sec:1discinflow}
The relatively large $|\theta-\rm{PA_{radio}}|$ ($81.4^{\circ}$ and $45.3^{\circ}$ for weighted and un-weighted, respectively) detected in \object{4C+04.11} is inconsistent with other targets (see Figure \ref{fig:PA_dis_hist}). This suggests that the nebula of \object{4C+04.11} is elongated in the East-West direction on both small ($<20$~pkpc, scope of jet) and large scales ($\gtrsim20$~pkpc, Sect. \ref{sec:2neb_mor_quan} and \ref{sec:1dis_env}). The tile with the largest distance from AGN position (\# 74, Fig. \ref{fig:smo_tess_4C04}) has a filamentary like structure. There is no known evidence that \object{4C+04.11} resides at the center of a dense cluster-like environment \citep[e.g.][]{Kikuta_2017} which makes it unlikely that the asymmetry is caused by contamination of nearby halos as discussed in Sect. \ref{sec:1dis_env}.

High velocity shocks \citep[$\gtrsim 100 \,\rm{km\,s^{-1}}$, ][]{Allen2008} can heat the gas which will then cool by radiating UV photons. Shocks could be caused by inflows and power the observed Ly$\alpha$ emission. We estimated the dark matter halo of \object{4C+04.11} is on the order of $M_{\rm DM}\sim10^{13}\,\mathrm{M_{\odot}}$ from it stellar mass, $M_{\star}\sim10^{11}\,\mathrm{M_{\odot}}$ with an empirical $M_{\star}-M_{\rm DM}$ relation \citep[][]{wang2021}. The surface brightness measured in the farthest tile \#74 is $\sim1.2\times10^{-17}\,\mathrm{erg\,s^{-1}\,cm^{-2}\,arcsec^{-2}}$ which is similar to (or slightly higher than) the simulated value from \citet{Rosdahl2012}. We converted our measurement to $z=3$ which is redshift reported in the simulation. Our measurement and the simulation are consistent given that dark matter halo of \object{4C+04.11} is likely more massive than the one in the simulation ($M_{\rm DM}\sim10^{12}\,\mathrm{M_{\odot}}$).

Given the estimated dark matter halo mass, the virial radius and virial velocity can then be calculated as $r_{\rm vir} \sim 100$ kpc and $v_{\rm vir}\simeq 580\,\mathrm{km\,s^{-1}}$. From the centre of the tile \#74, we can derive its distance to the central AGN as $\sim 60$ kpc.  We note that this is the projected distance averaged for spaxels in the tile. The actual distance of the filament could be farther. Even though it is closer, \citet{Nelson_2016} simulated gas accretion into $10^{12}\,M_{\odot}$ halos at $z=2$ and found that the accretion flow structure can remain filamentary within $r_{\rm vir}$ ($\sim0.5r_{\rm vir}$). The projected $v_{50}$ (velocity offset) of tile \#74 is $-536\,\mathrm{km\,s^{-1}}$ which is consistent with $v_{\rm vir}$. Thus, our measurements for the projected distance of the tile and its velocity offset are consistent with a scenario where the Ly$\alpha$ emission in tile \#74 may be tracing shock driven by inflows. If the distance we observe is indeed $<r_{\rm vir}$, then the post-virial accretion could be more complicated with multi gas phase components and fragmentation involved \citep[e.g.][]{Cornuault_2018}. The broad line width of \object{4C+04.11} at large radii (Fig. \ref{fig:v50_W80_r}) may indicate the disturbed nature of the presumed accretion flow.  

The discussion in this section based on the $|\theta-\rm{PA_{radio}}|$ and morphology of bin \#74 of \object{4C+04.11}. While the angle difference is a clear outlier in the sample, the tile shape may depend on the implemented method for nebula extent  (Section \ref{sec:2max_neb_ext}). However, we checked the Ly$\alpha$ narrow band image ($6690\,\AA$ to $6707\,\AA$) directly collapsed from the data cube and confirm the indication of emission from this region. As for the nearby potential emission (north to bin \#74, Fig. \ref{fig:smo_tess_4C04}), we confirm by spectral extraction that there is no S/N$>3$ detection. Even if the Ly$\alpha$ emission is detected in this additional region, it would still be consistent with the accretion scenario. Complex filamentary structures are seen in simulations \citep[see][and their Fig. 7]{Rosdahl2012} for accretion in higher mass dark matter halo. Nevertheless, we caution that this analysis only considers one possibility. As the data is near the detection limit, deeper observation are needed before more conclusive results can be derived.

\section{Conclusions}\label{sec:0conclusions}
In this paper, we study the intrinsic Ly$\alpha$ nebulae around a sample of eight high-redshift radio galaxies using optical IFU observations obtained with MUSE. We link our observations to results from the literature for other quasars (mainly type-1 quasars) at similar redshifts.

We have developed a new method to measure the maximum extent of the nebulae with improved sensitivity to low surface brightness emission. We also have developed a new method to tessellate the Ly$\alpha$ maps (Sect. \ref{sec:1nebuseltes}). We have detected the Ly$\alpha$ emission at scales $\gtrsim100\,\rm{pkpc}$ from the central AGN, down to an observed surface brightness of $\sim2-20\times10^{-19}\,\rm{erg\,s^{-1}\,cm^{-2}\,arcsec^{-2}}$.

We summarise our results as follows: The Ly$\alpha$ emission line profiles of all sources are affected by multiple and deep absorption troughs out to the edge of the nebulae. We have corrected for this \ion{H}{i} absorption (Sect.\ref{sec:1specfitmeth}) and constructed maps of the intrinsic Ly$\alpha$ (Sect. \ref{sec:0intmaps}) emission and also measured the kinematic properties of the Ly$\alpha$ emission spatially resolved. 

We first investigated radial dependencies of the surface brightness, velocity shift and velocity width of our sample. We found that circularly averaged profiles of the intrinsic (absorption-corrected) Ly$\alpha$ surface brightness are brighter and more extended than type-1 quasar samples (Sect. \ref{sec:2rad_sb_pro}). We did not find major differences when we investigate the radial profiles along the approaching and receding direction of the radio jets, respectively (Sect.  \ref{sec:2dir_sb_prof}). The surface brightness radial profiles can generally be described by an exponential drop-off for the inner high surface brightness part and a power law for the more extended part (Sect. \ref{sec:2sb_prof_fit}). We did not find evidence of ordered gas bulk motion from the $v_{50}$ radial profile (Sect. \ref{sec:2kin_prof}). For four targets, the $v_{50}$ profiles at radii within the jet hot-spots range show a similar gradient as the jet-driven outflow (Fig. \ref{fig:v50_W80_r}). The $W_{80}$ profiles show relatively large values ($\gtrsim 1500\,\mathrm{km\,s^{-1}}$, Fig. \ref{fig:v50_W80_r}, Sect. \ref{sec:2kin_prof}) at all radii and three targets show a decrease beyond the distance of jet hot spots is indicative of jet-gas interactions. 

We quantitatively studied the morphology of the HzRG nebulae (for both observed and intrinsic emission) and compared our results to other quasar samples (uncorrected for absorption, Sect. \ref{sec:1nebulaasy}). We found that our nebulae are, in general, more asymmetric than nebulae of type-1 quasars and are more similar to type-2 quasars (Sect. \ref{sec:2neb_asymm}), although, our sampled sources differ in their measure of asymmetry as a function of radius (Sect. \ref{sec:2asym_rad}). Furthermore, we found that the more asymmetric and larger the nebulae are, the greater the offset between the centroid of the nebulae and AGN positions ($d_{\rm max} - d_{\rm AGN-neb}$ and $e_{\rm weight} - d_{\rm AGN-neb}$ in Sect. \ref{sec:2mor_rela}, Fig. \ref{fig:asy_2comp}). 

 Ly$\alpha$ is a complicated emission line that can be heavily affected by absorption and resonant scattering which, as we demonstrated, needs to be accounted for. Assuming type-1 and type-2 quasars have similar intrinsic shapes of their nebulae, the difference of the nebulae asymmetry morphology between our sample and other quasars can be explained using the AGN unification model where the orientation of the ionisation cone is the fundamental parameter (Sect. \ref{sec:1dis_nebtype12}). Resonant scattering of the Ly$\alpha$ emission can result in the observed $d_{\rm max} - d_{\rm AGN-neb}$ and $e_{\rm weight} - d_{\rm AGN-neb}$ relation in our sample and partially explain the shape of the radial profile and the kinematic profiles (Sect. \ref{sec:1scatter_dis}). We also found evidence in our HzRGs that the extended nebulae are affected by Ly$\alpha$ emissions from nearby companions (Sect. \ref{sec:1dis_env}) and that CGM gas has higher density distribution along a specific direction which is coincident with the direction of the radio jet (Sect. \ref{sec:1dis_gaslarge}). There is also a possibility that, in our observations, we are capturing inflows from cosmic web (Sect. \ref{sec:1discinflow}).

In this paper we have shown that measurements of Ly$\alpha$ nebulae around high-redshift AGN can act as a probe of the environment of AGN and their host galaxies. We have found fundamental differences between the extended ionised nebulae of different types of quasars at Cosmic Noon and beyond.
Due to its resonant nature, it is a challenge to use Ly$\alpha$ as a tracer of gas kinematics especially out to the $\sim100$ pkpc scale. Nevertheless, Ly$\alpha$ line observations offer some insight into the state of CGM gas at a time when it is simultaneously being impacted by more than one physical mechanism (e.g. quasar outflow, jet-gas interaction and cosmic inflow). These kind of observations will be particularly useful for future simulation works. The MUSE data only reveals the tip of the iceberg.  In upcoming papers, the \ion{H}{i} absorbers will be reported together with the high-resolution spectroscopy data. In addition, all the UV emission lines covered by MUSE will be studied in detail and this will provide crucial information on the ionisation, metallicity and AGN feedback in the quasar nebulae. Furthermore, scheduled \textit{JWST} observations will be available for four HzRGs in our sample and this will provide unprecedented details of the AGN and host galaxy connection on scales of $\lesssim 1$ kpc ($\sim0.05$\arcsec).


\begin{acknowledgements}
We would like to thank Bjorn Emonts and Zheng Cai for valuable discussion which improved this work.

This work uses the NASA's Astrophysics Data System and a number of open source software other than the aforementioned ones such as Jupyter notebook \citep[][]{kluyver2016jupyter}; \texttt{matplotlib} \citep[][]{Hunter_2007}; \texttt{SciPy} \citep[][]{virtanen2020scipy}; \texttt{NumPy} \citep[][]{harris2020numpy}; \texttt{Astropy} \citep[][]{astropy2018}; and \texttt{LMFIT} \citep[][]{newville2016}.

BG acknowledges support from the Carlsberg Foundation Research Grant CF20-0644 `Physical pRoperties of the InterStellar Medium in Luminous Infrared Galaxies at High redshifT: PRISM- LIGHT’. MVM acknowledges support from grant    PID2021-124665NB-I00  by the Spanish Ministry of Science and Innovation/State Agency of Research MCIN/AEI/ 10.13039/501100011033 and by "ERDF A way of making Europe". GN acknowledges funding support from the Natural Sciences and Engineering Research Council (NSERC) of Canada through a Discovery Grant and Discovery Accelerator Supplement, and from the Canadian Space Agency through grant 18JWST-GTO1. PL (contract DL57/2016/CP1364/CT0010) is supported by national funds through Funda\c{c}\~ao para a Ci\^encia e Tecnologia (FCT) and 
the Centro de Astrof\'isica da Universidade do Porto (CAUP).
\end{acknowledgements}

%
%
\bibliographystyle{aa}
\bibliography{references}

\begin{appendix} 

\section{Detection map and tessellation procedures}\label{app:selmasktess}
\subsection{Procedure of detection map for nebulae extent}\label{app:detec_map}
\begin{table}[!h]
\caption{Optimal detection map parameters combination.}\label{tab:inmaspar}
\centering
\begin{tabular}{ccccc}
\hline
HzRG & $T_{S/N}$ & $n_{\lambda}$ & $2.355\sigma_{\rm max}$\tablefootmark{\rm{\dag}}  & $n_{\rm c}$\\
\hline
MRC 0943-242 & 3.5 & 4 & 25 & 2\\ 
MRC 0316-257 & 3.5 & 5& 30 & 3\\
TN J0205+2242 & 3.5 & 5& 25& 2\\
TN J0121+1320 & 2.5 & 3& 35& 4\\
4C+03.24 & 3.5 & 5& 20& 2\\
4C+19.71 & 3.5& 2& 30& 3\\
TN J1338-1942 & 3.5& 2& 30& 2\\
4C+04.11 & 3.5& 4& 25 &2\\
\hline
\end{tabular}
\tablefoot{
\tablefoottext{\rm{\dag}}{Maximum Gaussian kernel scale in FWHM. The unit is in pixel unit.} }
\end{table}
   \begin{figure*}
  \centering
        \includegraphics[width=16.4cm,clip]{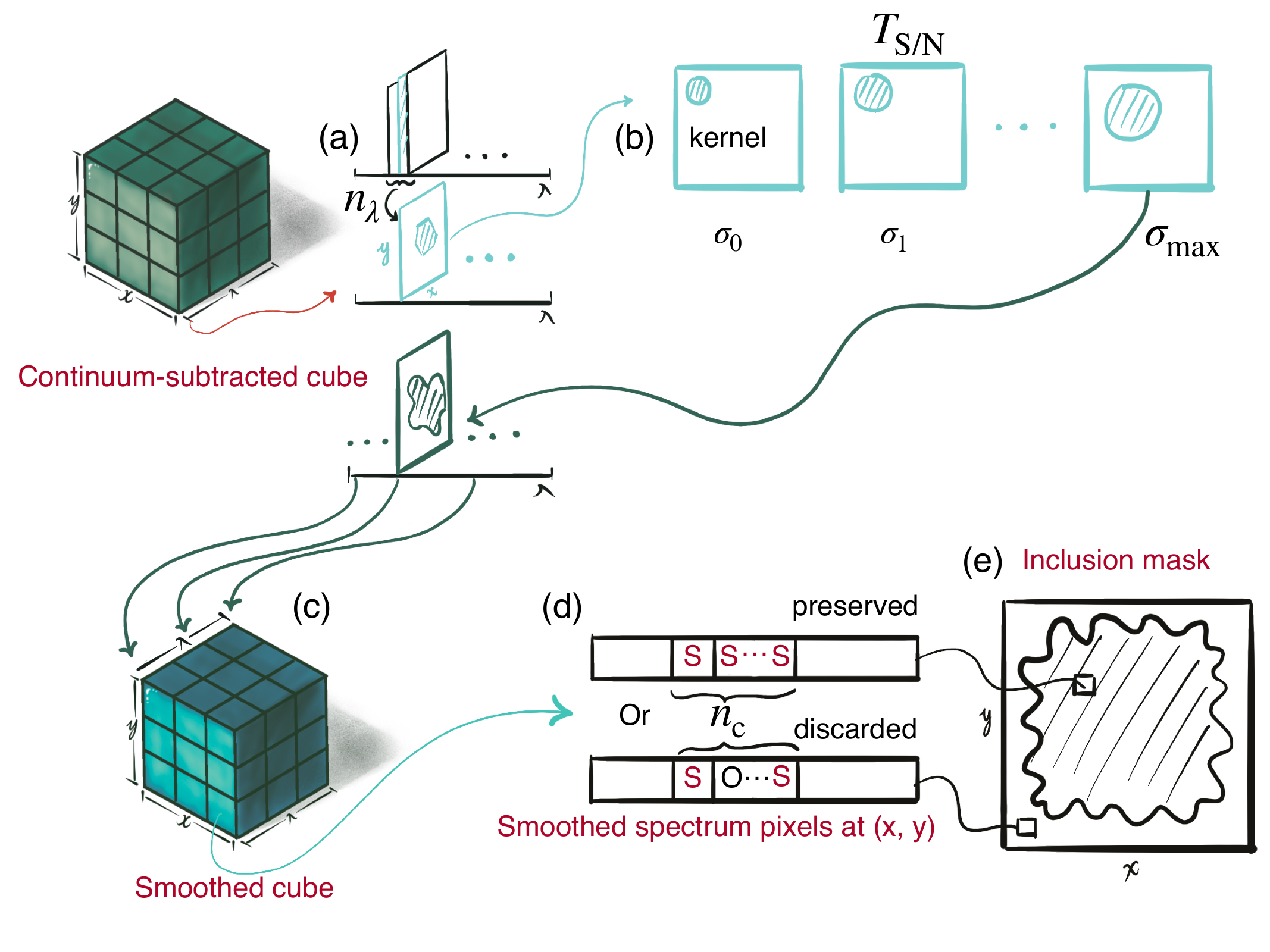}
      \caption{Schematic presentation of the detection map construction. \textbf{(a)} Average $n_{\lambda}$ number of images around each of the image slices (cyan) from the continuum-subtracted cube (dark green cube). \textbf{(b)} Spatially smooth each of the averaged image with Gaussian kernel. The algorithm will increase the kernel size ($\sigma_{0} < \sigma_{1}$) to smooth the spaxels that not passing the $T_{\rm S/N}$, S/N threshold, at each steps until the maximum, $\sigma_{\rm max}$, has been reached. \textbf{(c)} Combine the adaptively smoothed images into the smoothed cube (lighter green cube). \textbf{(d)} Check the smoothed spectrum (long black rectangular box) at location (x, y). The position is preserved when there are $n_{\rm c}$ consecutive number of pixels selected with signal detection (red 'S') in previous steps. \textbf{(f)} Construct the detection map.    }
         \label{fig:smoothproce}
   \end{figure*}


The basic idea for detection map is to use the 3D information of the MUSE data cube to determine the maximum spatial extent of the Ly$\alpha$ emission. Smoothing technique is the key process in this procedure to guarantee the faint structures are captured. To assist with conveying the procedure, the process is shown schematically in Fig. \ref{fig:smoothproce}. In this procedure, we can select the faint end of the emission nebula to the surface brightness detection limit.

Before the generation of detection map, we do a step of continuum subtraction for each spectra spatially centred around the AGN and spectrally around their Ly$\alpha$ wavelength range ($-5000-5000\,\rm{km\,s^{-1}}$ or $-6000-6000\,\rm{km\,s^{-1}}$ for \object{4C+03.24} which has broader line width) with the emission line masked. If there are sky-lines, emission or absorption features from foreground targets in the unmasked range, we do further masking. Continuum from the host galaxy, central AGN and foreground objects need to be subtracted to minimise their contamination. In this way, we exclusively focus on the line-emission nebula. The choice of flat or linear continuum do not have significant impact on the Ly$\alpha$ fitting \citep[][]{wang2021}. Hence to better account for the slope from foreground stars, we subtract a first-order polynomial from the cube. 

The 3D adaptive smoothing follows the \citet{martin2014b} procedure with adaptation \citep[e.g.][]{vernet2017}. As described in Sect. \ref{sec:2max_neb_ext}, it smooths each of the image planes of the continuum-subtracted cube adaptively with a Gaussian kernel over a wavelength range of $\sim15\,\AA$ around the Ly$\alpha$ emission. For the image slice at each wavelength, the algorithm first takes average of adjacent $n_{\lambda}$ number of slices around this image slice along the wavelength dimension (Fig. \ref{fig:smoothproce}a). Then it adaptively smooths the averaged image with a Gaussian kernel starting from the smallest kernel size, $\sigma_{0}=3/2.355$ pix, until the maximum, $\sigma_{\rm max}$, is reached (Fig. \ref{fig:smoothproce}b). Specifically at each step, the algorithm smooths the spaxels that are not passing the pre-set S/N threshold, $T_{\rm S/N}$. In the end, the algorithm set the voxels, after being maximally smoothed, that not pass the $T_{\rm S/N}$ to 0 as no-signal containing (i.e. the voxels that potentially contain Ly$\alpha$ signals have positive value). In this way, we preliminarily detect where we have Ly$\alpha$ in the cube (Fig. \ref{fig:smoothproce}c).

To generate a detection map working on the spatial dimensions and guarantee line fitting, we perform a further check along the wavelength dimension for the smoothed cube. Specifically, if the smoothed spectrum at one spatial location (x, y) has $n_{\rm c}$ numbers of consecutive wavelength pixels preserved (i.e. have positive values), then we flag this spatial location (x, y) as signal-contained (see Fig. \ref{fig:smoothproce}d as an example). The others that do not pass this check are discarded. In this way, we can construct the detection map (Fig. \ref{fig:smoothproce}e).

The question left is to find a combination of the four parameters ($T_{\rm S/N}$, $n_{\lambda}$, $\sigma_{\rm max}$ and $n_{\rm c}$) which returns an optimised detection map. For this, we generate a series of detection maps for each targets with different combination of the four parameters. Then we choose the one that optimises the spatial selection (i.e. captures large scale extent while avoid island-like structures). The best combination of the four parameters for each targets are presented in Table \ref{tab:inmaspar}. We check the detection maps constructed using different sets of parameters around our optimal combination (Table \ref{tab:inmaspar}) and find that the bulk ($\sim80\%$) of the selected spaxels remains the same. If using a stricter set of parameters (e.g. higher $T_{\rm S/N}$ and/or $n_{\lambda}$), we will miss the low S/N part of the nebula by comparing with previous individual target studies \citep[][]{swinkbank2015,vernet2017}. If using a more relaxed set, the peripheral regions, mostly disconnected to the bulk of the nebula, with pure noise (after checking spectrum) will likely be selected. Thus, we are assured that the constructed detection map represented the observation to the detection limit and the method is robust against objectiveness. After constructing the map with the best parameter combination, we perform a final manual selection to exclude island-like regions ($\gtrsim1$ arcsec$^2$ in size) which are detached from the the main nebulae ($>1"$) and could be due to noise or companion. A further check of the spectra extracted from these regions is also conducted. Around 50\% of the island regions only contain noise. The ones with potential Ly$\alpha$ signal detected are presented in Appendix \ref{app:smimage} (Fig. \ref{fig:smo_tess_mrc0316} and \ref{fig:smo_tess_tnj0205} ). Since these are detached from the extended nebula around the quasar, we do not include them in this analysis but point out they may trace companion emissions. We note that we refer to the smoothed cube obtained in this way (using $T_{S/N}$, $n_{\lambda}$ and $\sigma_{\rm max}$ in Table \ref{tab:inmaspar}) as 'optimally smoothed cube'. The smoothed cube are only used in constructing the detection map and not being further used in other analysis. In Appendix \ref{app:smimage} (Fig. \ref{fig:smo_tess_mrc0943}-\ref{fig:smo_tess_4C04}a), we present the smoothed nebula image extracted from the optimally smoothed cube.

\subsection{Tessellation procedure}\label{app:tesstll}

The Ly$\alpha$ nebulae are irregular in morphology \citep[e.g.][]{swinkbank2015,vernet2017}, affected by absorption and having high S/N gradient from center to outskirt and multiple dynamical components. Hence, neither the simple adoption of Voronoi tessellation \citep{cappellari2003} nor the tessellation method from \citet{swinkbank2015} \citep[which was invented for studying \object{TN J1338-1942} and has also been applied to \object{4C+04.11} but only works for the central part, see][]{wang2021} can be adopted without adaptation. We describe here the tessellation used in this work. 

The \citet{swinkbank2015} tessellation is not optimal for some of our target with significantly irregular nebula shape (e.g. \object{MRC0316-257}) since it works with rectangular binning. Hence, we decide to adopt the Voronoi binning \citep{cappellari2003}, which is a sophisticate adaptive spatial binning algorithm implemented by various studies of IFU data, for this work. Directly performing Voronoi binning on the image will return detached regions (i.e. regions of spaxels belong to the same tile but are physically separated), as the images are S/N limited for the outer nebulae. The solution could be (i) turning off \textit{Weighted Voronoi Tessellation} or \textit{Centroidal Voronoi Tessellation} \citep[see][for details]{cappellari2003}; or (ii) performing a twofold Voronoi binning by using a S/N cut for doing separate binning for high and low S/N parts (M. Cappellari private communication).  Solution (i) will return wedge shape tiles which is a known result \citep{cappellari2003}. It is, however, not desirable for our case since it smears out the radial structures which is one of the key interest of this work. Therefore, we adapt solution (ii); the twofold Voronoi binning for our sample.

Firstly, we apply the detection map (Sect. \ref{sec:2max_neb_ext} and Appendix \ref{app:detec_map}) to the continuum-subtracted un-smoothed cube to preserve only the Ly$\alpha$ signal detected spaxels for tessellation. To avoid complication and keep consistency for the whole sample, we perform the tessellation on a single narrow band image for each target. Since the Ly$\alpha$ spectra may have different observed peaks at different spatial locations, we choose the wavelength range, over which the narrow band image for tessellation will be produced (averaged by the number of wavelength pixels), to enclose as much of line emissions as possible and avoid adding pure noise. This range is $\sim10\AA$ wide. We note that for \object{4C+19.71}, we select two wavelength ranges at both the blue and red sides of 5577 \AA (sky-line), which is at roughly the systemic redshift of Ly$\alpha$. In this way, we construct the S/N map from the narrow band image.

Secondly, a Gaussian smooth is performed to the S/N map to minimise the impact of randomly located spaxels dominated by noise (i.e. further avoid detached region problem). We use a Gaussian kernel with $\sigma_s = 3$ in pixel unit. Then, we apply a S/N cut of 3 to the S/N map to select the high S/N regions for the first Voronoi binning. 

Thirdly, for the selected S/N $>3$ part, we reassign spaxels with S/N>6 to S/N=6 and perform the Voronoi binning with a target (S/N)$_{\rm vorbin,1st}=$ 15 (12 for \object{4C+03.24}, see later). The reason for the S/N reassignment is to control the minimum size of the tiles to avoid single spaxel tile and account for the seeing effect. Because the Ly$\alpha$ nebulae studied here have high S/N gradient from center to outskirt, performing  Voronoi binning with a high S/N threshold (avoid single spaxel bin in the high S/N region) will result in tiles with large size for the low S/N part which is an overkill for the fitting and smears out resolution information. After reassigning S/N>6 spaxel and using (S/N)$_{\rm vorbin,1st}=$15, the minimum number of spaxels for one tile is 6 ($=3\times2$ pix$^{2}$ or $0.6\times0.4$ arcsec$^2$) which is roughly half of the seeing disk. This is a compromise for being consistency for the whole sample and also physically connects the neighbour tiles which is useful for the implemented fitting procedure (Sec. \ref{sec:0dataana}). As for \object{4C+03.24} which was observed in the AO mode, we lower its (S/N)$_{\rm vorbin,1st}=$12 to have a minimum number of spaxels for one tile is 4 ($=2\times2$ pix$^{2}$ or $0.4\times0.4$ arcsec$^2$).

Finally, for the S/N$\leq$3 spaxels left in the first step, we apply a Voronoi binning with (S/N)$_{\rm vorbin,2nd}=$8 (11 for \object{4C+19.71} due to the impact of sky-line).

In this way, we obtain the tessellation maps where each tile has S/N$>5$. We note that the S/N stated here is calculated based on the narrow band image (see above) collapsed $\sim14$ wavelength elements (may vary for different targets) and averaged by the number of wavelength elements. This narrow band image is chosen to enclose as much of the Ly$\alpha$ emission as possible which is a compensation for the different emission peak due to kinematics. Hence, a S/N$>5$ in each tile selected in this way is feasible for further fitting as there will be other signal contained wavelength pixels outside the range given the broadness of the emission line. For targets where the extent of the nebula is overlapped with bright foreground stars (\object{TN J1338-1942} and \object{4C+03.24}), we manually assign a $d=1\,\rm arcsec$ circular mask at the position of the star to minimise its impact. For \object{MRC0316-257}, we mask the region where a known Ly$\alpha$ emitting galaxies at the similar redshift overlaps in spatial location with the nebula \citep[see][Arrow galaxy]{vernet2017}. These manually masked regions are marked by purple hatches in Fig. \ref{fig:map_int_1}bcd. The tessellation maps are shown in Appendix \ref{app:smimage}. Through the tessellation, we reach $\sim2\sigma$ surface brightness limit (as reported in Table \ref{tab:sampleinfo}) in the faintest tile.

For the convenience of the spatial fitting (Sect. \ref{sec:2spatialfit} and Appendix \ref{app:spatialfit}), we run an automatic re-numbering algorithm to make physically attached tiles to be as consecutive in number as possible. In each tile, the spatial spectra from every spaxel are then extracted and summed according to the tessellation from the original cube (i.e. the one produced by the reduction, Sect. \ref{sec:1datared}, without further continuum subtraction).

\subsection{Smooth nebula images and tessellation maps}\label{app:smimage}
\begin{figure*}
  \centering
        \includegraphics[width=16.4cm,clip]{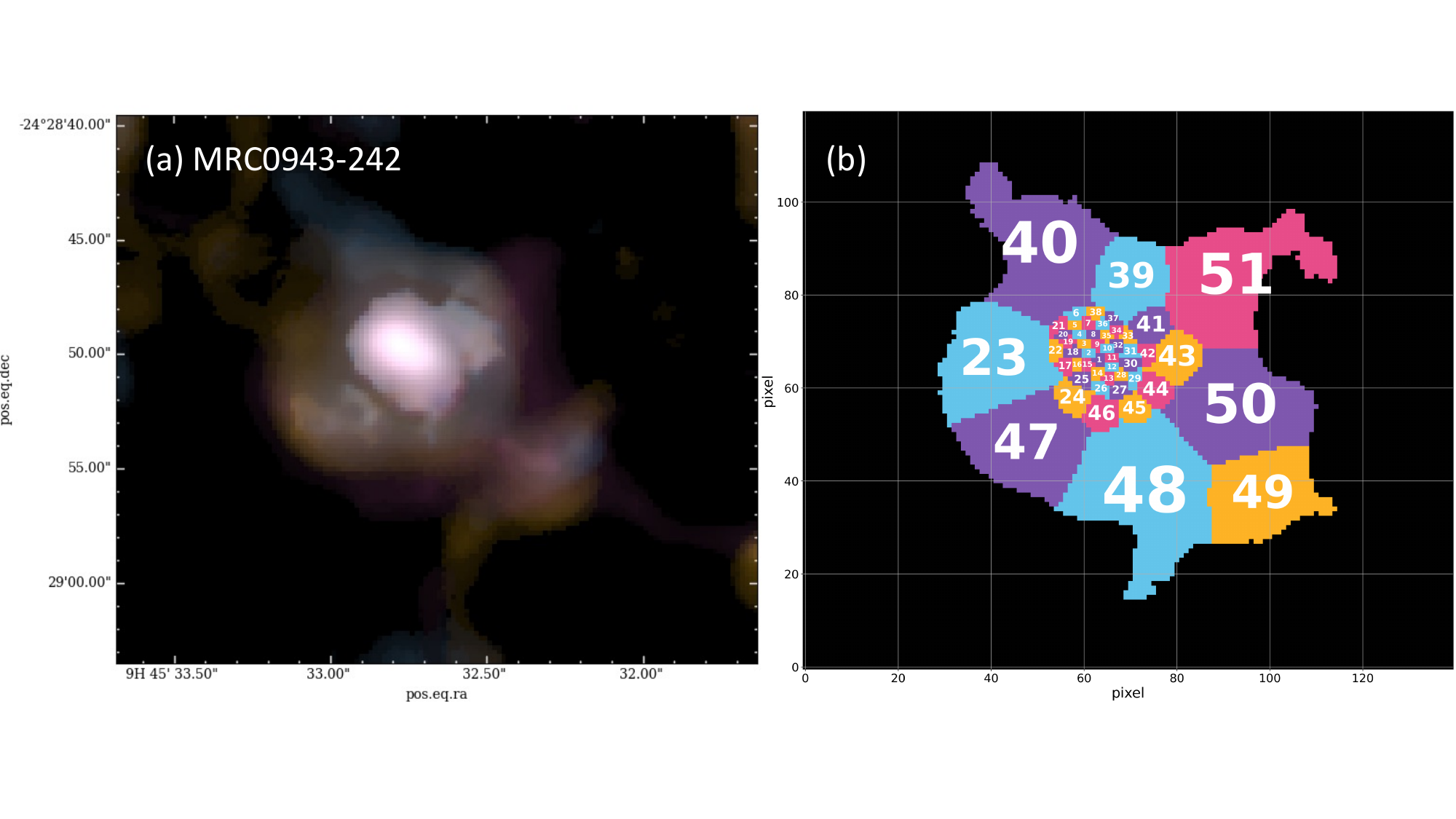}
      \caption{Smoothed nebula image (a) and tessellation map (b) of \object{MRC0943-242}. (\textit{a}) The purple, yellow and blue colours represent median pseudo-narrow band Ly$\alpha$ images collapsed arbitrarily from red part, middle and blue part, respectively, of the smoothed cubes.}
         \label{fig:smo_tess_mrc0943}
   \end{figure*}

\begin{figure*}
  \centering
        \includegraphics[width=16.4cm,clip]{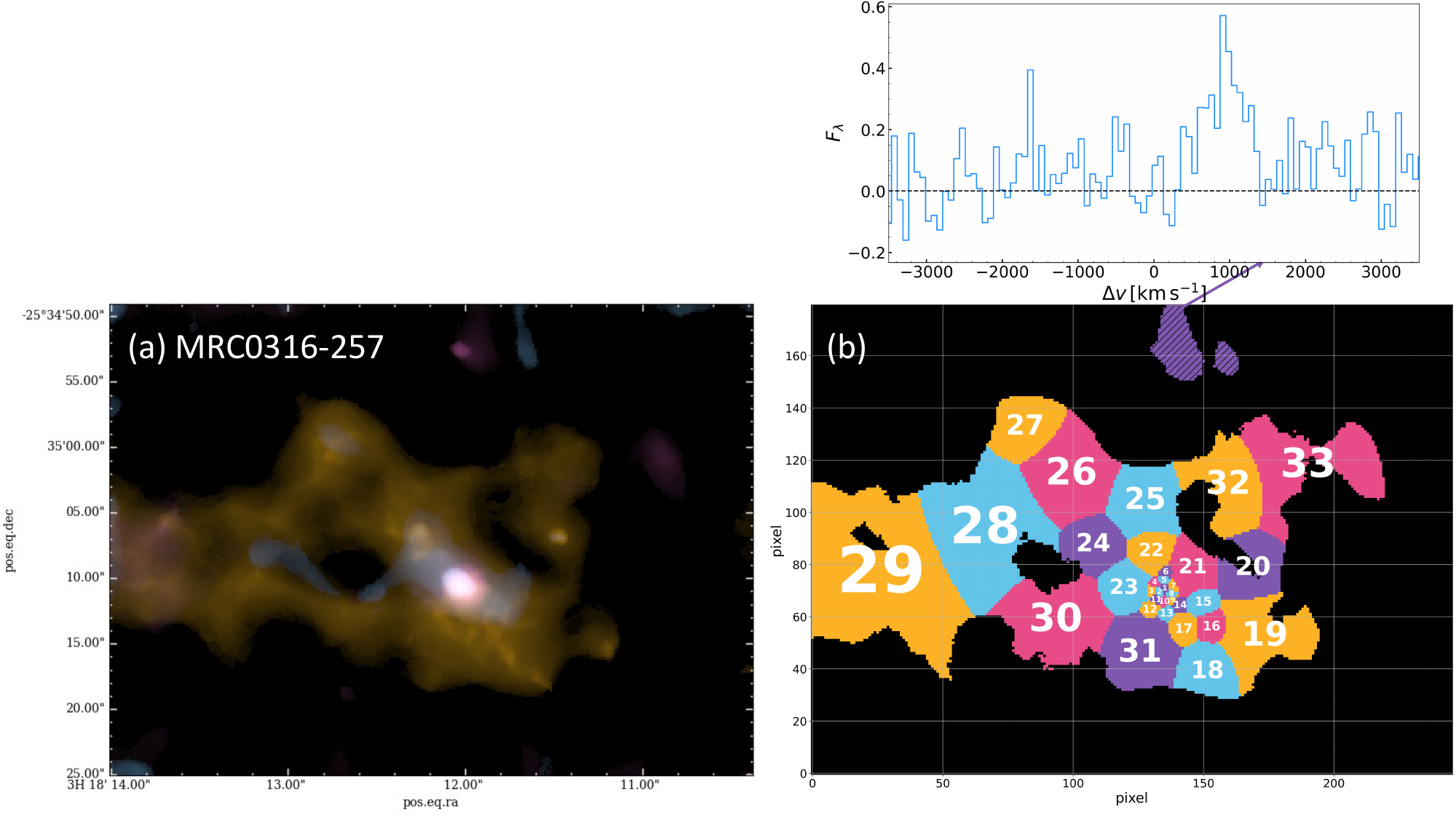}
      \caption{Smoothed nebula image (a) and tessellation map (b) of \object{MRC0316-257}. The inset spectrum is extracted from the detached regions (hatched) which is selected having Ly$\alpha$ emissions but left out from the analysis following Sect. \ref{sec:2max_neb_ext}. (\textit{a}) The color-composed image is created in a manner similar to that of Figure \ref{fig:smo_tess_mrc0943}a.}
         \label{fig:smo_tess_mrc0316}
   \end{figure*}

\begin{figure*}
  \centering
        \includegraphics[width=16.4cm,clip]{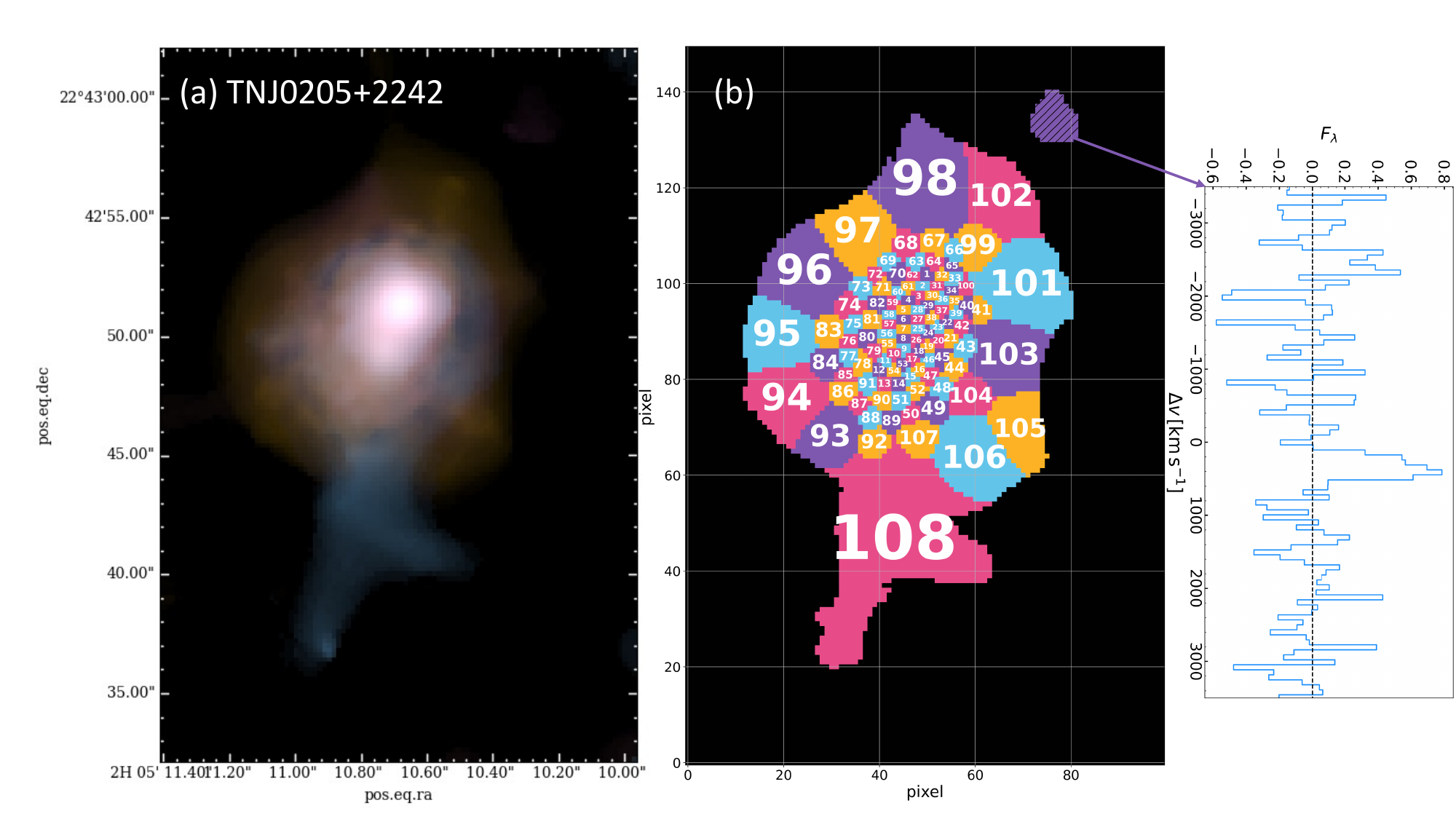}
      \caption{Smoothed nebula image (a) and tessellation map (b) of \object{TN J0205+2242}. The inset spectrum is extracted from the detached region (hatched) which is selected having Ly$\alpha$ emissions but left out from the analysis following Sect. \ref{sec:2max_neb_ext}. (\textit{a}) The color-composed image is created in a manner similar to that of Figure \ref{fig:smo_tess_mrc0943}a.}
         \label{fig:smo_tess_tnj0205}
   \end{figure*}

\begin{figure*}
  \centering
        \includegraphics[width=16.4cm,clip]{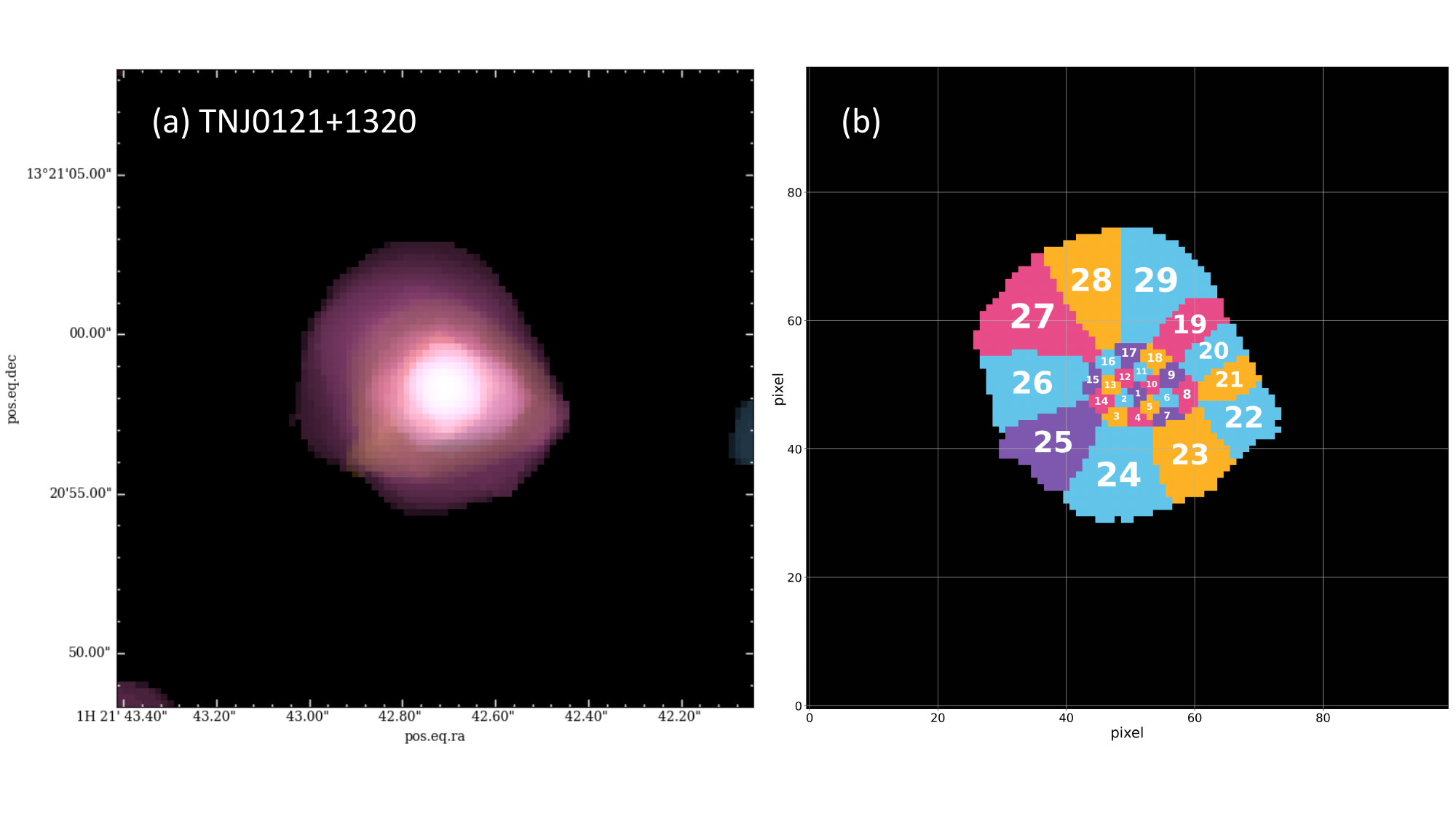}
      \caption{Smoothed nebula image (a) and tessellation map (b) of \object{TN J0121+1320}. (\textit{a}) The color-composed image is created in a manner similar to that of Figure \ref{fig:smo_tess_mrc0943}a. }
         \label{fig:smo_tess_tnj0121}
   \end{figure*}

\begin{figure*}
  \centering
        \includegraphics[width=16.4cm,clip]{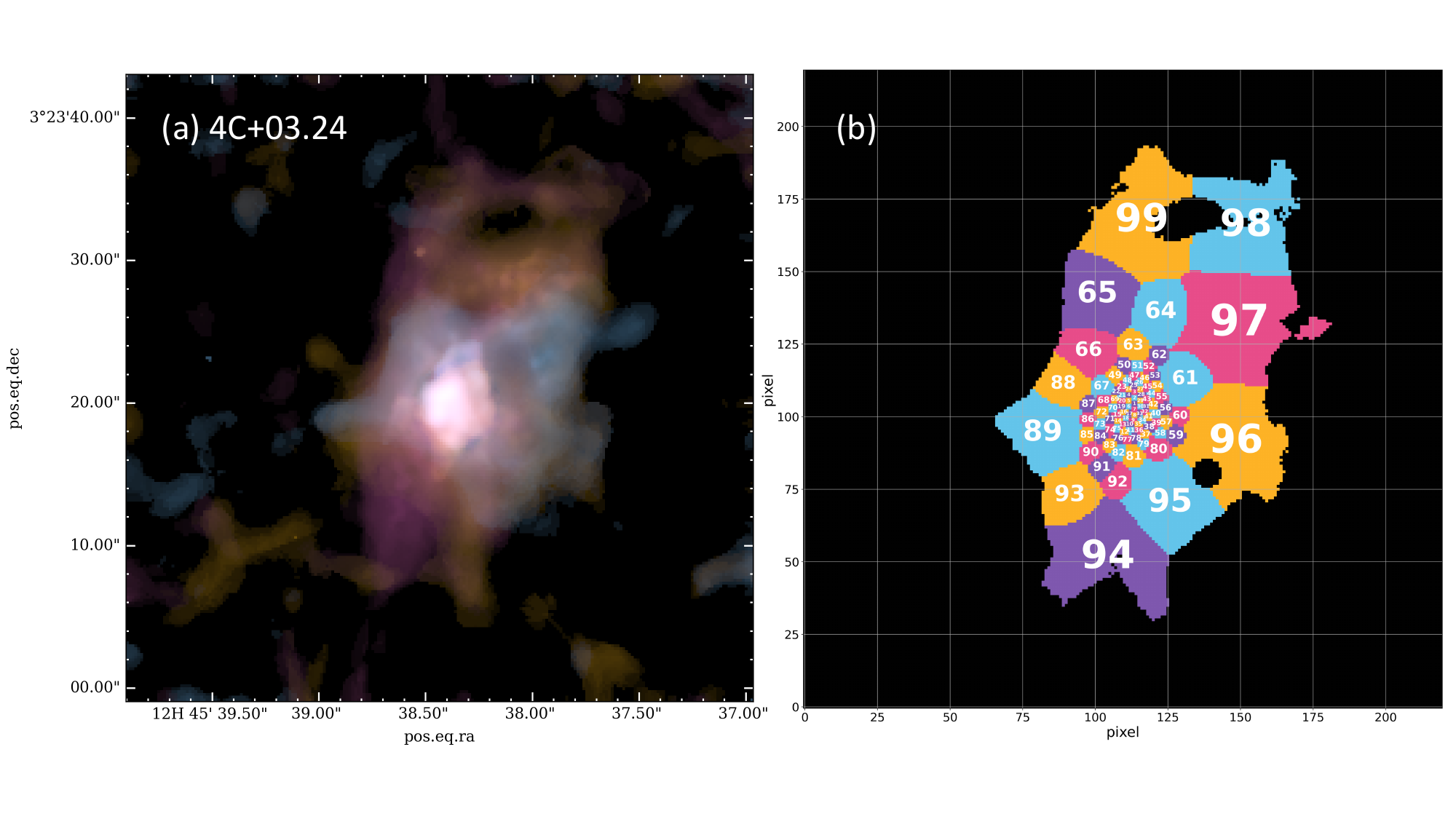}
      \caption{Smoothed nebula image (a) and tessellation map (b) of \object{4C+03.24}. (\textit{a}) The color-composed image is created in a manner similar to that of Figure \ref{fig:smo_tess_mrc0943}a.}
         \label{fig:smo_tess_4C03}
   \end{figure*}

\begin{figure*}
  \centering
        \includegraphics[width=16.4cm,clip]{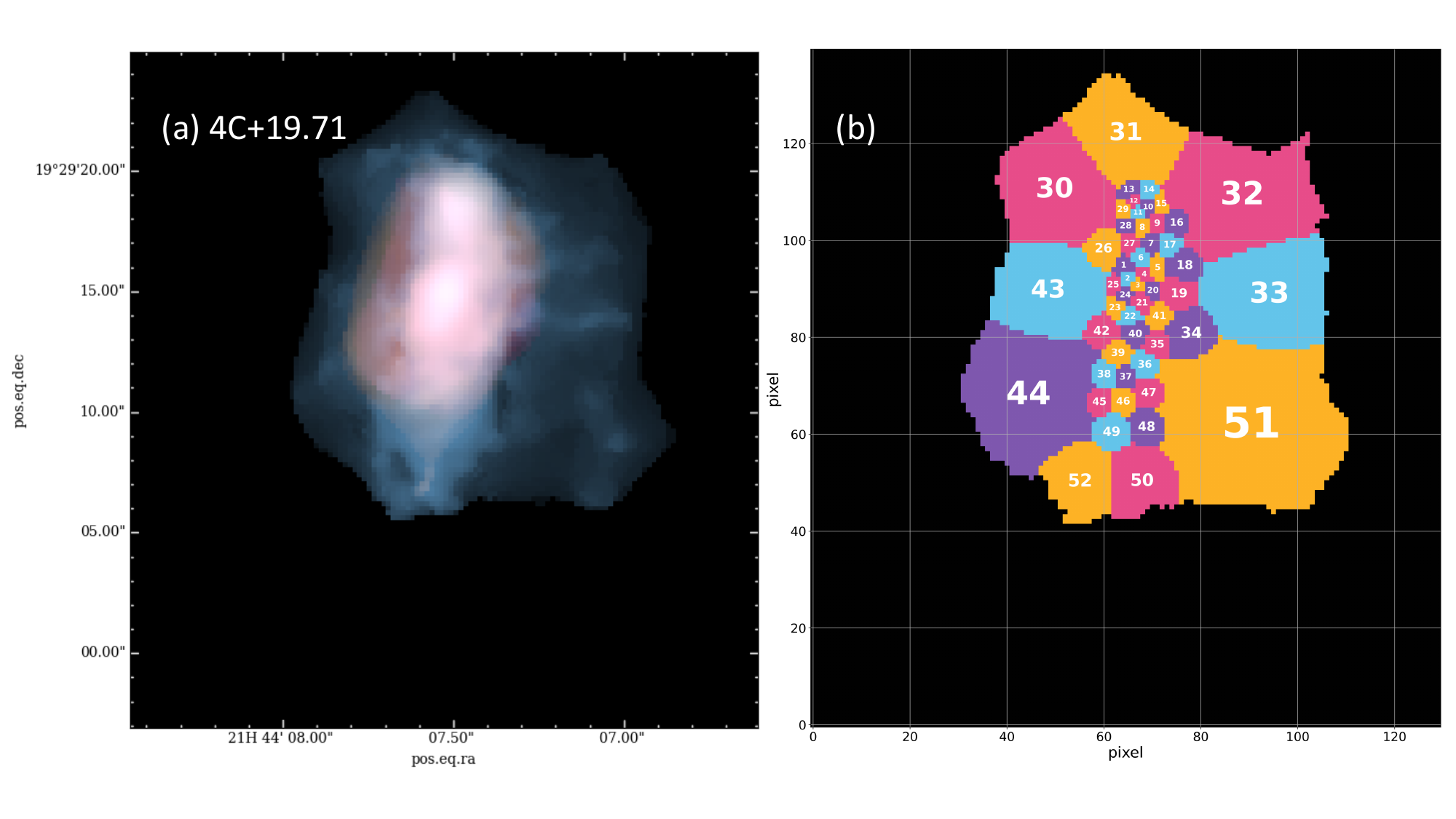}
      \caption{Smoothed nebula image (a) and tessellation map (b) of \object{4C+19.71}. (\textit{a}) The color-composed image is created in a manner similar to that of Figure \ref{fig:smo_tess_mrc0943}a.}
         \label{fig:smo_tess_4c19}
   \end{figure*}

\begin{figure*}
  \centering
        \includegraphics[width=16.4cm,clip]{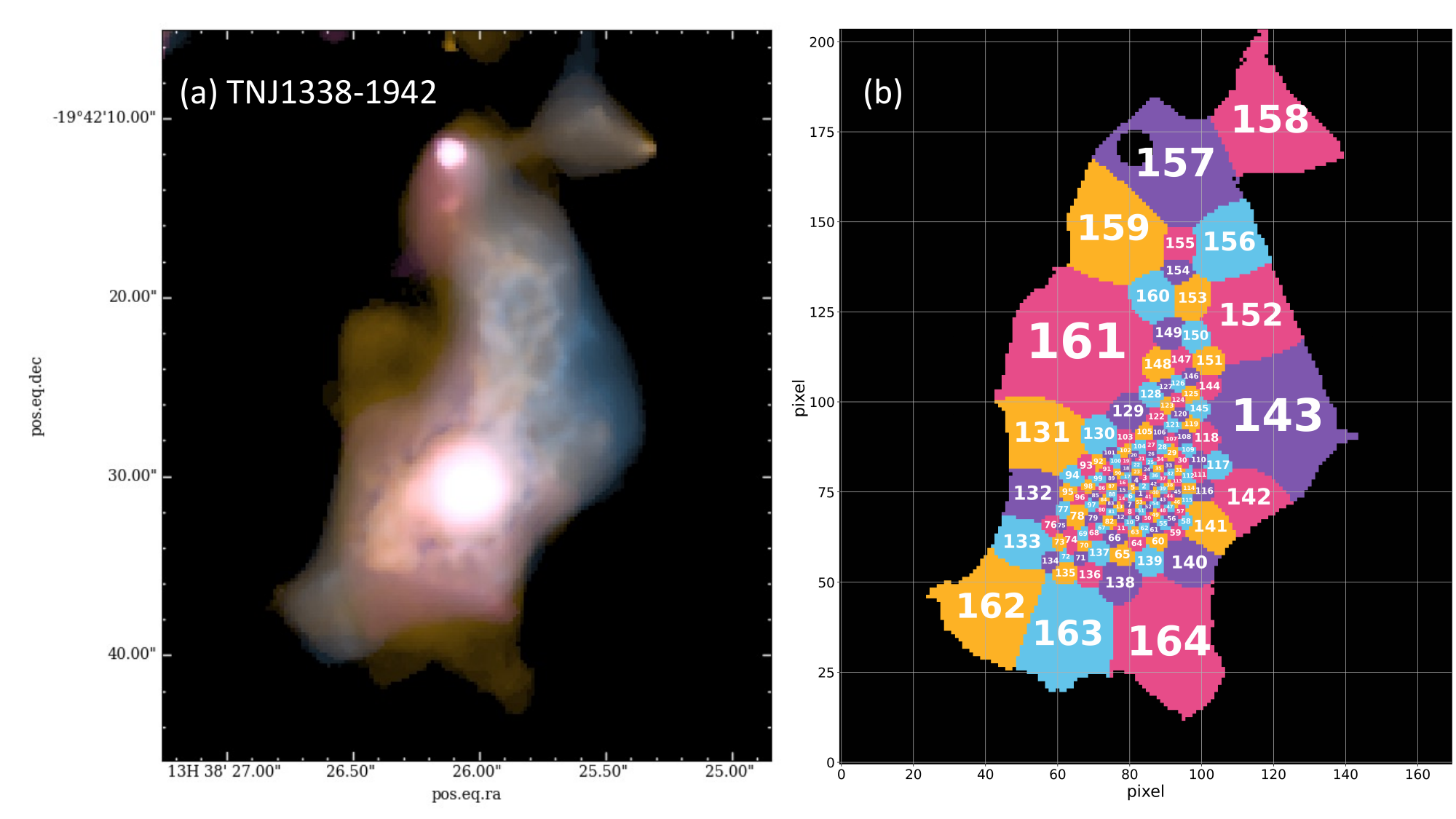}
      \caption{Smoothed nebula image (a) and tessellation map (b) of \object{TN J1338-1942}. (\textit{a}) The color-composed image is created in a manner similar to that of Figure \ref{fig:smo_tess_mrc0943}a.}
         \label{fig:smo_tess_tnj1338}
   \end{figure*}

\begin{figure*}
  \centering
        \includegraphics[width=16.4cm,clip]{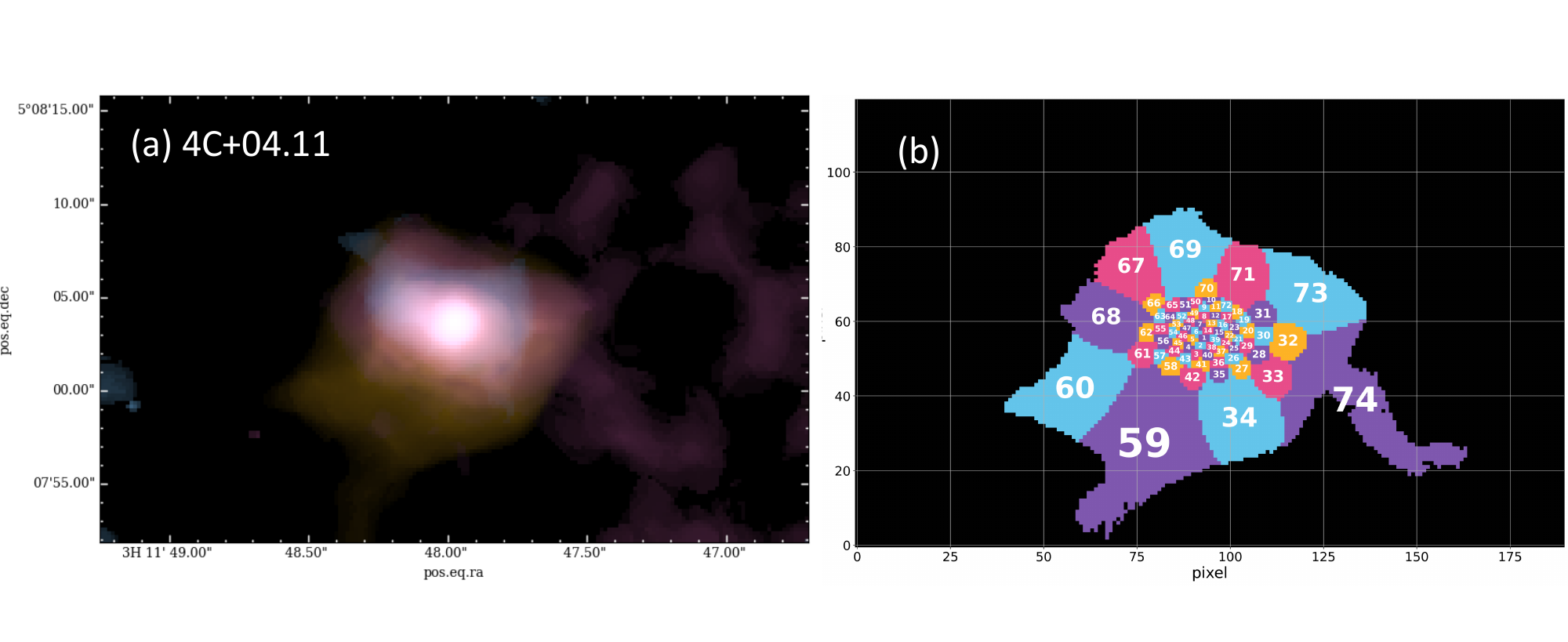}
      \caption{Smoothed nebula image (a) and tessellation map (b) of \object{4C+04.11}. (\textit{a}) The color-composed image is created in a manner similar to that of Figure \ref{fig:smo_tess_mrc0943}a.}
         \label{fig:smo_tess_4C04}
   \end{figure*}

In this Appendix, we present the smoothed nebula images (produced based on the optimally smoothed cube, Fig. \ref{fig:smo_tess_mrc0943}-\ref{fig:smo_tess_4C04}a) and the tessellation maps (Fig. \ref{fig:smo_tess_mrc0943}-\ref{fig:smo_tess_4C04}b)  constructed following Appendix \ref{app:detec_map} (Sect. \ref{sec:2max_neb_ext}) and \ref{app:tesstll} (Sect. \ref{sec:2lyatesmethod}), respectively. The false-colour smooth images are generated using multicolorfits \citep{cigan2019}. Each colour represents a pseudo-narrow band Ly$\alpha$ image constructed from the optimally smoothed cube (Appendix. \ref{app:detec_map}) in arbitrary wavelength range with the goal to show different kinematic structures. Blue, yellow and red are relatively from blue wing, middle and red wing of the Ly$\alpha$ emission, respectively. We note that the smoothed nebula images are only used as representation and demonstration of how the algorithm in Appendix\ref{app:detec_map} works. They are not included in the analysis of this work. For \object{MRC0316-257} and \object{TN J0205+2242}, we detected line emissions around their Ly$\alpha$ wavelength at isolated regions from the main nebula. We did not include these regions in to account due to its possible origin from companion galaxies, but show the spectra and region where they are extracted in Fig. \ref{fig:smo_tess_mrc0316} and Fig. \ref{fig:smo_tess_tnj0205}. 

\section{Fitting procedure}\label{app:spatialfit}
To minimize the uncertainties introduced by fitting the spectra in neighbour bin independently, and impact from foreground targets and the UV continuum, we have developed the following spatial fitting procedure after extracting spectra from the tessellation in Appendix \ref{app:tesstll}. In this way, we can link the fitting between adjacent bins to avoid un-physical jump in parameters:

To estimate the UV continuum , we firstly extract a narrow band image between the rest-frame \ion{N}{v}$\lambda1240$ and \ion{O}{i} $+$ \ion{Si}{ii}$\lambda1305$ from the data cube leaving a rest-frame $10\,\AA$ buffer at each end of the central wavelength of the emission line to avoid line emission contamination. This step is implemented to select which tessellated bins are affected by the continuum such that we can model it in the fitting. Hence, this wavelength range is chosen because it is a line-free region closet to Ly$\alpha$ which could be a relatively accurate estimation of the UV continuum emission of the radio galaxy: since the blue wing of Ly$\alpha$ is suffered from absorption. We use this continuum image and the tessellation map (Sect. \ref{sec:2lyatesmethod}) to determine which tiles have continuum emissions.

To determine the initial value of the spatial fitting, we secondly extract the 1D Ly$\alpha$ master spectrum from a $d=1$ arcsec (25 spaxels) at the position of the AGN center (Table \ref{tab:sampleinfo}) and fit it with Gaussian $+$ Voigt model (Sect. \ref{sec:2fitmodel}, Fig.\ref{fig:map_int_1}a). In this way, the initial information of the absorbers (number, column density and positions) are determined. We note that for this fit a 0th-order polynomial for continuum and two Gaussian for both the systemic emission and broad (or blue/redshifted) components are used.

To minimise the impact of undesired features, we thridly mask the strong $5577\, \AA$ sky-line, which affects \object{4C+03.24} and \object{4C+19.71} most, by 5 spectral resolution units. We also mask the foreground or background objects line emission wavelength range (none of them overlapped with the Ly$\alpha$ of the radio galaxy to our current knowledge) and replace with the noise from the cube variance extension \citep[for example of the reported objects in the vincity of Ly$\alpha$ nebula see, e.g.][]{falkendal2021}. 

To get a the first spatial fit, we fourthly fit the spatial spectra from each tiles in the order determined in Appendix \ref{app:tesstll} with the Least-square method. In the fitting, we pass the results from the previous tile fit to the next as initials to minimize the potential randomness introduced by fitting each spectra freely without any constrain. We note that (i) the constrains from the previous fit may be relaxed more if the ratio of the integrated observed fluxes between the tiles are significant ($\sim50\%$) and/or the distance ratio is large (>1.1)\footnote{Distance ratio, $\frac{r_{1}+r_{2}}{d_{1,2}}$, is the ratio between the sum of the radii, $r_{1}+r_{2}$, of the two tiles (estimated by $r=\sqrt{A/\pi}$, where A is the area of the tile) and the distance between the geometric center of the two tiles, $d_{1,2}$.} and/or the area of the tile is large ($>9\pi$ arcsec$^{2}$); (ii) the Doppler parameter, $b$, \citep[][]{teppergarcia2006} is set to a broad range \citep[][$40<b<400\,\rm{km\,s^{-1}}$]{kolwa2019,wang2021} because of the $b-N$ degeneracy \citep[e.g.][]{silva_2018a}. We only constrain the range of column density, $N_{\rm{\ion{H}{i}}}$. The continuum model is included in the fitting with a 0th-order-polynomial for tiles overlapped with continuum detected region determined in first step. A fixed step function with the step at the wavelength of the systemic Ly$\alpha$ is used for the highest redshift target, \object{4C+04.11}, due to the heavily absorbed continuum on the blue-side which may be due to Ly$\alpha$ forest, e.g. \citet{rauch1998}.

Due to the scattering nature of Ly$\alpha$ and presence of broad (outflow) components, it is important to select which region contains more than one emission component to better study the broader wing. We fit the set of spatial spectra two times: (i) first using 1-Gaussian only for the emission and (ii) then using 2-Gaussian for all spectra. In this way, we can do a simple $\chi^{2}$ ratio selection between the 2-Gaussian vs. the 1-Gaussian fitting results and select the tiles favor 2-Gaussian fit with a threshold of $\sim0.80-0.98$ (depends on different targets). We point that the $\chi^{2}$ value is not a robust measurement of the fitting quality \citep[e.g.][]{andrae2010}. For a quick test in our case, however, it is good enough to the first order. We note that for \object{4C+03.24}, we stay with the 1-Gaussian fit for all tiles due to the complicated spatial variance of spectral shapes. We note that this selection is not crucial in this work since we do not distinguish and separately interpret different velocity components in the analysis. The purpose for this step is to consistently ensure that the non-single-Gaussian line shape is considered without missing flux.

Fifthly, we fit the spatial spectra with 2-Gaussian models for the tiles determined in the last step and 1-Gaussian for the others. The results from the previous steps are passed to the next as initial guesses. To keep consistence, we choose to use the same number of absorbers for the whole map due to the difficulties in determining where one absorber disappears. This is a good assumption to first order given that we observe most absorbers on large extent. For example in Fig. \ref{fig:4c03_spa}, we see the tiles at the nebula edge are also affected by absorbers. The column density of the absorbers is a free parameter during the fit. Given the degeneracy between the column density and Doppler parameter, $b$ \citep[][]{Silva_2018b}, we leave the constraint of $b$ to a broad range following \citet{kolwa2019} ($40$ to $400\,\mathrm{km\,s^{-1}}$). The centroid (redshift) of the absorbers are also allowed to vary ($\Delta z \approx\pm0.001$) except for \object{4C+04.11}\footnote{For this case, we follow the procedure in \citet[][]{wang2021} and fix the positions of five absorbers on the blue low S/N wing to the value determined from the Master spectrum.}. For spectra extracted from inner tiles ($\lesssim 10\arcsec$, with high S/N), we constrain their column densities to vary within a $\sim2$ dex range from the initial input. We ease the column density constrain for the absorbers from outer tiles with distance to AGN $\gtrsim10\arcsec$ such that they can be given low column density ($\sim 10^{13}\,\mathrm{cm^{-2}}$). In this way, the absorbers at low S/N regions can have negligible impact ($<0.1$ \%) on the reconstructed flux. We use the same number of absorbers reported in previous works for some of our targets \citep[\object{MRC 0943-242}, \object{TN J0121+1320}, \object{TN J1338-1942}][]{wilman2004,swinkbank2015}. Otherwise, we use the number determined in the first step of 1D spectra fitting. For \object{TN J0121+1320}, we use 3 absorbers instead of 2 as in \citet{wilman2004}. For \object{4C+04.11}, we use 9 absorbers (instead of 7 as in \citet{wang2021}) with 7 of them fixed to the redshifts determined in 1D spectra fit.

Finally, we run MCMC sampling \citep[][]{foremanmackey2013} using the results from last step as initials and with larger boundaries to probe the probability distribution of the fitted parameters and uncertainties.

 We point out one caveat that the low S/N (especially for the tiles at the edge) will affect the reconstructed intrinsic Ly$\alpha$ flux. We implemented a test where we artificially decrease the S/N of the master spectrum and do the absorption correction fit. It shows that the reconstructed intrinsic flux can vary by a factor of $\sim2$ when the S/N decreases by one dex (e.g. $\sim100$ to $\sim10$). This result is based on the fact that we have a relatively good initial guess for the fit. We note that in the aforementioned fitting procedure the fitting parameters are constrained by the neighbouring tiles to have them physically linked. There may still be the chance that the low S/N will affect the fitted flux leading to over-correction. This may be the case for \object{4C+04.11} (Fig. \ref{fig:hist_FOV_tile}). We present the flux ratio maps between intrinsic and observed in Appendix \ref{app:suppmap} which indicate this possibility ($\sim50$ for the tiles at the edge).
\section{Supplementary maps}\label{app:suppmap}

  \begin{figure*}
  \centering
        \includegraphics[width=\textwidth,clip]{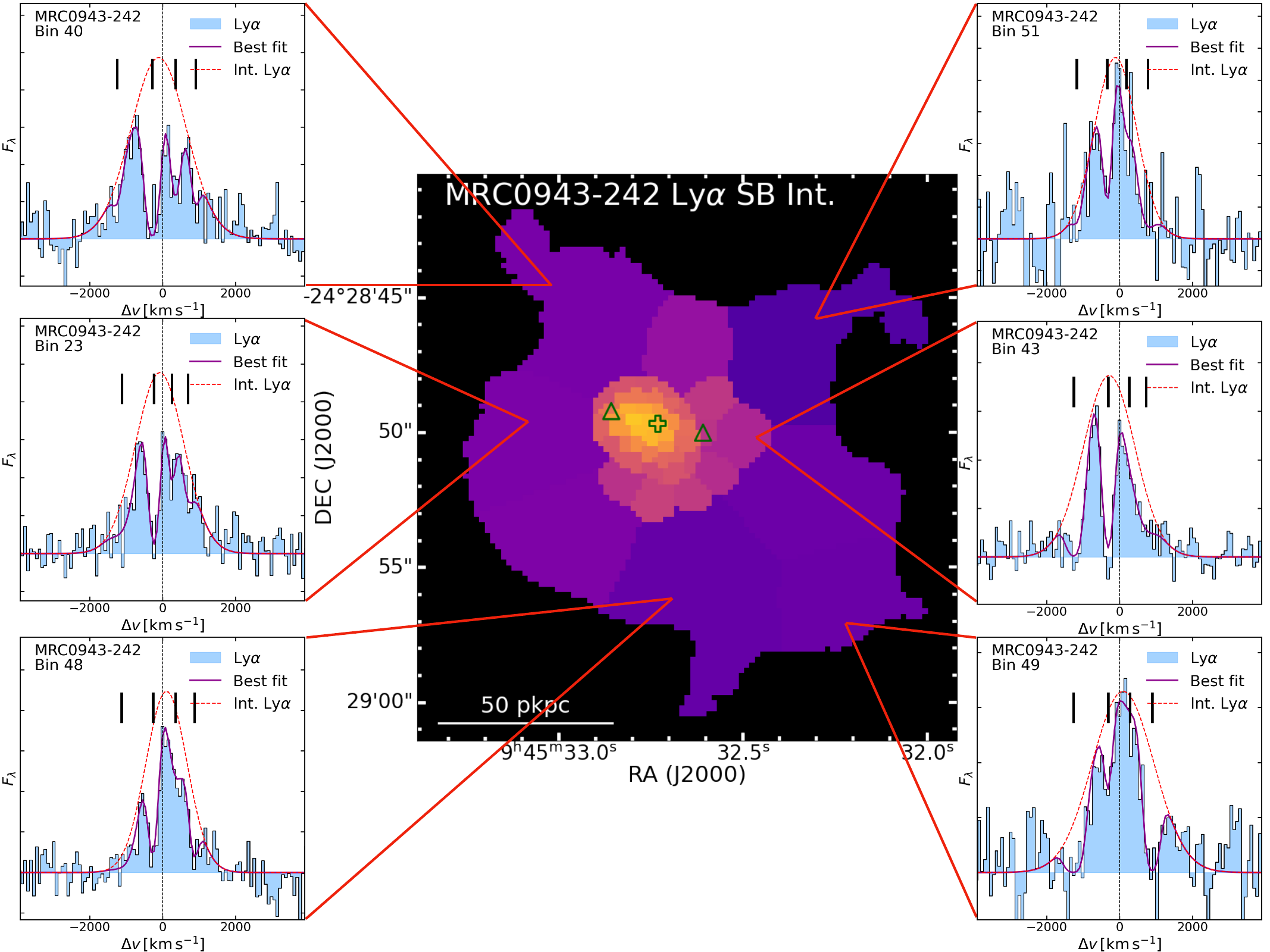}
      \caption{Similar figure as Fig. \ref{fig:tnj0205_spa}$-$\ref{fig:4c03_spa} for \object{MRC0943-242}.}
        \label{fig:mrc0943_spa}
         
  \end{figure*}

  \begin{figure*}
  \centering
        \includegraphics[width=\textwidth,clip]{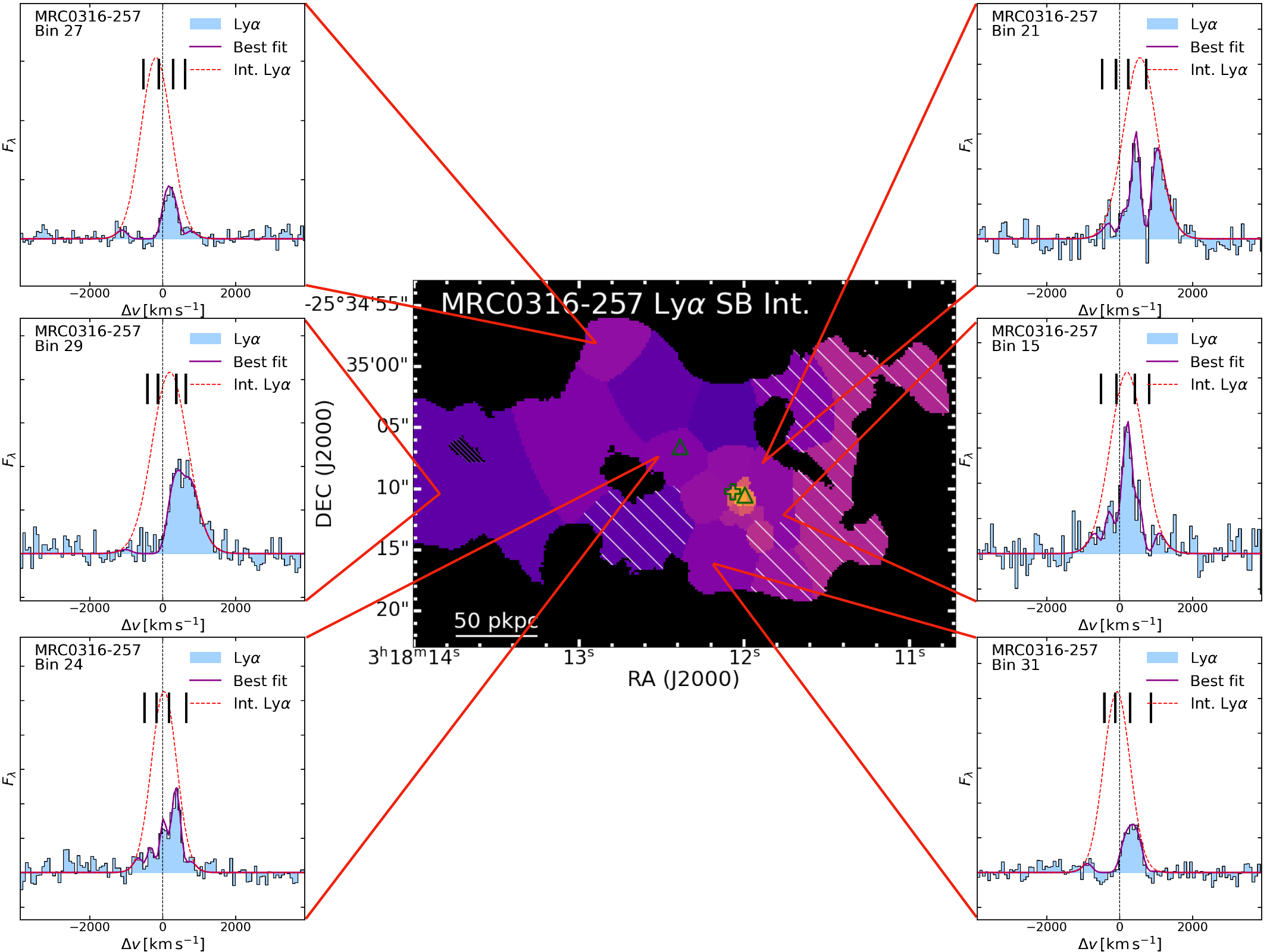}
      \caption{Similar figure as \ref{fig:tnj0205_spa}$-$\ref{fig:4c03_spa} for \object{MRC0316-257}.}
        \label{fig:mrc0316_spa}
         
  \end{figure*}

  \begin{figure*}
  \centering
        \includegraphics[width=\textwidth,clip]{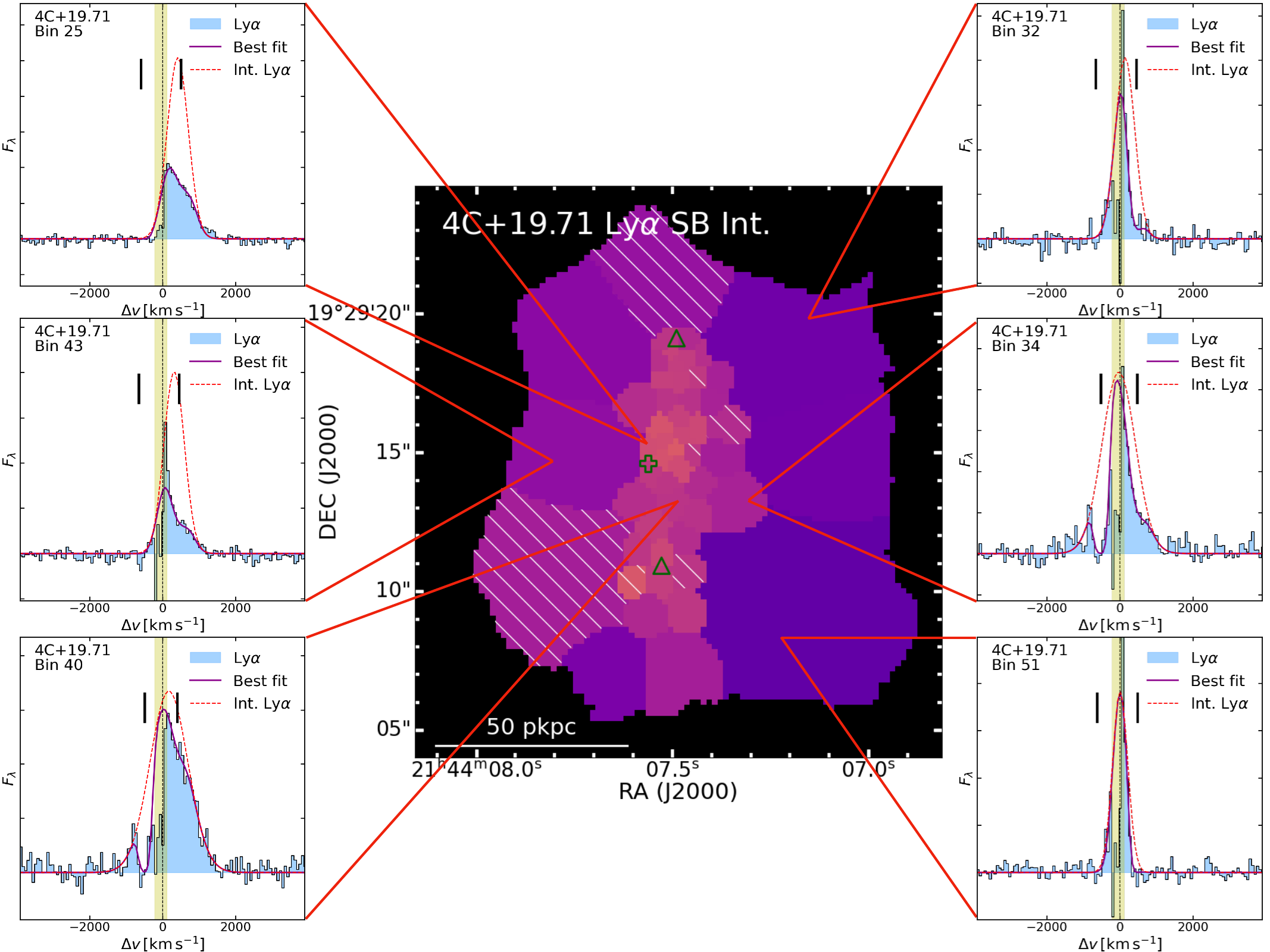}
      \caption{Similar figure as \ref{fig:tnj0205_spa}$-$\ref{fig:4c03_spa} for \object{4C+19.71}.}
        \label{fig:4C19_spa}
         
  \end{figure*}

  \begin{figure*}
  \centering
        \includegraphics[width=\textwidth,clip]{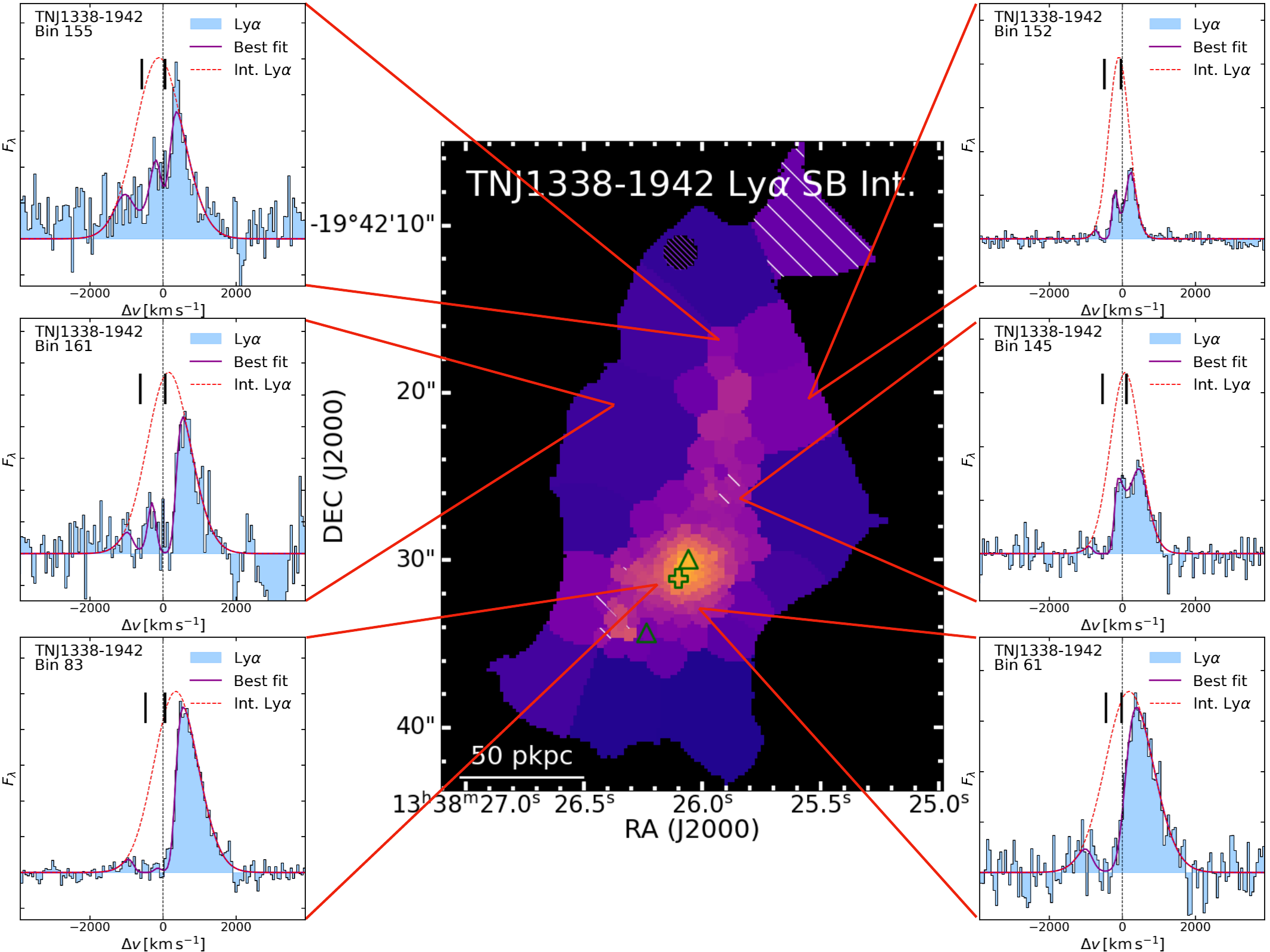}
      \caption{Similar figure as \ref{fig:tnj0205_spa}$-$\ref{fig:4c03_spa} for \object{TNJ1338-1942}.}
        \label{fig:tnj1338_spa}
         
  \end{figure*}

  \begin{figure*}
  \centering
        \includegraphics[width=\textwidth,clip]{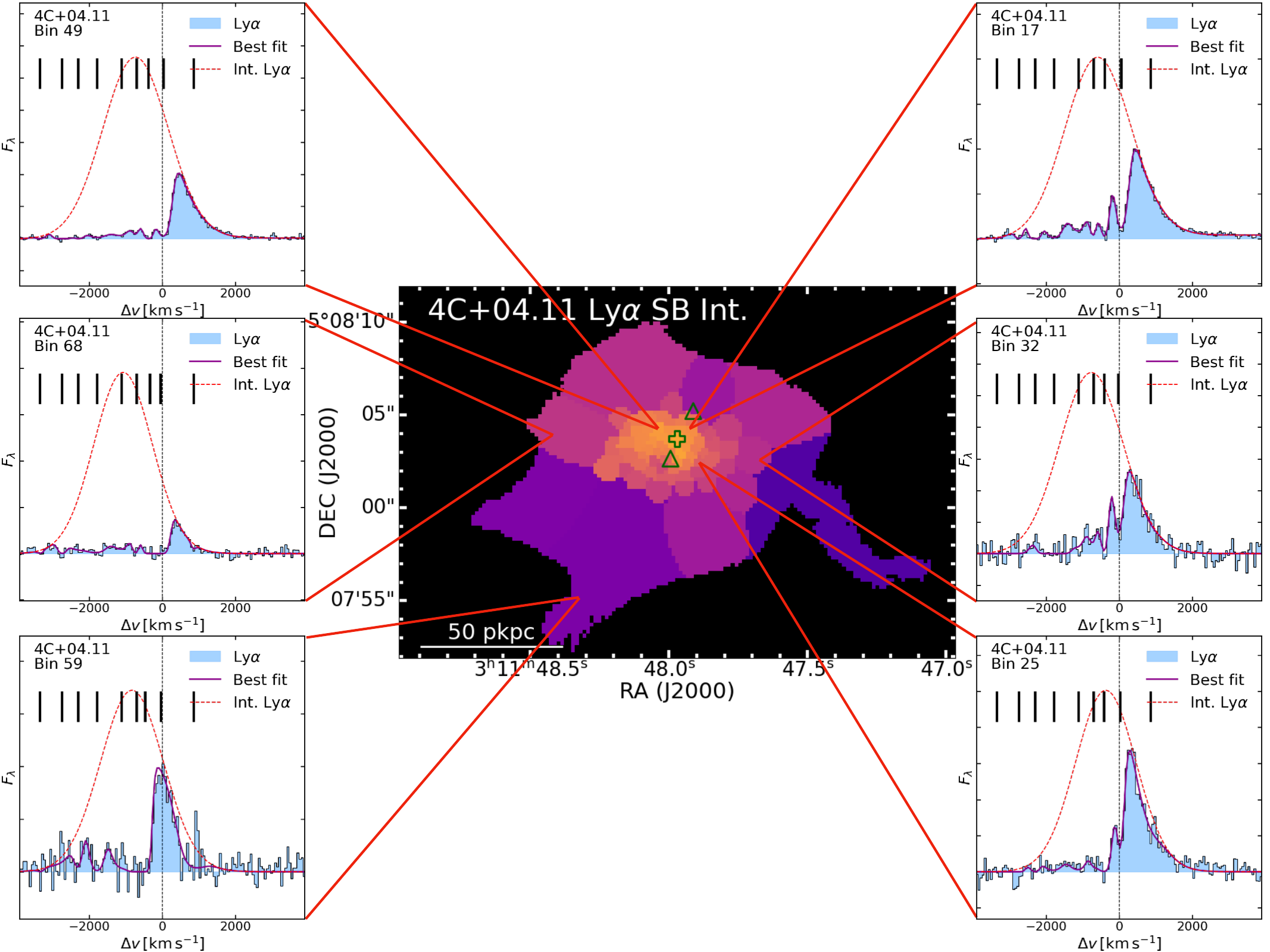}
      \caption{Similar figure as \ref{fig:tnj0205_spa}$-$\ref{fig:4c03_spa} for \object{4C+04.11}.}
        \label{fig:4c04_spa}
         
  \end{figure*}


   \begin{figure*}
  \centering
        \includegraphics[width=\textwidth,clip]{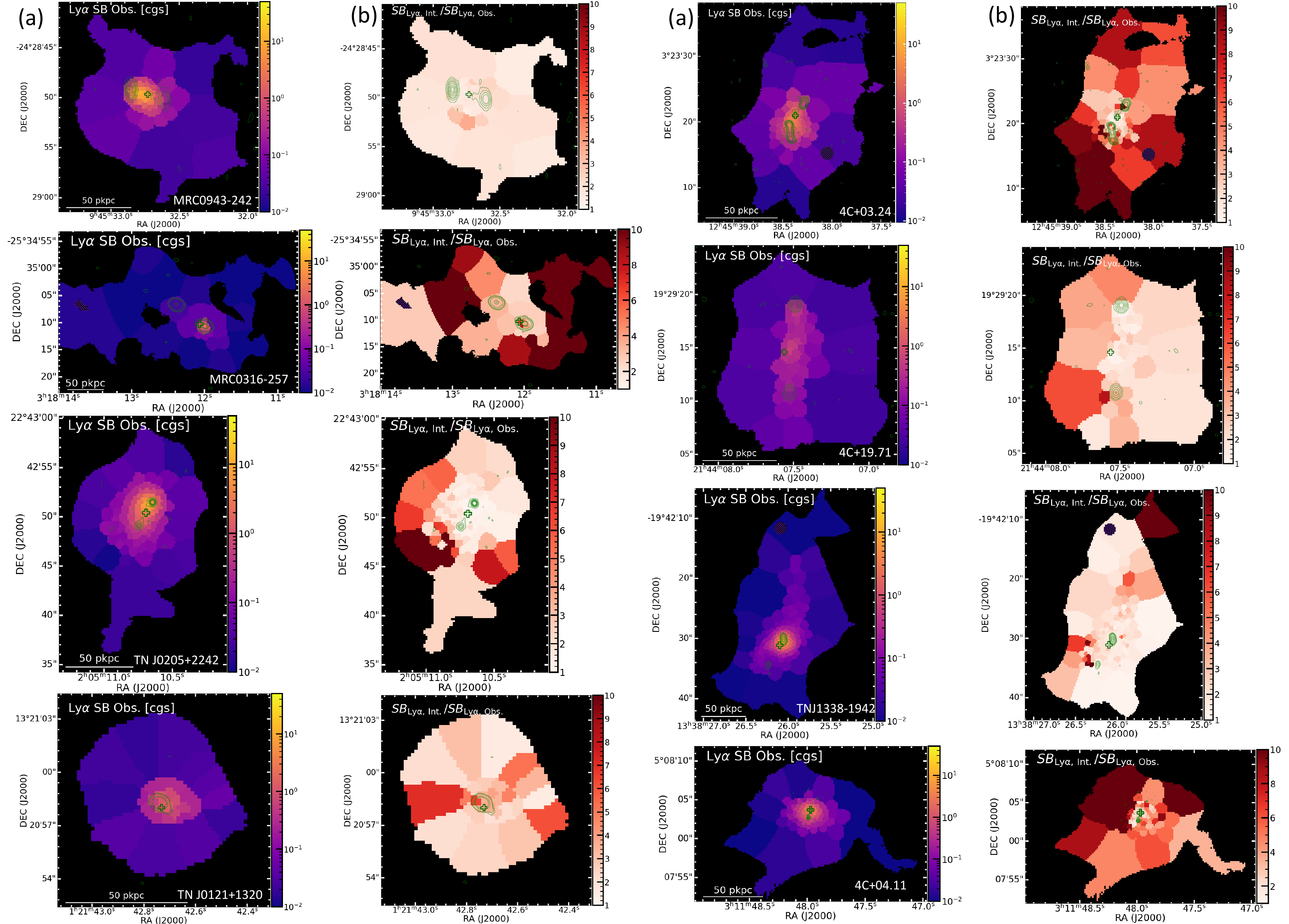}
      \caption{(a) Observed surface brightness maps in the unit of $10^{-16}\,\mathrm{erg\,s^{-1}\,cm^{-2}\,arcsec^{-2}}$. (b) Flux ratio maps of intrinsic and observed surface brightness. The green cross and contours are the same as Fig. \ref{fig:map_int_1} for individual targets, respectively.}
      \label{fig:suppmap}
   \end{figure*}

We show the intrinsic Ly$\alpha$ maps with selected spectra for \object{MRC0943-242}, \object{MRC0316-257}, \object{4C+19.71}, \object{TNJ1338-1942} and \object{4C+04.11} in Fig. \ref{fig:mrc0943_spa},  \ref{fig:mrc0316_spa}, \ref{fig:4C19_spa}, \ref{fig:tnj1338_spa} and \ref{fig:4c04_spa}, respectively. The spatial study of the Ly$\alpha$ nebulae for these sources with MUSE were published previously \citep[e.g.][]{Gullberg_2016a,vernet2017,falkendal2021,swinkbank2015,wang2021}. In this paper, we performed a consistent analysis for the full sample with the new method of correcting for absorption. With the individual spectra mostly showing the tiles at larger distance away from the AGN, we can find that the absorption features are observed nearly across the entire nebula.

To better show the difference before and after absorption correction, we present the observed surface brightness maps and flux ratio maps between the intrinsic flux and observed flux. For the observed flux in each tiles, we use the flux integrated from $v_{05}$ to $v_{95}$ and show in Fig. \ref{fig:suppmap}a for each target. For consistency, the $v_{05}$ (and $v_{95}$) is determined based on the intrinsic line since the $v_{05}$ (and $v_{95}$) of the observed lines in low S/N regions are affected by noise more. To ensure the intrinsic and observed surface brightness are comparable as in Fig. \ref{fig:suppmap}b, we take the ratio between the intrinsic and observed flux all integrated between $v_{05}$ and $v_{95}$. Generally, the tiles at the outskirt of the nebulae having larger ratios ($>10$). The observed nebular properties derived in the main text are based on the $v_{05}-v_{95}$ observed maps obtained here.

\section{Supplements of nebula radial profile and morphology analysis}\label{app:nebradinfo}
\begin{table}[!h]
\caption{Position angle and hot spot distance of the radio jet.}\label{tab:radioPA}
\centering
\begin{tabular}{ccll}
\hline
HzRG & PA$_{\rm radio}$\tablefootmark{\rm{\dag}}  &$r_{\rm app.}$\tablefootmark{\rm{*}}  & $r_{\rm rec.}$\tablefootmark{\rm{*}} \\
&deg & arcsec & arcsec\\
\hline
MRC 0943-242 & 74 & 1.79 & 1.76 \\
MRC 0316-257 & 53 & 5.56 & 1.15\\
TN J0205+2242 & 150 & 1.47 & 1.24\\
TN J0121+1320 & 90 & 0.82\tablefootmark{a} & 0.82\tablefootmark{a}\\
4C+03.24 & 146 & 2.00\tablefootmark{b} & 1.75\tablefootmark{b}\\
4C+19.71 & 176& 3.82 & 4.58\\
TN J1338-1942 & 150& 3.78 & 1.18\\
4C+04.11 & 158& 1.15\tablefootmark{c} & 1.70\tablefootmark{c}\\
\hline
\end{tabular}
\tablefoot{\tablefoottext{\rm{\dag}}{Jet position angle. This is measured east from north. We note that we do not quantify the uncertainties which could be $\pm1$ deg.} 
\tablefoottext{\rm{*}}{Distance between the hot spot and the AGN position. $r_{\rm app.}$ and $r_{\rm rec.}$ are distance for approaching and receding hot spot, respectively.}
\tablefoottext{\rm{a}}{The radio emission of \object{TN J0121+1320} is spatially unresolved. The `hot spot' distance is represented by the distance from the host galaxy position to the either side of $3\sigma$ contour of the radio image along the east-west direction.}\tablefoottext{\rm{b}}{Approaching hot spot is A1 and receding hot spot is B1 \citep[named after][]{vanojik1996}.} \tablefoottext{\rm{c}}{Approaching hot spot is knot8 and receding hot spot is knot1 \citep[named after][]{parijskij2014}.} }

 
\end{table}
   \begin{figure*}
  \centering
        \includegraphics[width=\textwidth,clip]{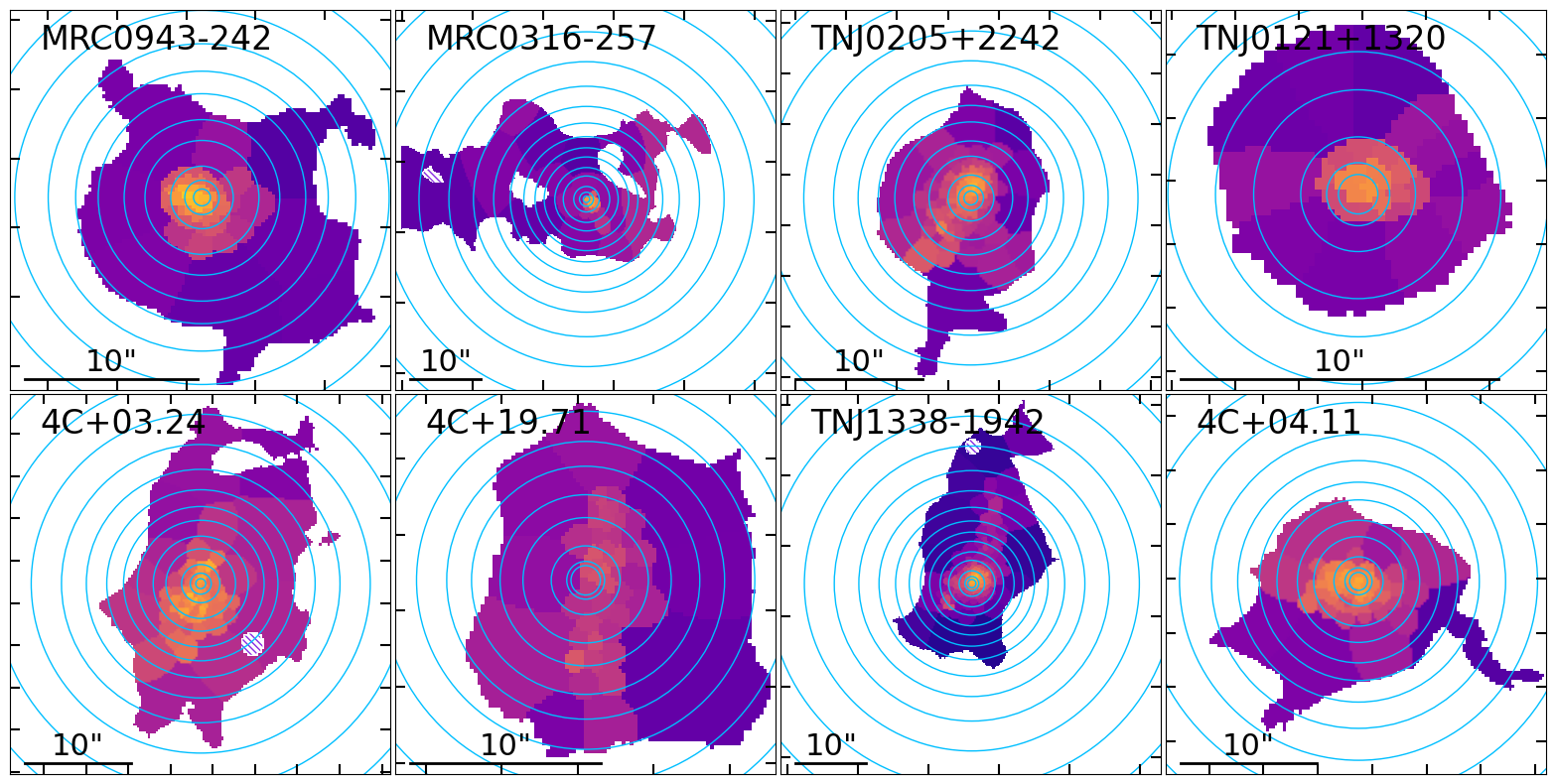}
      \caption{Circular apertures for radial profile (Sect. \ref{sec:2rad_sb_pro}) extraction on top of intrinsic surface brightness maps (Fig. \ref{fig:map_int_1}a) of the HzRGs sample. In each panel, the blue concentric annuli show the apertures where the radial profile is extracted. The annuli are centered around the host galaxy (central AGN) position (Table \ref{tab:sampleinfo}). The radii of the annuli in each panel are the same in the unit of arcsec for consistence except the smallest radius which is set to be $0.75\times$ seeing for each target. The black bar at the bottom left corner of each panel indicate the 10 arcsec scale.}
         \label{fig:mapradcir}
   \end{figure*}


   \begin{figure*}
  \centering
          \includegraphics[width=\textwidth,clip]{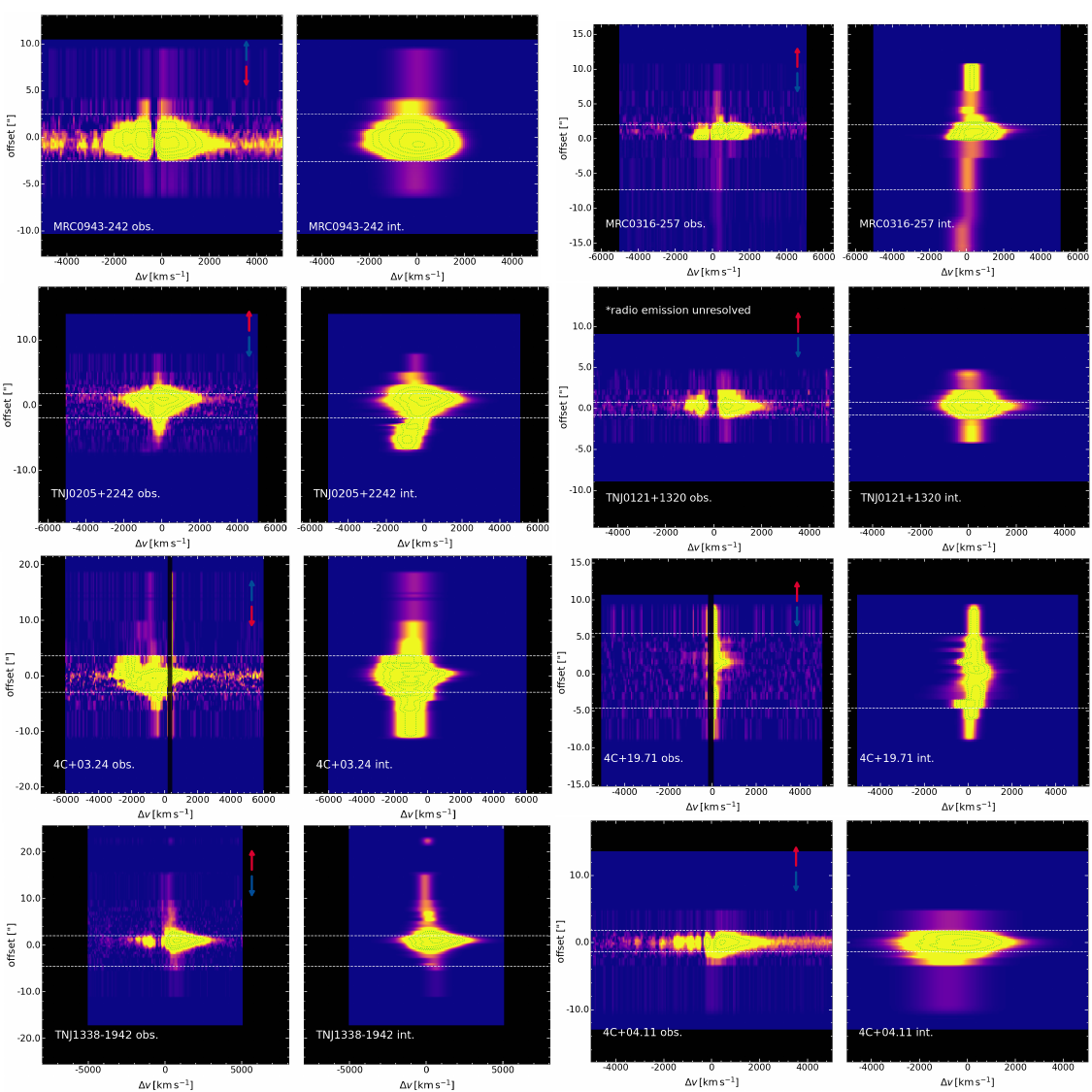}
      \caption{Ly$\alpha$ position-velocity diagrams (i.e. 2D spectra) of our sample targets extracted along the radio jet axis. For each target, the left panel shows the diagram constructed from tessellated observed cube (not continuum-subtracted for the host galaxy) and the right panel shows the diagram from absorption-corrected intrinsic cube. The zero offset is set to the position of the central AGN (Table \ref{tab:sampleinfo}). The direction of approaching and receding sides of the jet (see Sect. \ref{sec:2jetk}) are marked in the left panel by the blue and red arrows, respectively. The white horizontal dashed lines represent the furthest extent of the jet. The dotted green contours are given in arbitrary steps which are used to guide the eye for the high brightness part. The vertical black shaded regions in the observed position-velocity diagrams of \object{4C+03.24} and \object{4C+19.71} indicate the wavelength ranges affected by the $5577\,\AA$ sky-line.}
         \label{fig:pvdiagram}
   \end{figure*}

In this appendix, we show the supplementary information accompanied with studying nebulae radial profiles and morphology. 

We present the annuli used in Sect. \ref{sec:2rad_sb_pro} in Fig. \ref{fig:mapradcir}. For each targets, the smallest one aperture has the radius equals to 0.75 of its seeing. The radii are in logarithmic steps from centre to outskirt, and the values extract from the annuli are presented in Table \ref{tab:Int_SB}.

For the directional radial surface brightness profile analysed in Sect. \ref{sec:2dir_sb_prof}, it is extracted in half annuli along approaching and receding directions (Fig. \ref{fig:mapradind}) with the same step as Fig. \ref{fig:mapradcir}. 

The position angle of the radio axis is obtained from the two jet hot-spot positions \citep[e.g.][and unpublished radio maps]{vanojik1996,Carilli1997,Pentericci2000,parijskij2014} and presented in Table \ref{tab:radioPA}. For \object{TN J0121+1320} which has a compact radio emission, we assign an east-west jet position angle to it. We also present the distance of the radio hot spot from the AGN position in Table \ref{tab:radioPA}. This is the value shown in Fig. \ref{fig:rad_pf_fit} and \ref{fig:v50_W80_r} (after converted to ckpc). The hot spot is determined to be located at position of the brightest radio emission. \object{4C+03.24} and \object{4C+04.11} show multiple radio flux peaks in their radio data. We calculate the jet position angle and hot spot distance based on A1 and B1 knots for \object{4C+03.24} \citep[named after][]{vanojik1996} and knots 1 and 8 for \object{4C+04.11} \citep[named after][]{parijskij2014}. We note that the radio jet of \object{TN J0121+1320} is unresolved. Hence, we use the farthest distance reached by the jet (i.e. $3\sigma$ contour of the radio image) along the jet direction as a proxy. 

We show the exponential and power law fitted parameters (Sect. \ref{sec:2sb_prof_fit}) of the circularly averaged and directional surface brightness profiles in Table \ref{tab:SBfit}. In Fig. \ref{fig:pvdiagram}, we show the position-velocity diagram of the observed and intrinsic cubes along the jet direction. This can be used as a direct comparison with \citet{villarmartin2003}. On each target, we also mark the largest extent of the radio lobes in both directions in horizontal dashed lines. This is determined as the distance from the AGN position along the direction of the jet position angle to the farthest position reached by the $3sigma$ radio flux contour. We note that the broader line width observed visually is due to the contrast between the high and low surface brightness parts. We find in some cases (e.g. \object{4C+04.11} and \object{TN J1338-1942}) that there is a sharp surface brightness drop at the distance of the radio jet boundary. A discontinuity in the line width and surface brightness of the Ly$\alpha$ nebulae across the extent of the radio source has been previously reported in \citet{villarmartin2003}. The changes is not as sharp as shown in \ref{fig:pvdiagram}. The sharpness seen in the intrinsic panels is due to tessellation as it is the transition from high to low surface brightness part. Hence, we assure that these are not all artificial effect due to our analysis methods.

\begin{sidewaystable*}
\caption{Intrinsic surface brightness from circular aperture}\label{tab:Int_SB}
\centering
\begin{tabular}{crrrrrrrr}
\hline             
R & MRC0943-242 & MRC0316-257 & TN J0205+2242 & TN J0121+1320 & 4C+03.24 & 4C+19.71 & TN J1338-1942 & 4C+04.11 \\
$[\rm{arcsec}]$ & \multicolumn{8}{c}{$(1+z)^{4}\rm{SB}\,[10^{-16}\,{\rm erg\,s^{-1}\,cm^{-2}\,arcsec^{-2}}]$} \\
\hline
0.0-$r_{\rm s}$\tablefootmark{\rm{\dag}}  & 2621.4 & 1534.1 & 1886.6 & 1452.9 & 2955.1 & 261.9 & 5177.9 & 6464.6 \\
$r_{\rm s}$\tablefootmark{\rm{\dag}} -1.0 & 1792.0 & 833.8 & 1764.7 & 889.7 & 2102.8 & 224.9 & 3617.2 & 4686.2 \\
1.0-1.8 & 623.3 & 390.1 & 953.4 & 250.2 & 1884.4 & 195.4 & 1829.5 & 2659.6 \\
1.8-3.3 & 95.3 & 60.9 & 232.4 & 48.5 & 1020.3 & 130.1 & 385.0 & 950.4 \\
3.3-4.5 & 24.6 & 36.4 & 120.3 & 38.6 & 465.3 & 105.4 & 106.4 & 367.7 \\
4.5-6.0 & 15.2 & 31.5 & 97.2 & 25.4 & 227.4 & 68.3 & 58.4 & 229.6 \\
6.0-7.3 & 14.7 & 32.2 & 87.1 & 0.0 & 194.8 & 51.3 & 29.2 & 168.9 \\
7.3-8.9 & 13.1 & 35.9 & 45.0 & 0.0 & 142.9 & 43.0 & 25.0 & 97.6 \\
8.9-10.8 & 11.9 & 38.5 & 23.3 & 0.0 & 117.3 & 25.1 & 26.3 & 65.5 \\
10.8-13.2 & 13.0 & 37.2 & 23.3 & 0.0 & 82.8 & 20.7 & 33.9 & 58.3 \\
13.2-16.0 & 0.0 & 38.3 & 23.3 & 0.0 & 65.0 & 0.0 & 33.9 & 31.2 \\
16.0-19.5 & 0.0 & 33.0 & 0.0 & 0.0 & 45.5 & 0.0 & 17.0 & 0.0 \\
19.5-23.8 & 0.0 & 14.5 & 0.0 & 0.0 & 0.0 & 0.0 & 29.2 & 0.0 \\
23.8-28.9 & 0.0 & 12.3 & 0.0 & 0.0 & 0.0 & 0.0 & 36.6 & 0.0 \\
\hline
\end{tabular}
\tablefoot{\tablefoottext{\rm{\dag}}{$r_{\rm s} = 0.75\times$ seeing which is specified for each target Table \ref{tab:sampleobs}. The median seeing is 0.75 arcsec.}\\
Intrinsic surface brightness values extracted from circular annuli following Fig. \ref{fig:mapradcir}. To be consistent, the first column gives the radii of each annulus in the unit of arcsec instead of ckpc. The reported surface brightness values are corrected for cosmological dimming. 
}
\end{sidewaystable*}

\begin{table*}
\caption{Fit parameters of surface brightness profiles.}\label{tab:SBfit}
\centering
\begin{tabular}{ccccc}
\hline
HzRG & $r_{\rm h}$\tablefootmark{\rm{\dag}} & $C_{\rm e}$\tablefootmark{\rm{*}} & $r_{\rm b}$\tablefootmark{\rm{\ddag}} & $\alpha$\tablefootmark{\rm{$\mathsection$}}\\
     & pkpc        & $10^{-16}$ cgs & pkpc        & \\
\hline
MRC 0943-242 & 6.9$\pm$0.2  & 20.7$\pm$1.0  & 39.8$\pm$1.5  & -0.24$\pm$0.18\\ 
" " app.     & 5.6$\pm$0.2  & 18.4$\pm$1.1  & 32.0$\pm$1.3  & -0.13$\pm$0.15\\ 
" " rec.     & 7.5$\pm$0.2  & 25.8$\pm$1.0  & 42.9$\pm$1.3  & -0.15$\pm$0.19\\ 
MRC 0316-257 & 7.0$\pm$0.6  & 8.9$\pm$1.2   & 27.3$\pm$3.2  & -0.62$\pm$0.24 \\
" " app.     & 1.3$\pm$0.3  & 69.0$\pm$70.0 & 7.7$\pm$1.0   & -0.46$\pm$0.07\\
" " rec.     & 7.5$\pm$3.3  & 14.2$\pm$15.7 & 31.6$\pm$14.4 & -0.01$\pm$0.49 \\
TN J0205+2242& 48.9$\pm$30.4& 5.7$\pm$0.4   & 10.9$\pm$21.4 & -1.90$\pm$0.58 \\
" " app.     & 8.8$\pm$2.0  & 6.2$\pm$0.7   & 13.4$\pm$10.8 & -1.52$\pm$0.26 \\
" " rec.     & 10.8$\pm$3.1 & 8.3$\pm$0.3   & 48.7$\pm$0.7  & -0.97$\pm$0.14\\
TN J0121+1320& 9.8$\pm$1.9  & 5.6$\pm$1.2   & 6.2$\pm$10.0  & -2.05$\pm$0.71\\
" " app.     & 4.4$\pm$1.0  & 8.9$\pm$2.3   & 19.6$\pm$6.6  & -0.67$\pm$0.76 \\
" " rec.     & 5.1$\pm$1.3  & 10.2$\pm$2.1  & 10.9$\pm$9.3  & -2.16$\pm$0.75 \\
4C+03.24      & 17.6$\pm$1.8 & 7.9$\pm$0.6   & 20.0$\pm$14.9 & -1.64$\pm$0.21 \\
" " app.     & 11.2$\pm$0.9 & 10.3$\pm$0.9  & 38.5$\pm$4.2  &-0.87$\pm$0.18 \\
" " rec.     & 27.7$\pm$2.0 & 6.5$\pm$0.4   & 23.4$\pm$10.8 &-1.67$\pm$0.18 \\
4C+19.71      & 29.5$\pm$7.2 & 0.8$\pm$0.1   & 23.1$\pm$9.7  &-1.20$\pm$0.16 \\
" " app.     & 63.6$\pm$20.3& 0.4$\pm$0.1   & 38.1$\pm$7.8  &-1.88$\pm$0.37 \\
" " rec.     & 23.3$\pm$24.5& 1.0$\pm$0.2   & 15.1$\pm$15.3 &-1.24$\pm$0.35 \\
TN J1338-1942& 7.3$\pm$0.3  & 13.6$\pm$0.7  & 41.0$\pm$2.7  &-0.12$\pm$0.26 \\
" " app.     & 3.6$\pm$7.0  & 20.7$\pm$8.8  & 7.5$\pm$1.7   &-2.10$\pm$0.07 \\
" " rec.     & 7.3$\pm$8.0  & 19.0$\pm$5.1  & 40.0$\pm$14.7 &-0.52$\pm$0.77 \\
4C+04.11      & 8.2$\pm$1.5  & 11.8$\pm$1.3  & 15.2$\pm$11.7 &-2.04$\pm$0.24 \\
" " app.     & 9.9$\pm$1.0  & 11.0$\pm$0.9  & 38.9$\pm$12.0 &-1.77$\pm$0.37 \\
" " rec.     & 7.5$\pm$1.6  & 11.6$\pm$1.6  & 16.0$\pm$8.3  &-2.14$\pm$0.28 \\
\hline
\end{tabular}
\tablefoot{
See Section \ref{sec:2sb_prof_fit} for the fitting equations (Eq. \ref{eq:exp_pl_piece}):
\tablefoottext{\rm{\dag}}{Scale length of the exponential profile.}
\tablefoottext{\rm{*}}{Normalisation parameter of the exponential profile. The cgs unit is $\mathrm{erg\,s^{-1}\,cm^{-2}\,arcsec^{-2}}$.}
\tablefoottext{\rm{\ddag}}{Distance where inner exponential profile changes to power law profile.}
\tablefoottext{\rm{$\mathsection$}}{Power law index.}
}
\end{table*}

\section{MRC0316-257 systemic redshift}\label{app:z_sys0316}
  \begin{figure*}
  \centering
        \includegraphics[width=16.4cm,clip]{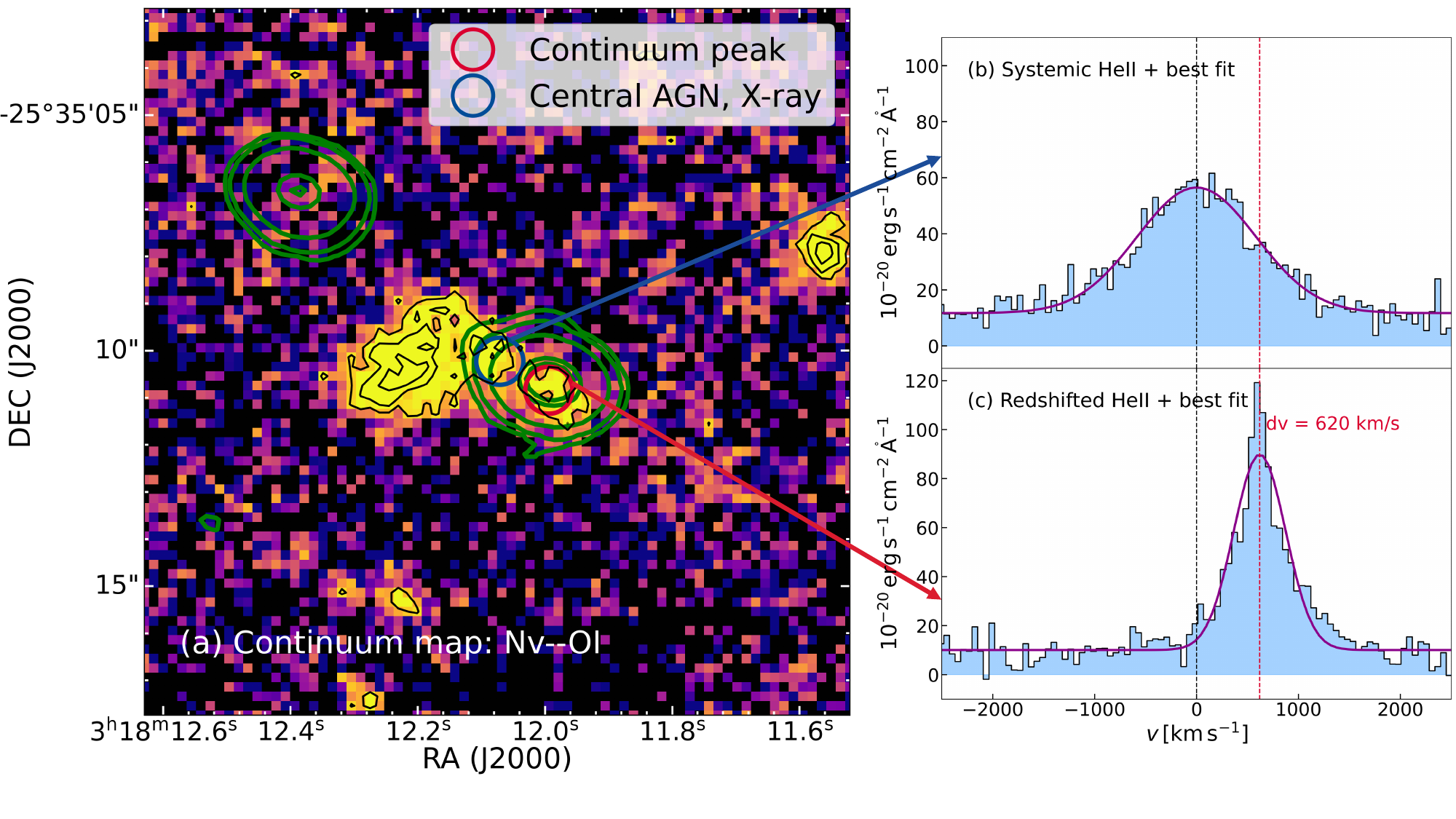}
      \caption{UV continuum map around \object{MRC0316-257} (a) and \ion{He}{ii} spectra from the X-ray position (central AGN, b) and UV continuum peak position (c). The UV continuum map is collapsed between the observed wavelength of \ion{N}{v}$\lambda1240$ and \ion{O}{i}+\ion{Si}{ii}$\lambda1305$. The green contours show the radio jet in the same format as Fig. \ref{fig:map_int_1}cd. The black contours in the step of [3$\sigma$, 5$\sigma$, 7$\sigma$, ...] trace the UV continuum emission, where $\sigma$ is the background standard deviation. Blue and red circular regions indicate the $r=0.5\,\rm{arcsec}$ apertures where the systemic and redshifted \ion{He}{ii} spectra are extract, respectively. The right panels (b, c) show the \ion{He}{ii} spectra (histogram) along with their best Gaussian fitting (dark magenta line) results. The velocity zero (vertical black dotted line) in both panels is the systemic redshift. In the panel(c), we also mark the velocity shift (vertical red dotted line) of the redshifted \ion{He}{ii} emission with respect to the systemic one.}
         \label{fig:mrc0316_heii}
  \end{figure*}

For \object{MRC0316-257}, two velocity components of the \ion{He}{ii} emission are detected in our MUSE observation which are also separated spatially. The one detected at the position of the X-ray emission peak from Chandra observation (Table \ref{tab:sampleinfo}) is believed to be the systemic one while the one at north-west position that is coincident with UV continuum emission peak may trace jet-gas interaction. We show the UV continuum map of \object{MRC0316-257} and the fits of the two \ion{He}{ii} spectra in Fig. \ref{fig:mrc0316_heii}. We note that the UV continuum map is constructed from the MUSE cube using the wavelength in observed frame between  \ion{N}{v}$\lambda1240$ and \ion{O}{i}+\ion{Si}{ii}$\lambda1305$ which is a emission-line-free region of HzRGs \citep[][]{McCarthy1993}. The bright continuum emission object east of the central AGN position peak is a foreground galaxy \citep[][]{vernet2017}. We report here that the systemic redshift detected is $z_{\rm sys} = 3.1238\pm0.0002$ and the redshift of the component at the UV continuum peak is $z_{\rm red} = 3.1323\pm0.0002$ which is redshifted of $v=620\,\rm{km\,s^{-1}}$ from the systemic one. The velocity gradient of the \ion{He}{ii} agrees with the Ly$\alpha$ $v_{50}$ (Fig. \ref{fig:map_int_1}c) and [\ion{O}{iii}]$\lambda5007$ \citep[][]{nesvadba2008b} within the scope of the jet. The UV continuum at this position may suggest the younger stellar population distribution. Combine with the jet kinematics, we may seeing jet induced star forming activities. However, there is also the possibility that the UV continuum could be produced by the inverse Compton processes. The redshifted \ion{He}{ii} near the west radio lobe could then be due to the ionisation emission from the shock region exerting on the un-shocked gas. This is supported by the relatively narrow with of the redshift \ion{He}{ii} (FWHM $\sim600\,\mathrm{km\,s^{-1}}<1000\,\mathrm{km\,s^{-1}}$) which could indicate that is has not been impacted by shocks \citep{Best_2000d,Allen2008}. We note that a detailed verification for this scenario is beyond the scope of this paper which involves spectral aging inspection of the radio jet hot-spot \citep[e.g.][]{Harwood_2013}. Hence, we simply point these possibilities and leave them to future study with multi-wavelength observations combined.



\end{appendix}

\end{document}